\RequirePackage{fix-cm}
\documentclass[smallextended]{svjour3}       % onecolumn (second format)
\smartqed  % flush right qed marks, e.g. at end of proof
\usepackage{graphicx}
\usepackage{url}
\usepackage{ulem}
\usepackage{mathptmx}      % use Times fonts if available on your TeX system
\usepackage{epstopdf}
\usepackage{amsmath}
\usepackage{color}
\usepackage{multirow}
\usepackage{appendix}
\usepackage{bm}
\journalname{Journal of Fusion Energy}
\begin{document}

\title{MHD analysis on the physical designs of CFETR and HFRC%\thanks{Grants or other notes
}


\author{Ping Zhu$^{1,2,*}$     \and
%RWM        
        Li Li$^{3}$            \and
        Yu Fang$^{4}$          \and
        Yuling He$^{5}$       \and
        Shuo Wang$^{6}$        \and
        Rui Han$^{7}$          \and
        Yue Liu$^{4}$          \and
%ASIPP NTM
        Xiaojing Wang$^{8}$    \and
        Yang Zhang$^{8}$       \and
        Xiaodong Zhang$^{8}$   \and
        Qingquan Yu$^{9}$      \and
        Liqun Hu$^{8}$         \and
%ASIPP EF       
        Huihui Wang$^{8}$      \and
        Youwen Sun$^{8}$       \and
%DLUT NTM        
        Lai Wei$^{4}$          \and
        Weikang Tang$^{4}$     \and
        Tong Liu$^{4}$         \and
        Zhengxiong Wang$^{4}$  \and
%Ustc NTM        
        Xingting Yan$^{7}$     \and       
        Wenlong Huang$^{10}$    \and
        Yawei Hou$^{7}$        \and        
        Xiaoquan Ji$^{6}$      \and
%MGI            
        Shiyong Zeng$^{7}$     \and
        Zafar Abdullah$^{7}$   \and
        Zhongyong Chen$^{8}$   \and
        Long Zeng$^{8}$        \and
%HFRC
        Haolong Li$^{11}$      \and
        Zhipeng Chen$^{1}$     \and
        Zhijiang Wang$^{1}$    \and
        Bo Rao$^{1}$           \and
        Ming Zhang$^{1}$       \and
        Yonghua Ding$^{1}$     \and
        Yuan Pan$^{1}$         \and
        the CFETR Physics Team and the HFRC Physics Team
}
%\address{International Joint Research Laboratory of Magnetic Confinement Fusion and Plasma Physics, State Key Laboratory of Advanced Electromagnetic Engineering and Technology, School of Electrical and Electronic Engineering, Huazhong University of Science and Technology, Wuhan, Hubei 430074, China}
%\address{Department of Engineering Physics, University of Wisconsin-Madison, Madison, Wisconsin 53706, USA}


\institute{Ping. Zhu \at
              Tel.: +86-027-87793005-209\\
              Fax: +86-027-87793005\\
              \email{zhup@hust.edu.cn}  \\
\\
1.~International Joint Research Laboratory of Magnetic Confinement Fusion and Plasma Physics, State Key Laboratory of Advanced Electromagnetic Engineering and Technology, School of Electrical and Electronic Engineering, Huazhong University of Science and Technology, Wuhan, Hubei 430074, China \\
2.~Department of Engineering Physics, University of Wisconsin-Madison, Madison, Wisconsin 53706, USA  \\
3.~College of Science, Donghua University, Shanghai 201620, China \\
4.~Key Laboratory of Materials Modification by Laser, Ion, and Electron Beams (Ministry of Education), School of Physics, Dalian University of Technology, Dalian 116024, China \\
5.~Guangdong Provincial Key Laboratory of Quantum Engineering and Quantum Materials, School of Physics and Telecommunication Engineering, South China Normal University, Guangzhou 510006, China.  \\
6.~Southwestern Institute of Physics, P.O. Box 432, Chengdu 610041, China \\
7.~CAS Key Laboratory of Geospace Environment and Department of Engineering and Applied Physics, University of Science and Technology of China, Hefei 230026, China \\
8.~CAS Institute of Plasma Physics, Hefei, Anhui 230031, China \\
9.~Max-Planck-Institute for Plasma Physics, Garching, Germany\\
10.~School of Computer Science and Technology, Anhui Engineering Laboratory for Industrial Internet Intelligent Applications and Security, Anhui University of Technology, Ma'anshan, Anhui 243002, China \\
11.~College of Physics and Optoelectronic Engineering, Shenzhen University, Shenzhen 518060, China \\
}

\date{Received: / Accepted: }

\maketitle

\begin{abstract}
The China Fusion Engineering Test Reactor (CFETR) and the Huazhong Field Reversed Configuration (HFRC), currently both under intensive physical and engineering designs in China, are the two major projects representative of the low-density steady-state and high-density pulsed pathways to fusion. One of the primary tasks of the physics designs for both CFETR and HFRC is the assessment and analysis of the magnetohydrodynamic (MHD) stability of the proposed design schemes. Comprehensive efforts on the assessment of MHD stability of CFETR and HFRC baseline scenarios have led to preliminary progresses that may further benefit engineering designs. For CFETR, the ECCD power and current for full stabilization on NTM have been predicted in this work, as well as the corresponding controlled magnetic island width. A thorough investigation on RWM stability for CFETR is performed. For 80\% of the steady state operation scenarios, active control methods may be required for RWM stabilization. The process of disruption mitigation with massive neon injection on CFETR is simulated. The time scale of and consequences of plasma disruption on CFETR are estimated, which are found equivalent to ITER. Major MHD instabilities such as NTM and RWM remain challenge to steady state tokamak operation. On this basis, next steps on CFETR MHD study are planned on NTM, RWM, and SPI disruption mitigation. For HFRC, plasma heating due to 2D adiabatic compression has been demonstrated in NIMROD simulations. The tilt and rotational instabilities grow on ideal MHD time scale in single fluid MHD model as shown from NIMROD calculations. Two-fluid MHD calculations using NIMROD find FLR stabilizing effects on both tilt and rotational modes. Energetic-particle stabilization of tilt mode was previously demonstrated in C-2 experiments and NIMROD simulations. With stabilization on major MHD instabilities from two-fluid and energetic particle effects, FRC may promise to be an alternative route to compact magnetic fusion ignition. To explore such a potential, we plan on further perform analyses of the MHD instabilities in HFRC during magnetic compression process.
%Insert your abstract here. Include keywords, PACS and mathematical subject classification numbers as needed.
\end{abstract}


\section{Introduction}
\label{intro}

Based on the extrapolation from the empirical scaling laws obtained from decades of research efforts on tokamaks, for a sufficiently large size and a sufficiently strong magnetic field, the tokamak plasma may achieve steady-state self-sustained ignition. The complexity and economic cost associated with the tokamak size is expected to far exceed the existing tokamak devices in the world, often requiring collaborative efforts of multiple countries and unions for many years. For example, the ITER, jointly constructed by nine international governments, will reach a height of 30 meters after completion, costing more than 10 billion euros. Since the launch of the ITER program in 2006, after nearly 15 years of international cooperation and efforts, the construction of the main unit of the device is close to completion, and it is expected that the experimental operation will begin in 2025.

Besides being a partner in International Thermonuclear Experimental Reactor (ITER)~\cite{aymar2002ppcf}, China has recently proposed to design and potentially build China Fusion Engineering Test Reactor~(CFETR)~\cite{wan2014}. The goal is to address the physics and engineering issues essential for bridging the gap between ITER and DEMO (DEMOnstration Power Station), including achieving tritium breeding ratio (TBR) $>1$ and exploring options for DEMO blanket and divertor solutions~\cite{chan2015,shi2016,wanyuanxi2017nf}. During the past several years, significant progress has been made in CFETR conceptual physics and engineering design~\cite{song2014,wanyuanxi2017nf,lijiangang2019jfe}. Since 2018, a new design version of CFETR has been made, by choosing a larger machine with major radius $R=7.2m$, minor radius $a=2.2m$ and axis toroidal magnetic filed $B_T=5-7T$~\cite{lijiangang2019jfe,zhuang2019}. The primary missions of the CFETR project are proposed to demonstrate the fusion energy production of $200–1000 MW$, generate the steady-state burning plasmas with duty time of about $50\%$ and test the self-sustainable burning state with fusion gain, Q, about $20–30$.

%{\color{red}\sout{A conceptual engineering design of CFETR including different coils and remote maintenance systems was prepared in the beginning~\cite{song2014}. The initial design parameters of CFETR are based on a 0-D analysis~\cite{wan2014}, and later are optimized using several 1.5D transport codes~\cite{chan2015}. To achieve staged goals, the CFETR has been designed for two steady-state scenarios - baseline and advanced scenarios~\cite{shi2016}. The baseline scenario is designed for moderate fusion power (200MW) with a fully non-inductive current drive, giving more importance towards challenging annual duty factor of $0.3-0.5$. The advanced design is aimed at higher fusion power with a substantial challenging fraction of bootstrap current drive.}}

As the international fusion community is about to verify the scientific feasibility of fusion energy through the tokamak programs such as ITER and CFETR, in order to achieve the compactness and commercialization of future nuclear fusion reactors, developed industrial countries around the world are also carrying out research on and exploring alternative path to fusion. For example, the first fusion project supported by the US Department of Energy's Energy Advanced Research Projects Agency (ARPA-E, or Advanced Research Projects Agency-Energy) between 2015 and 2018 focused on developing potentially revolutionary fusion concepts and related technologies (i.e. pulsed, medium density fusion concepts), these technologies can significantly reduce the cost of fusion development and final deployment~\cite{nehl2019jfe}. In 2019, the US Department of Energy continued to newly establish the Breakthroughs Enabling THermonuclear-fusion Energy (BETHE) program~\cite{BETHE} and the Innovation Network for Fusion Energy (INFUSE) program~\cite{INFUSE} to support the encouragement of the private sector, industry, government, and academia. Collaborate and make full use of their respective advantages to jointly explore new fusion concepts and new configurations to accelerate the development of commercially viable fusion energy channels.

Among all the alternative paths, field reversed configuration (or FRC) has complete axisymmetry, its magnetic field structure is relatively simple, and the plasma $\beta$ is close to $1$, which is complementary to the tokamak configuration to a certain extent~\cite{steinhauer2011pop}. Since the concept of field reversed configuration was first proposed, FRC has been one of the preferred candidate configurations for fusion devices such as compact nuclear fusion reactors and neutron sources, which has received increasing attention from various countries. For example, the C-2 experiment in TAE Technology has been able to obtain and maintain an FRC with plasma density $n$ reaching $3\times 10^{19}m^{-3}$, the electron and ion temperature $Te$ and $T_i$ around $1 keV$ respectively, the energy confinement time $\tau_E$ about $1 ms$, and the plasma $\beta$ around $90\%$~\cite{guohouyang2015nc}. The Los Alamos National Laboratory and the ALPHA project in the United States successfully increased the temperature and density of an FRC plasma by $1$ order of magnitude using staged magnetic compression~\cite{nehl2019jfe}. In addition, Japan, Canada, Russia and other countries also have related experimental devices~\cite{tuszewski1988nf}. Recently, the Huazhong Field Reversed Configuration (HFRC) has been designed to explore a novel concept of ``two-staged'' magnetic compression of FRC as a path to achieving a compact and economic neutron source and potential fusion reactor.

The China Fusion Engineering Test Reactor (CFETR) and the Huazhong Field Reversed Configuration (HFRC), currently both under intensive physical and engineering designs in China, are the two major projects representative of the low-density steady-state and high-density pulsed pathways to fusion. One of the primary tasks of the physics designs for both CFETR and HFRC is the assessment and analysis of the magnetohydrodynamic (MHD) stability of the proposed design schemes. This includes the determination of the parameter boundary for major MHD instabilities, the prediction of pre-cursor signals and saturation level of nonlinear MHD instabilities, and the evaluation of their control and mitigation schemes. Over the past few years, a comprehensive efforts have been devoted to such a task as a part of the physical design activities for both CFETR and HFRC, and considerable progress has been made towards the goal of the tasks. This paper reports these collective progresses as a snapshot of their current status. In particular, on the side of CFETR, analyses on the error field tolerance, ECCD suppression of neoclassical tearing mode (NTM), stability of resistive wall mode (RWM), and the effectiveness of disruption mitigation using massive gas injection (MGI) are presented and discussed. On the side of HFRC, the linear growth rates of tilt and rotational instabilities, as well as their potential stabilization due to finite Larmor radius (FLR) effects are reported. The results reported here in the paper are by no means final; most of the results are preliminary in nature, partly because of constantly evolving scenarios of the designs, and partly because of the complexity of the tasks themselves. Nonetheless, the current report is meant to serve as an overall and initial assessment on one of key components, i.e. the MHD stability prospects, of the physical designs for both major magnetic fusion projects as the potential next step in the China fusion energy program.

The rest of the paper is organized as follows. In Sec.~\ref{sec:eq}, the equilibrium scenarios of CFETR and HFRC are introduced. In Sec.~\ref{CFETRmhd}, progresses on the analyses of error field tolerance, ECCD suppression of neoclassical tearing mode (NTM), stability of resistive wall mode (RWM), and the effectiveness of disruption mitigation using massive gas injection (MGI) are reported respectively. In Sec.~\ref{HFRCmhd}, recent analyses on the MHD instabilities of HFRC equilibrium and compression are presented. We conclude with a summary and discussion in Sec.~\ref{summary}.

\section{MHD equilibria of baseline scenarios for CFETR and HFRC}
\label{sec:eq}
\subsection{Steady-state scenarios and hybrid scenarios for CFETR}
\label{sec:eq_cfetr}
\begin{figure}
  \begin{center}
  \includegraphics[width=0.8\textwidth]{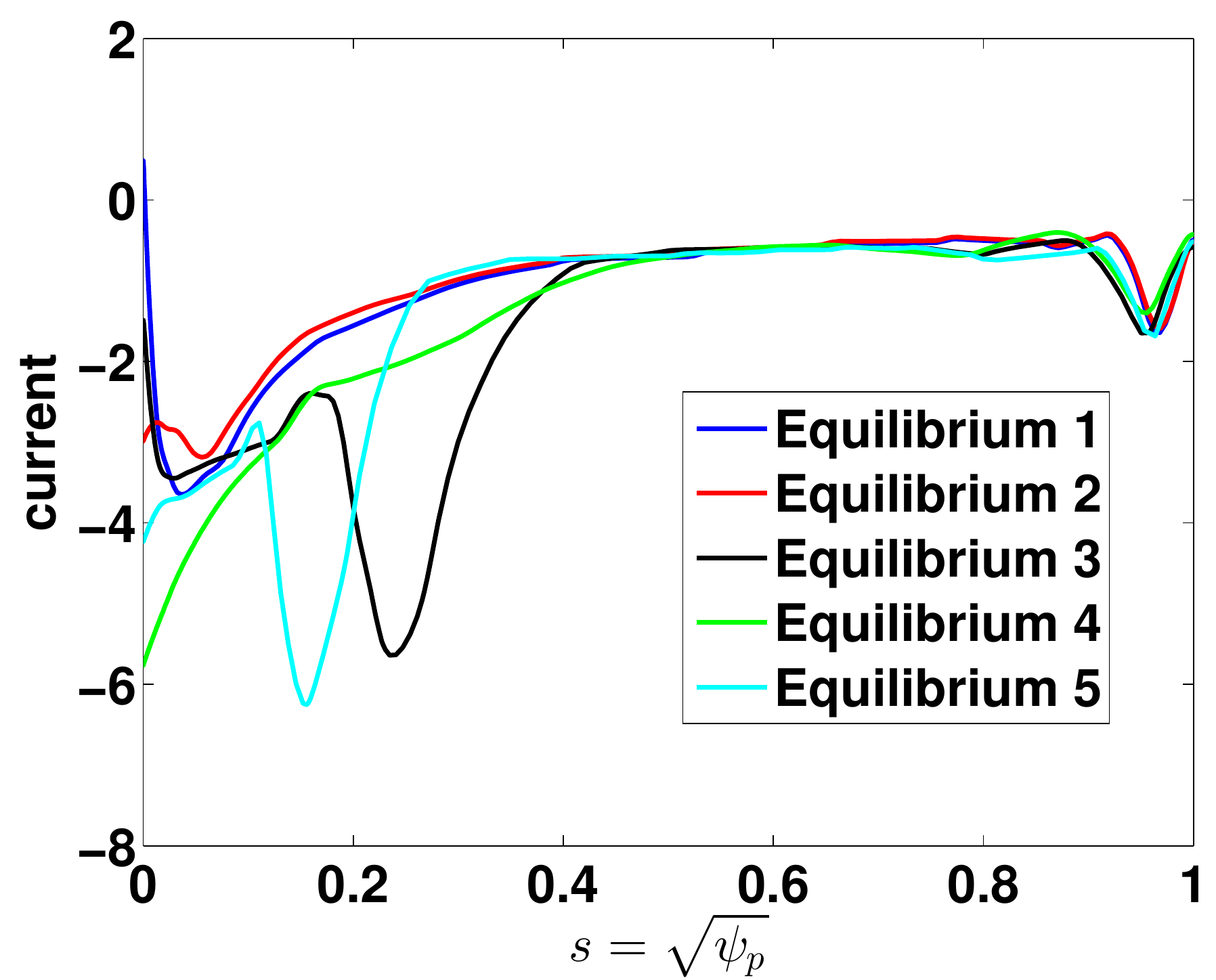}
  \put(-200,180){\textbf{(a)}}

  \includegraphics[width=0.8\textwidth]{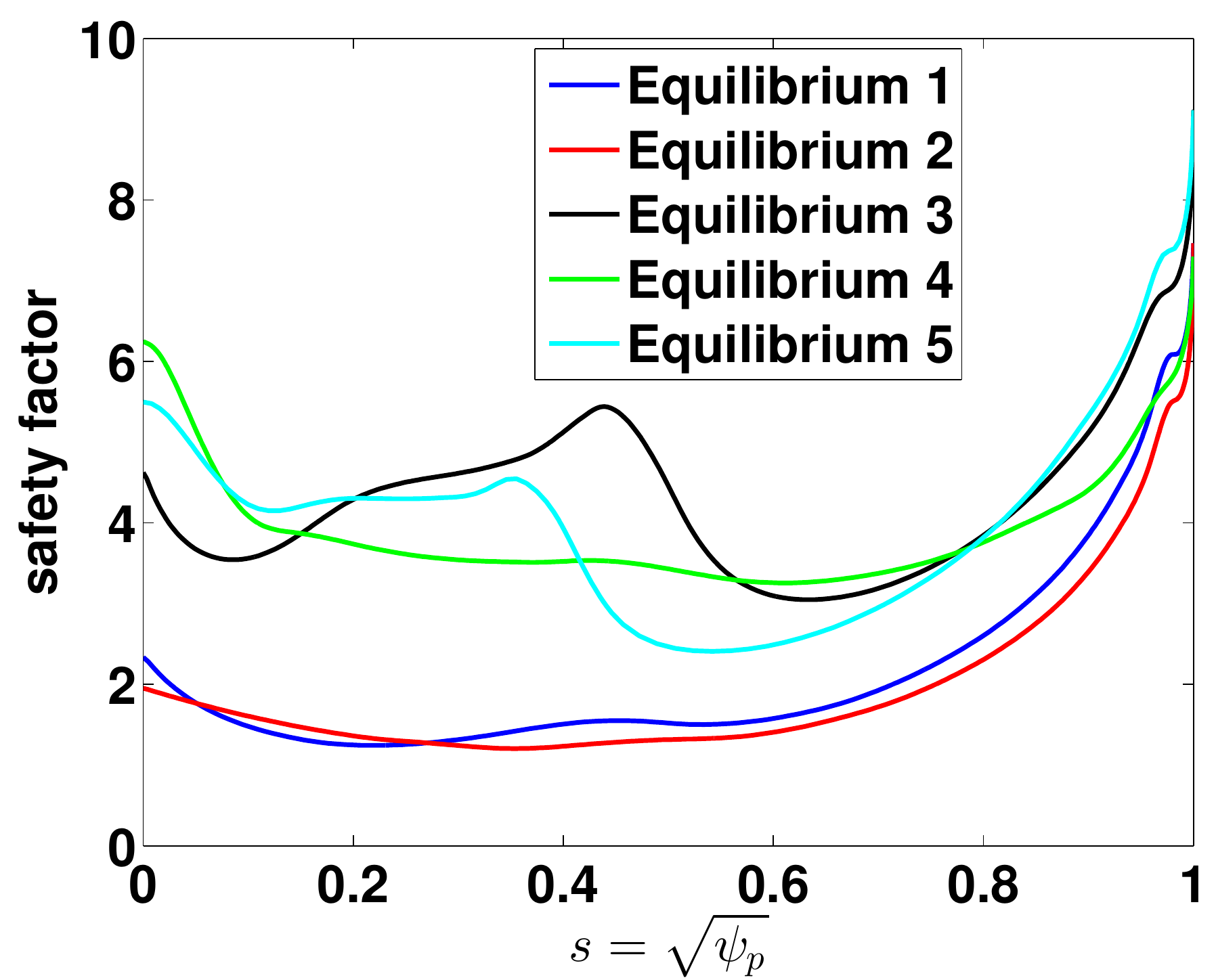}
  \put(-200,180){\textbf{(b)}}
  \caption{(a) Equilibrium plasma current density and (b) safety factor as functions of normalized minor radius in five CFETR steady-state scenarios (Other equilibrium parameters are shown in Table~\ref{tableRWMeq}).}
  \label{fig:cfe_sseq}      
  \end{center}
\end{figure}

\begin{figure}[ht]
  \begin{center}
  \includegraphics[width=0.9\textwidth]{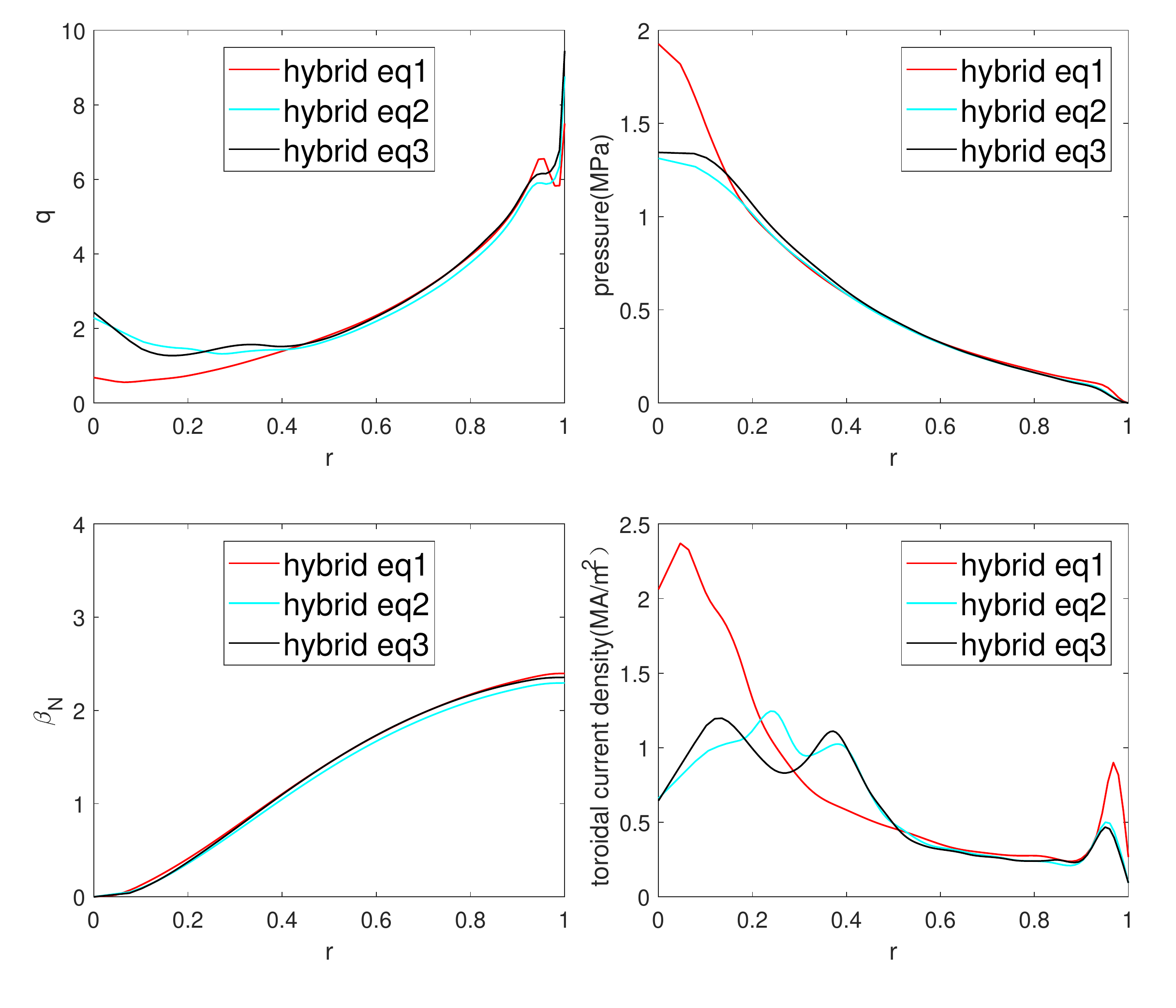}
  \put(-270,240){\textbf{(a)}}
  \put(-100,240){\textbf{(b)}}
  \put(-270,110){\textbf{(c)}}
  \put(-100,110){\textbf{(d)}}
  \end{center}
  \caption{(a) Safety factor, (b) plasma pressure, (c) normalized $\beta$, and (d) toroidal plasma current density as functions of normalized minor radius for the equilibria of the three CFETR hybrid scenarios considered in this study.}
\label{fig:cfe_hyeq}
\end{figure}
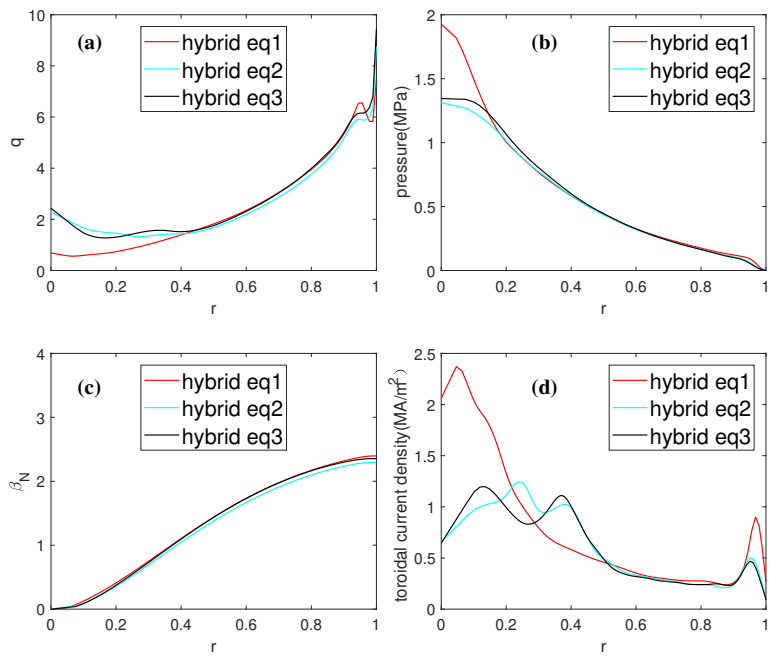

To achieve the mission goal of fusion power production of $1GW$, the self-consistent steady-state scenarios for CFETR with fully sustained non-inductive current drive and as well hybrid mode scenarios are developed using a multi-dimensional code suite with physics-based models as shown in~\cite{Jian2017,Chen2017}. The equilibrium profiles for the steady-state and hybrid scenarios presented in Figs.~\ref{fig:cfe_sseq} and~\ref{fig:cfe_hyeq} respectively are the bases of our MHD stability analyses for CFETR reported later in this paper. A dominant bootstrap current together with auxiliary extra heating current drive is required in the steady-state scenarios. As a result, safety factor $q$-profiles with reversed magnetic shear in the core region and a minimum $q$ value (i.e. $q_{min}$) larger than $2$ or $3$ characterize the equilibria of steady-state scenarios. This feature shall benefit the stability of the dangerous internal kink modes and tearing modes with low poloidal numbers, whereas the external kink modes may still grow. The hybrid scenarios, however, have much lower $q_{min}$, that may give rise to the internal MHD modes. 

\subsection{HFRC equilibrium designed for magnetic compression}
\label{sec:eq_hfrc}
\begin{figure}[ht]
  \begin{center}
    \includegraphics[width=0.45\textwidth, height=0.25\textheight]{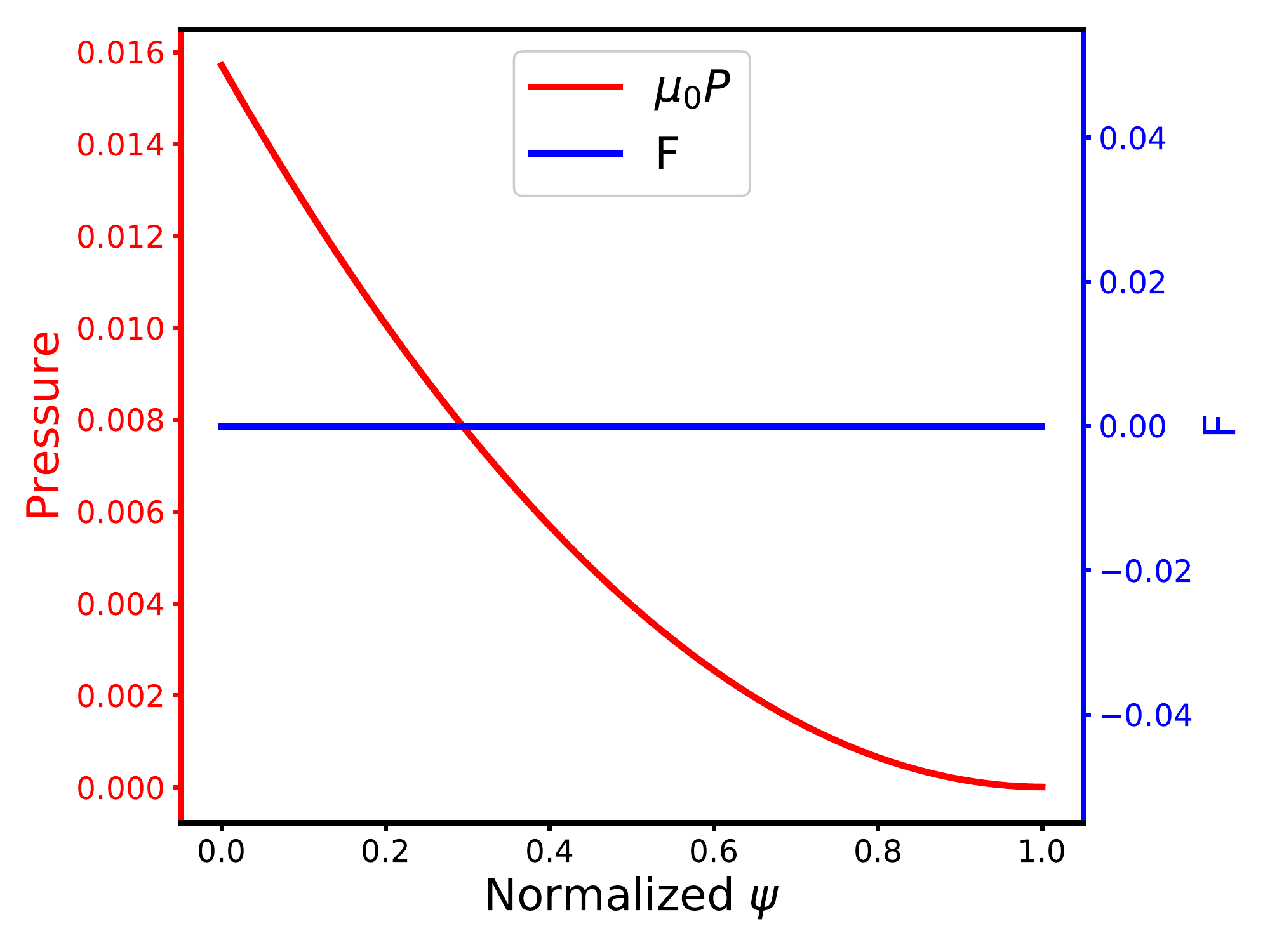}
    \includegraphics[width=0.45\textwidth, height=0.25\textheight]{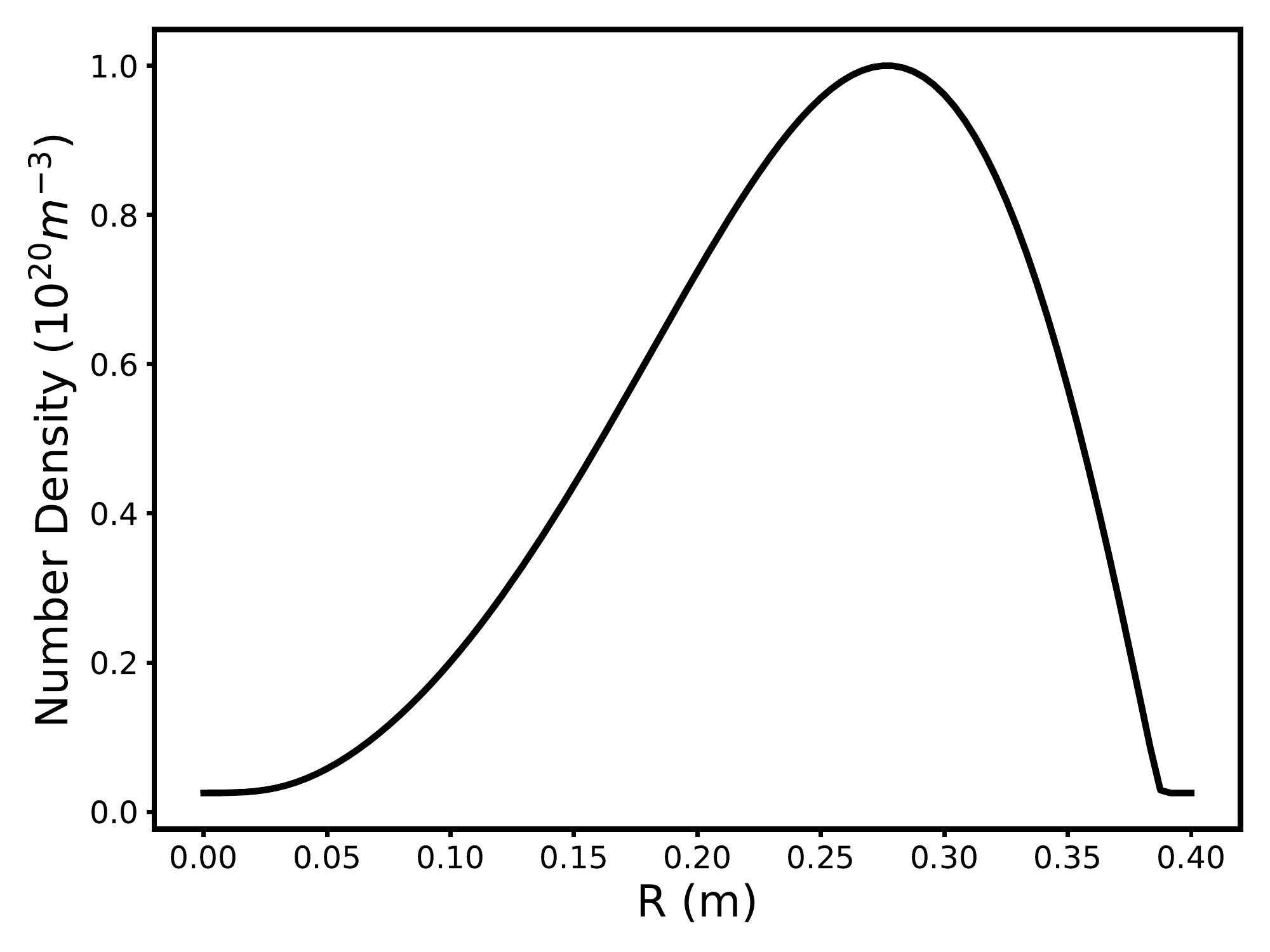}

    \includegraphics[width=0.9\textwidth]{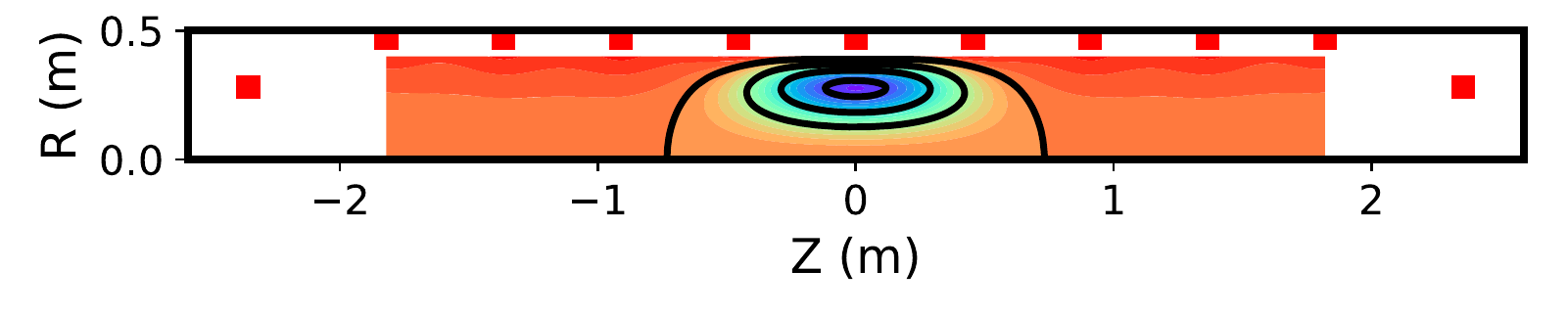}
  \end{center}
  \caption{HFRC equilibrium: the pressure and $F=RB_{\phi}$ profiles (upper left), density and temperature profiles (upper right), and pressure (color) and poloidal magnetic flux (lines) contours (lower panel).}
\label{fig:hfrc_eq}
\end{figure}

The HFRC is designed to achieve fusion reaction condition through the means of magnetic compression, whose feasibility and efficiency are subject to the constraints imposed from the MHD stability. As a first step, the MHD equilibrium for the initial FRC state prior to magnetic compression is generated using the NIMEQ code~\cite{howell14a,lihl20a} based on the engineering designs  (Fig.~\ref{fig:hfrc_eq}). The maximum strength of magnetic field at core is $0.1T$, and the magnetic mirror ratio is $3$. The axial length of magnetic confinement region is about $5m$, and the length between two separatrix X-points is around $2.5m$. The radius of vacuum vessel is $0.7m$, and the radius of confined region is around $0.4m$. For the number density and pressure profiles shown in Fig.~\ref{fig:hfrc_eq}, the line averaged density of plasma is approximately the order of $10^{19} m^{-3}$, and the combined ion and electron temperature $T=T_{i}+T_{e} \sim 800eV$ ($T_{i}\sim 3T_{e}$). 

\section{Assessments and analyses of the MHD stability of CFETR}
\label{CFETRmhd}
\subsection{Ideal MHD mode and RWM stability}
\label{RWM}
The steady-state plasmas in the large tokamak devices, such as ITER and CFETR, should operate in regimes where the resistive wall mode (RWM) is stable or marginal stable. Thus, an unstable RWM with low toroidal mode number limits the operational space of tokamak devices. The RWM can be viewed as the residual instability of the external kink mode, which is a low toroidal mode number MHD instability with global structure along the plasma torus, and can be driven by plasma current or pressure. For a pressure driven external kink mode, it becomes unstable when $\beta_N$ exceeds the Troyon no-wall limit~\cite{troyon1984ppcf}, where $\beta_N$ is the normalized $\beta$. Fortunately, this kind of external kink mode can be stabilized by a closed-fit perfect conducting wall outside the plasma torus. However, the presence of the resistivity in the conducting wall will lead to the penetration of the perturbed magnetic field, thus the external kink mode can still be unstable, and its growth rate depends on the field penetration time through the conducting wall, hence the name of RWM. Generally speaking, if no other control methods are considered, RWM is unstable, because no perfect conducting wall exists. Therefore, for the initial operation phase of CFETR, we shall design the plasma scenarios along with the consideration of control schemes for RWM.

%\subsubsection{CFETR plasma scenarios and computational model of wall}
%\textbf{\textit{--Plasma scenarios designed for CFETR}} \\
%\subsubsection{Plasma and wall models}% for CFETR ideal MHD mode analysis}
%The five 1GW SSO scenarios designed for CFETR are summarized in Table~\ref{tableRWMeq}, and their equilibrium plasma current density and safety factor profiles are compared in Fig.~\ref{fig:cfe_sseq}.
The ideal MHD instabilities are evaluated using the single-fluid MHD models implemented in the MARS-F~\cite{liuyueqiang2000pop} and the AEGIS~\cite{zheng2006jcp} codes. The details on the ideal MHD and the computational models in these two codes are briefly reviewed in~\ref{sec:mars} and~\ref{sec:aegis}.

\subsubsection{Instabilities of ideal MHD modes without wall}
\begin{figure}
\begin{center}  
  \includegraphics[width=0.49\textwidth]{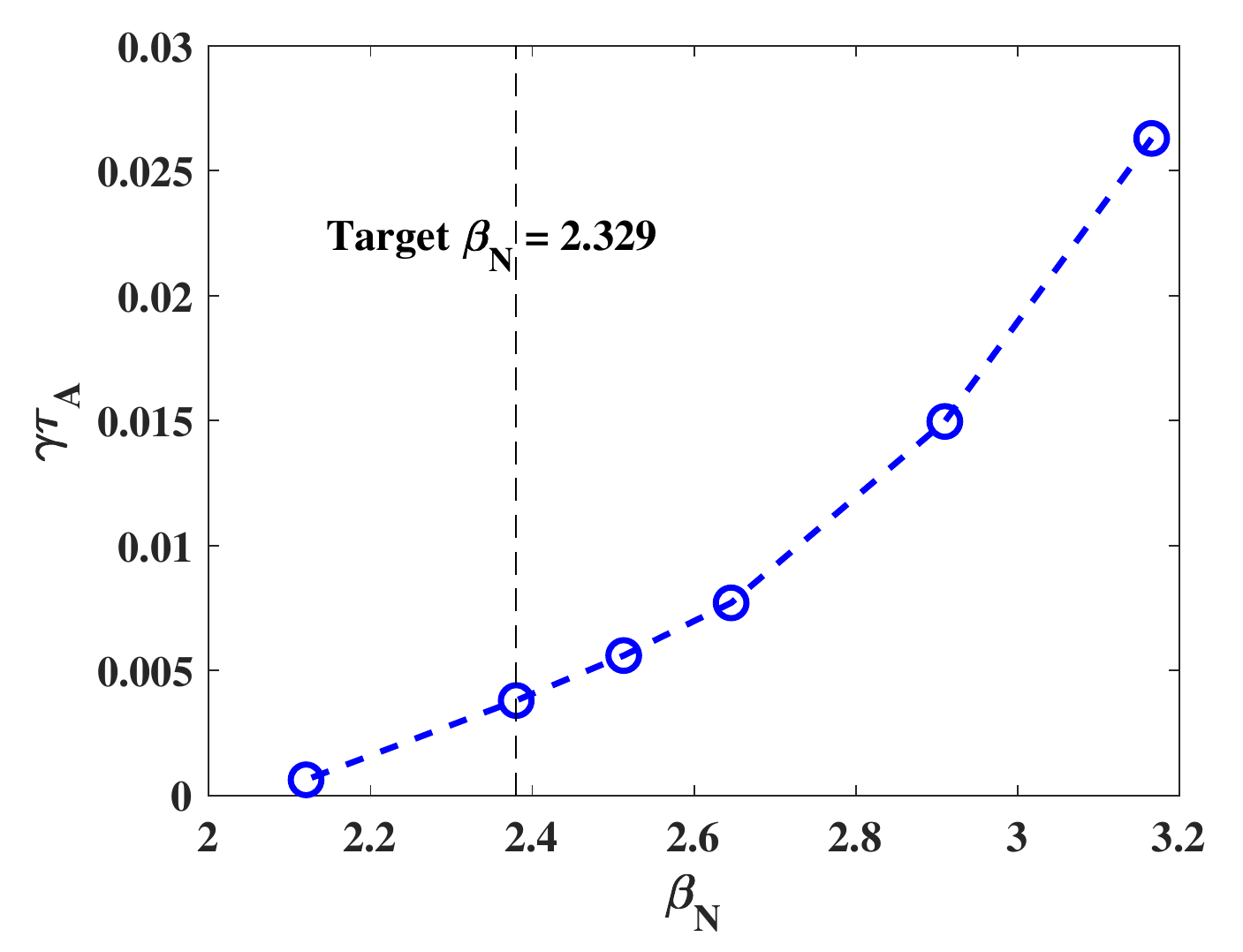}
  \put(-125,110){\textbf{(a)}}
  \includegraphics[width=0.49\textwidth]{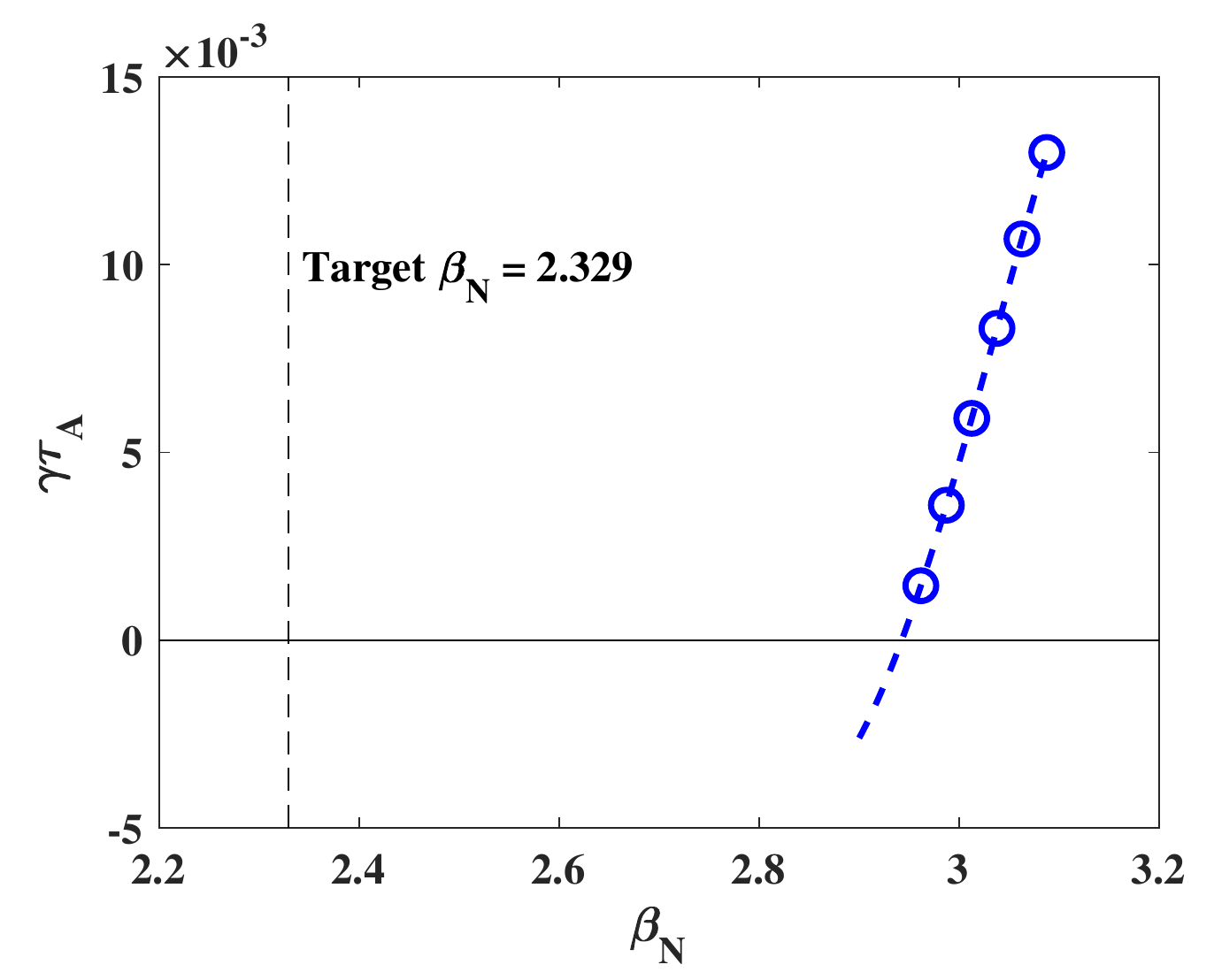}
  \put(-125,110){\textbf{(b)}}

  \includegraphics[width=0.49\textwidth]{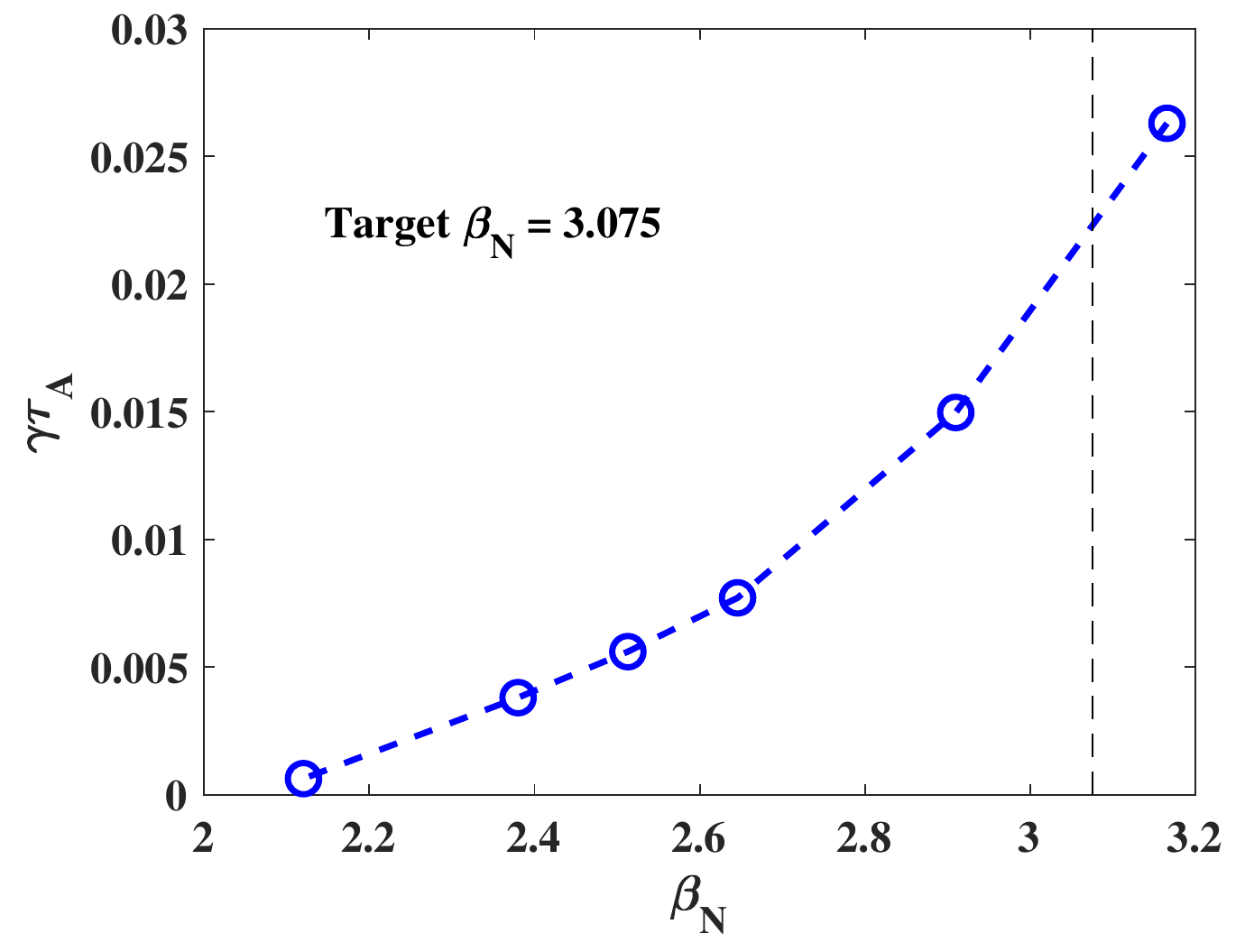}  
  \put(-125,110){\textbf{(c)}}  
  \includegraphics[width=0.49\textwidth]{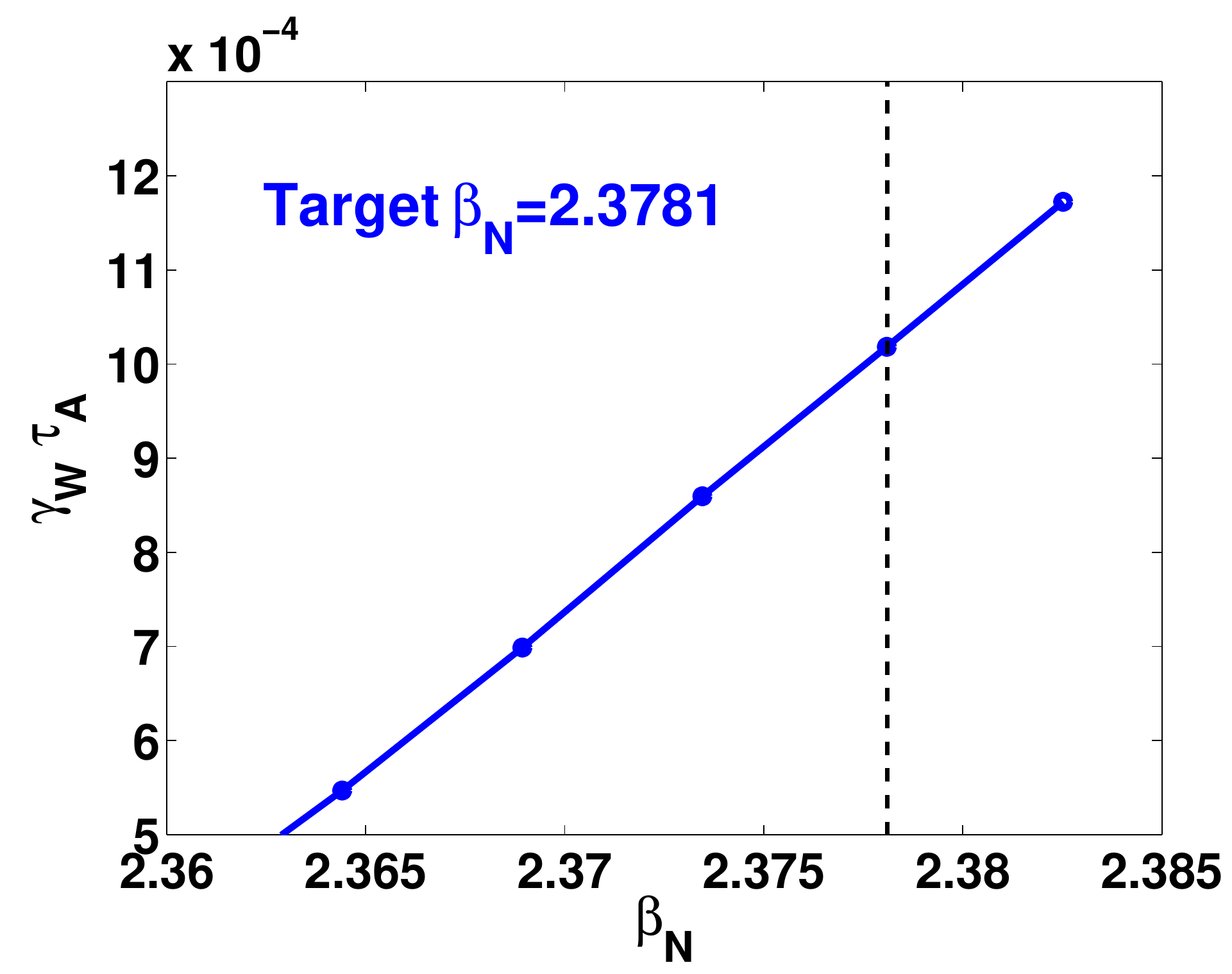}
  \put(-125,110){\textbf{(d)}}

  \includegraphics[width=0.49\textwidth]{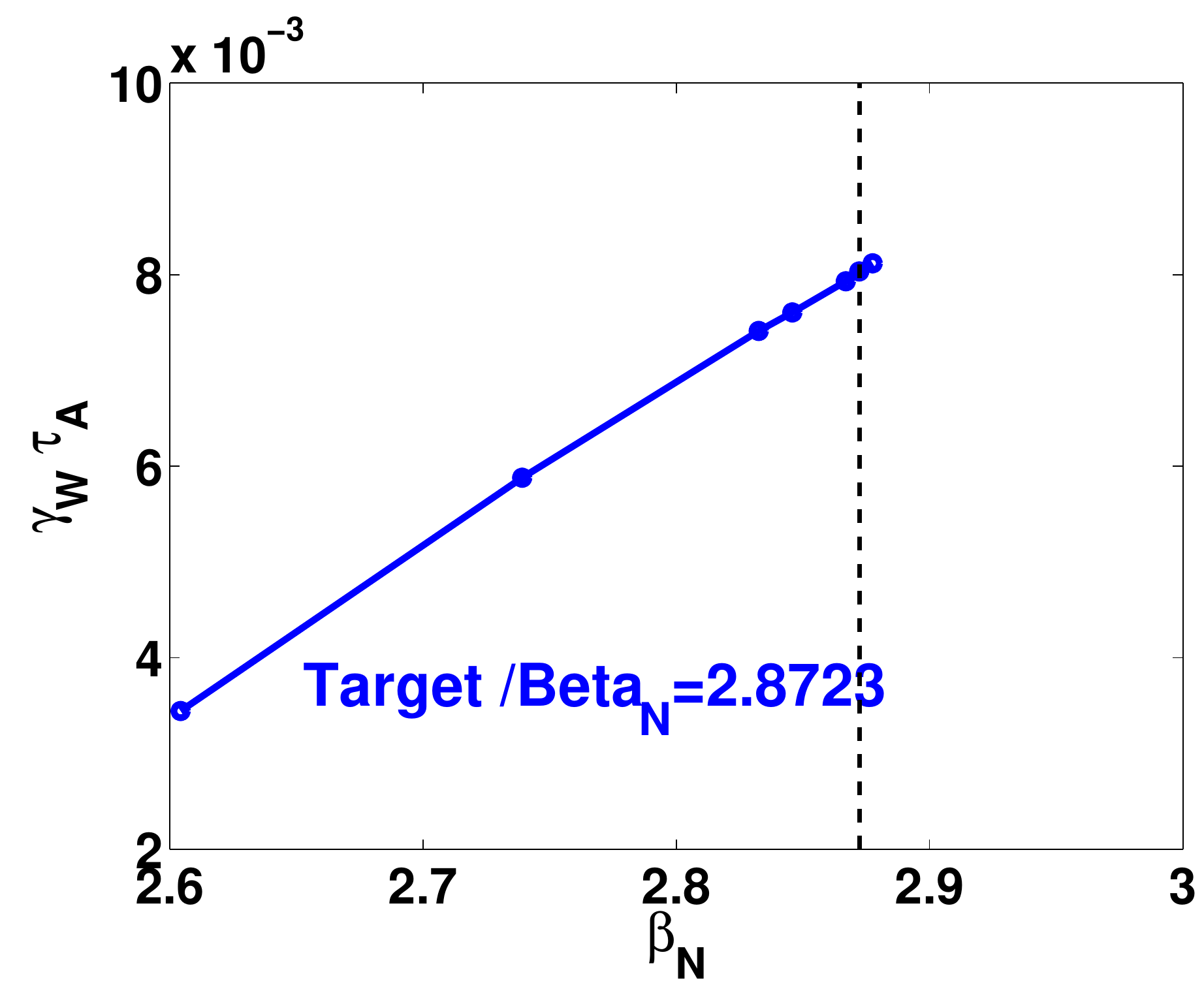}
  \put(-125,110){\textbf{(e)}}
\end{center}
\caption{(a-e) Growth rates of the ideal MHD mode as functions of the plasma pressure $\beta_N$ in absence of conducting wall for five different CFETR SSO scenarios from MARS-F calculations, where the vertical lines denote the designed plasma pressure $\beta_N$ for each corresponding equilibrium.}
\label{rwmfig3}      
\end{figure}

\begin{figure}
\begin{center}  
  \includegraphics[width=0.49\textwidth]{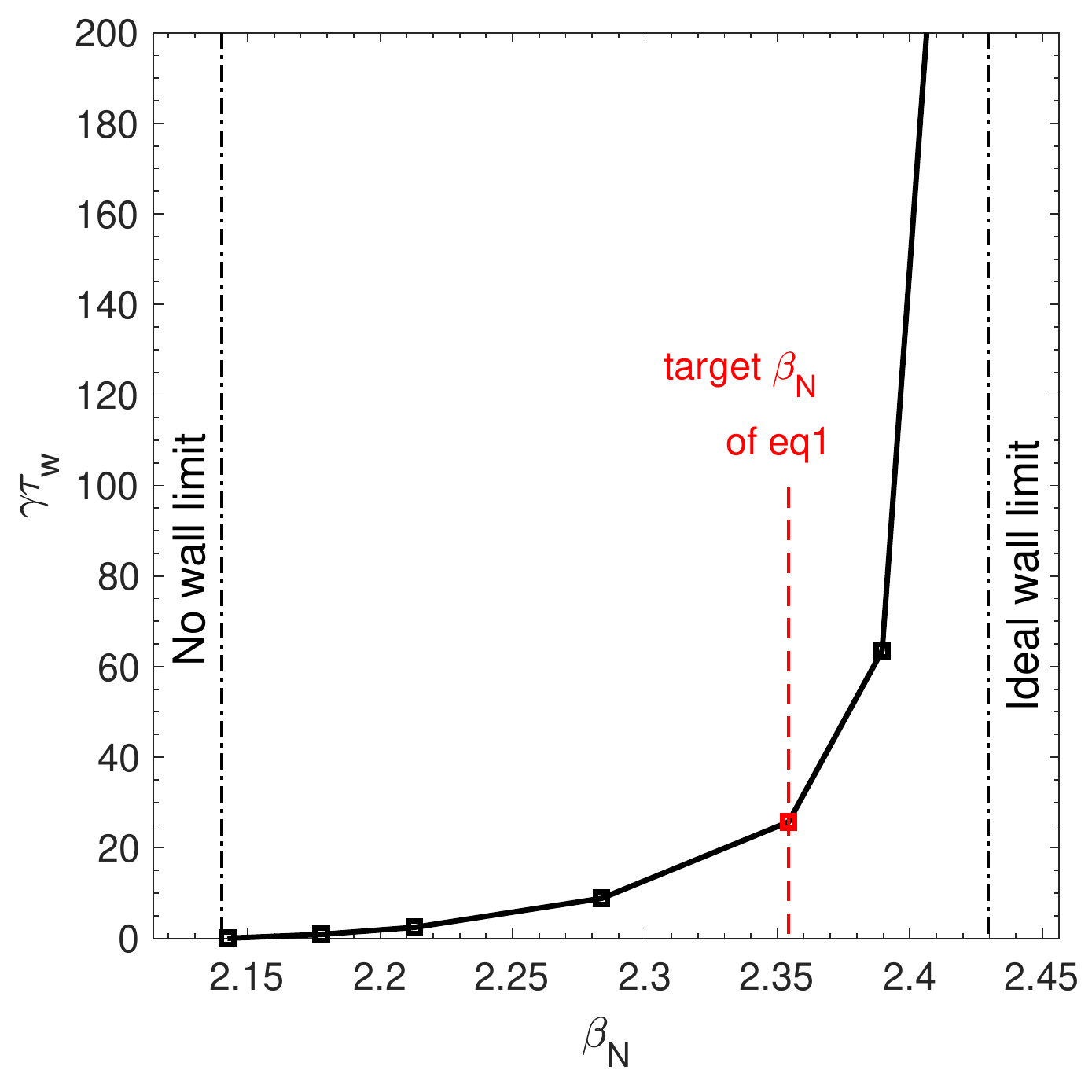}
  \put(-125,110){\textbf{(a)}}
  \includegraphics[width=0.49\textwidth]{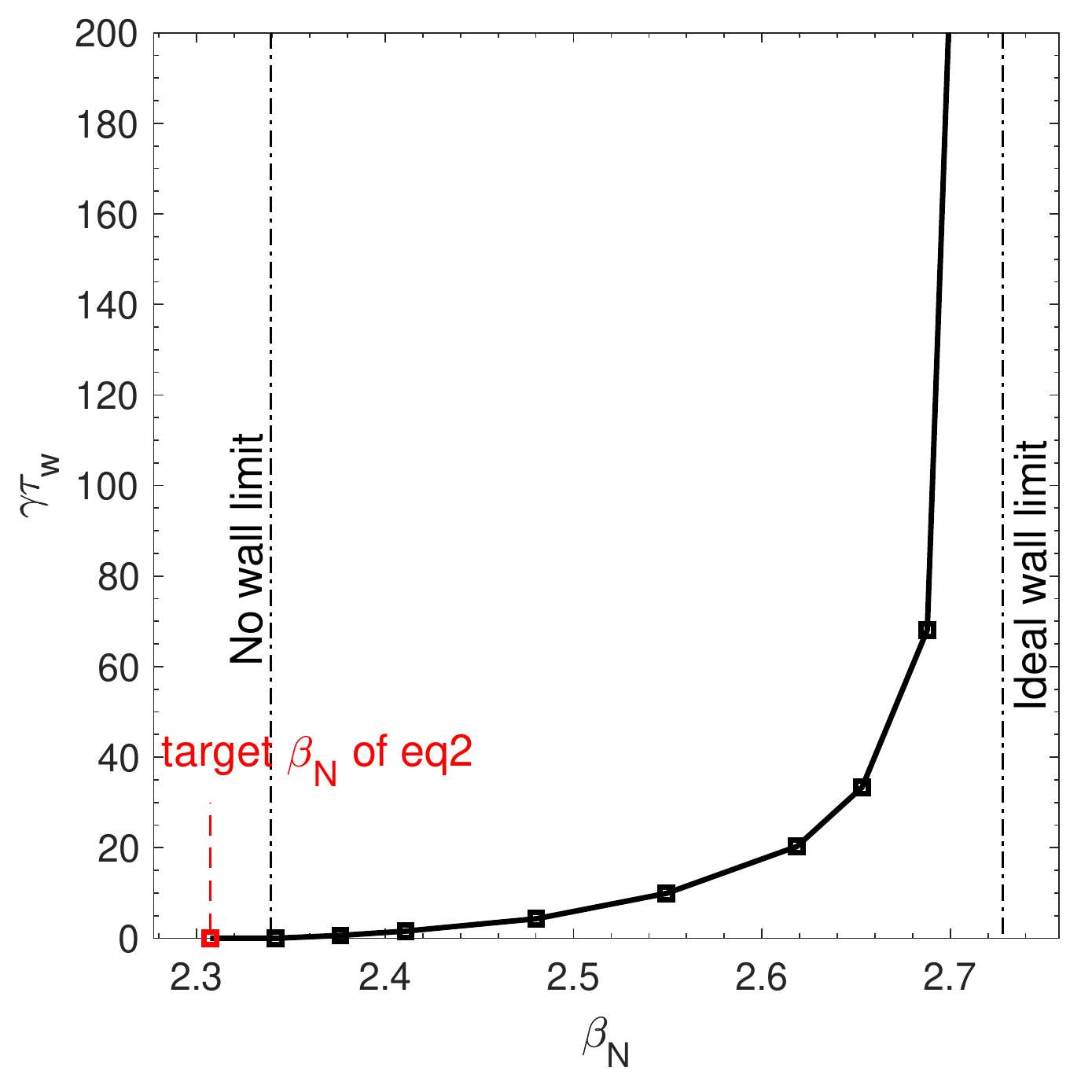}
  \put(-125,110){\textbf{(b)}}

  \includegraphics[width=0.49\textwidth]{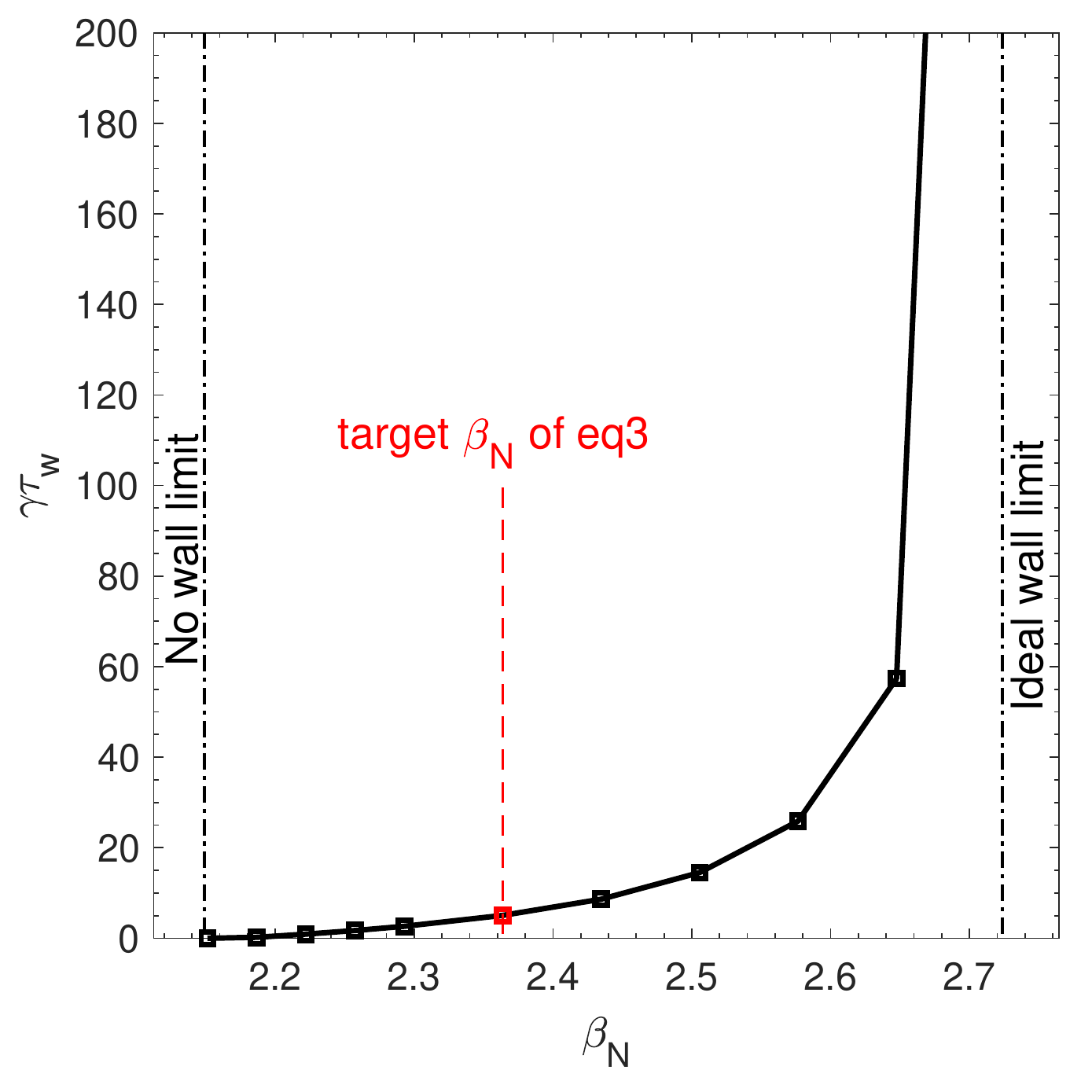}  
  \put(-125,110){\textbf{(c)}}  
  \includegraphics[width=0.49\textwidth]{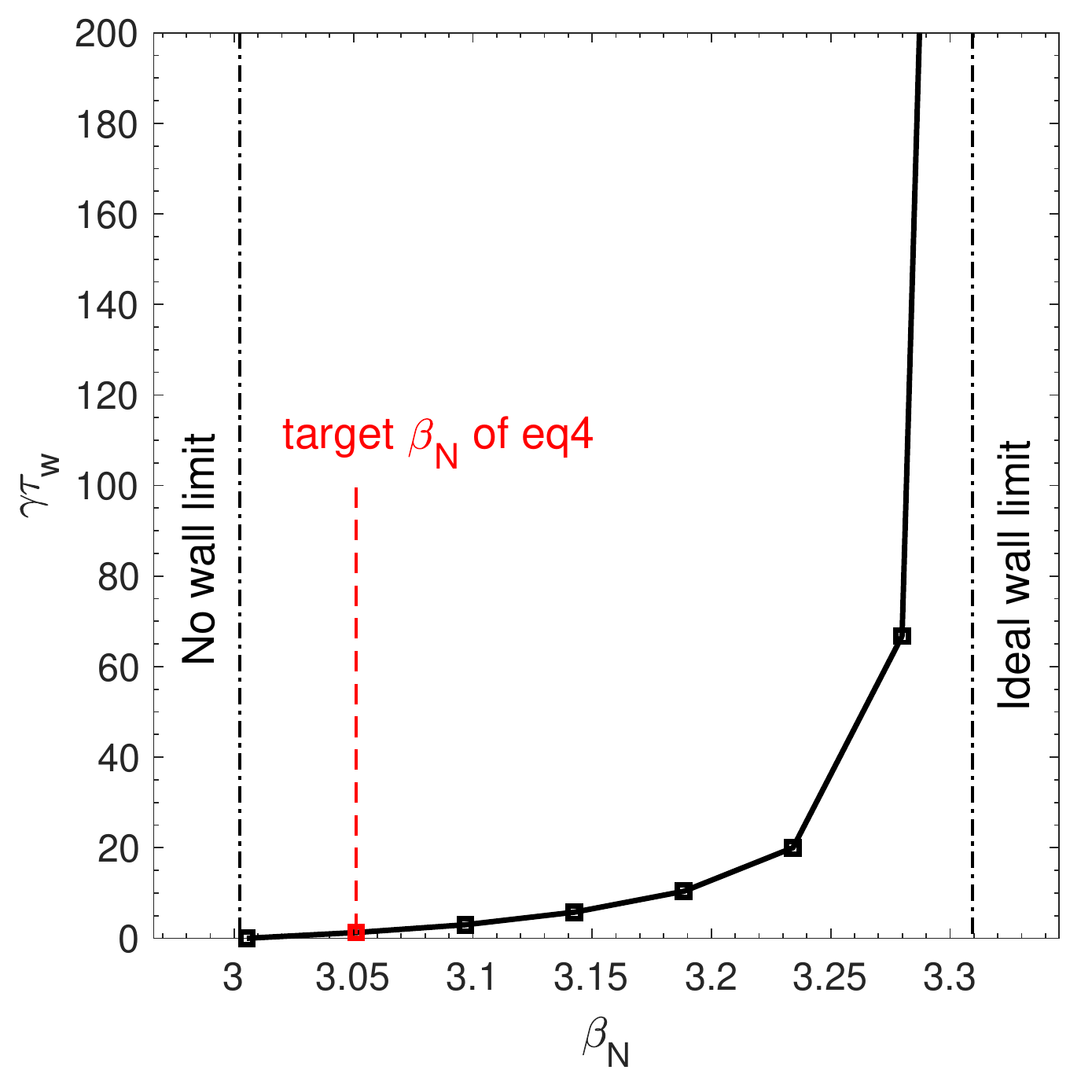}
  \put(-125,110){\textbf{(d)}}

  \includegraphics[width=0.49\textwidth]{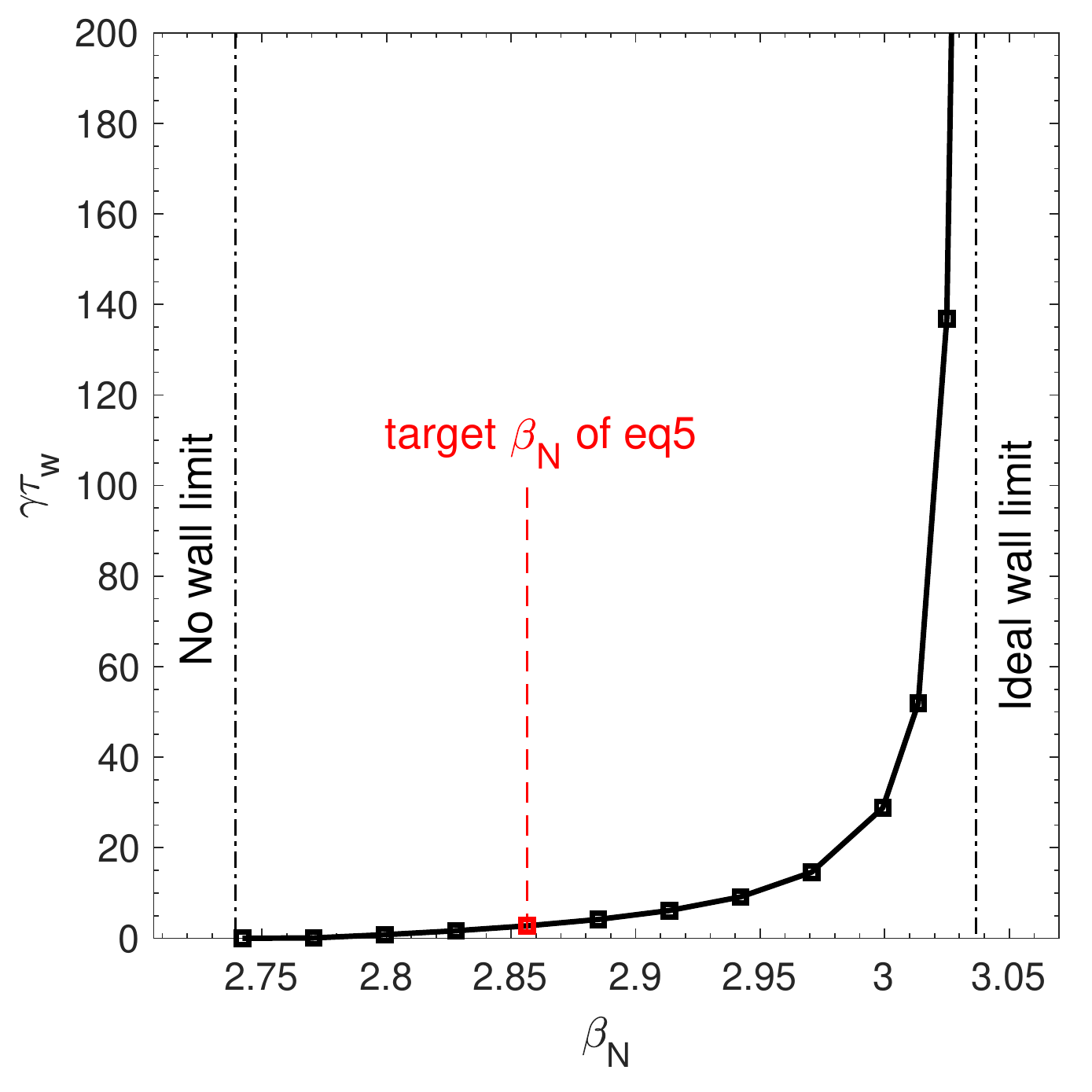}
  \put(-125,110){\textbf{(e)}}
\end{center}
\caption{(a-e) Growth rates of the ideal MHD mode as functions of the plasma pressure $\beta_N$ in absence of conducting wall for five different CFETR SSO scenarios from AEGIS calculations, where the vertical red lines denote the designed plasma pressure $\beta_N$ for each corresponding equilibrium.}
\label{rwmfig4}    
\end{figure}

As aforementioned, for the steady state CFETR operation with high plasma pressure, Troyon no-wall limit is one of the first critical factors to consider for the stability of the ideal MHD modes. The ideal MHD growth rates of the $n=1$ toroidal mode are scanned over $\beta_N$ for all five CFETR SSO equilibria in absence of flow or wall using both MARS-F and AEGIS codes, including the corresponding target plasma pressure or $\beta_N$ value of each equilibrium (Figs.~\ref{rwmfig3} and~\ref{rwmfig4}).
%The ideal MHD stabilities for all five CFETR equilibria are also evaluated using the AEGIS code.

Generally speaking, the results reported in Figs.~\ref{rwmfig3} and~\ref{rwmfig4} show a good agreement between MARS-F and AEGIS codes. Both results find that only the equilibrium 2 plasma with its design target $\beta_N$ is stable to ideal MHD modes in absence of perfect conducting wall. For all other four equilibria with their corresponding target $\beta_N$, the ideal MHD modes are unstable. Note that the no-wall $\beta_N$ limits of equilibria 1 and 2 are different, even though the two equilibria are similar, especially the $q_{95}$ and the target $\beta_N$-values. This difference may derive from the different values of internal inductance $l_i$ in scenario 1 and 2, which are $0.90$ and $0.96$ respectively, where the plasma internal inductance is computed using CHEASE code~\cite{lutjens96a} with the following definition,
\begin{equation}
l_i=\frac{4\pi}{I^2R_0}\int\frac{|\nabla\psi|^2}{R^2}\mathcal{J}d\psi d\chi
\end{equation}
with $\mathcal{J}$ being the Jacobian $\mathcal{J}=|(\nabla\psi\times\nabla\chi)\cdot\nabla\phi|^{-1}$.

\begin{figure}
\begin{center}
  \hspace{0.1in}
  \includegraphics[width=0.82\textwidth]{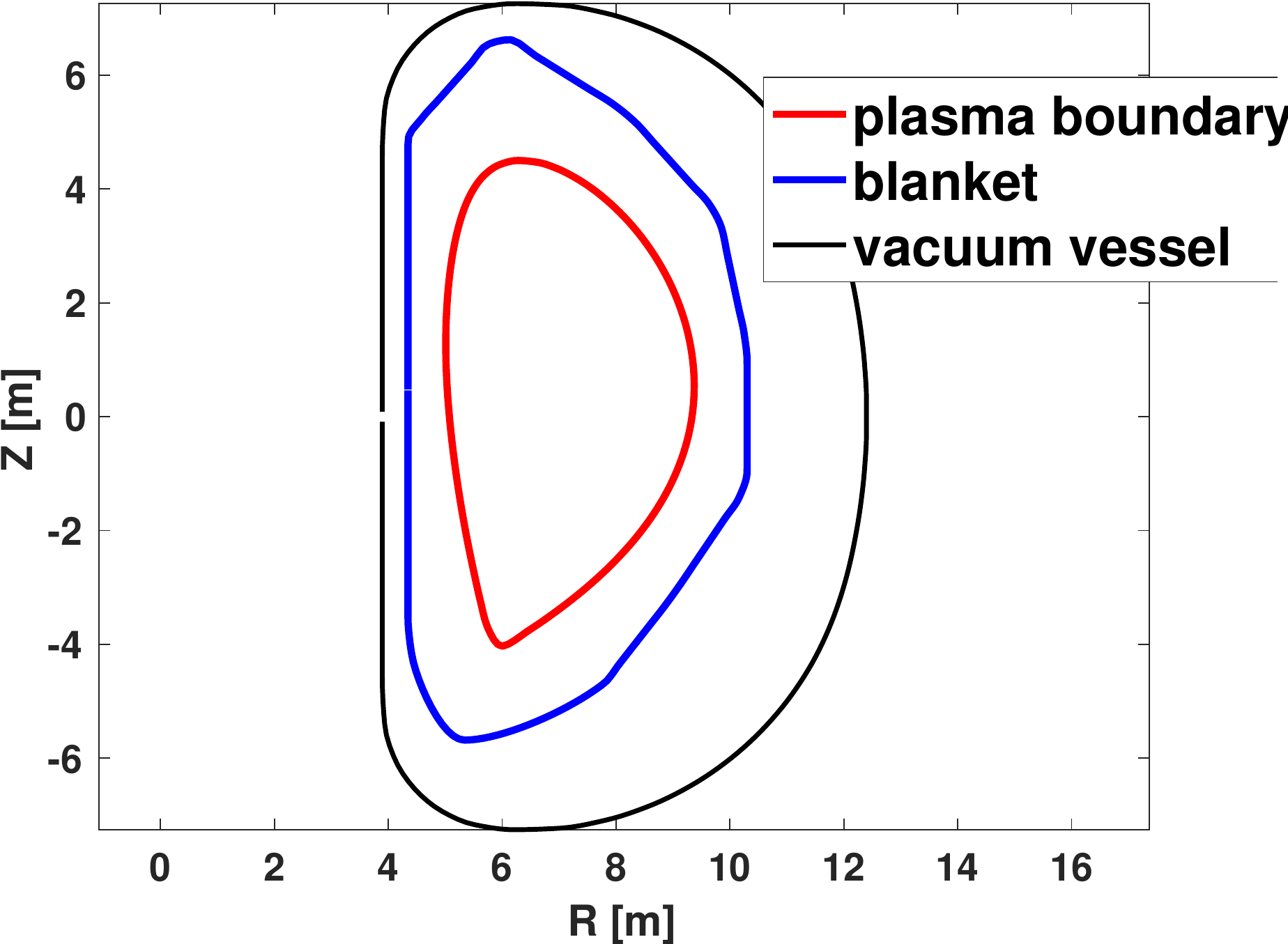}
  \put(-210,160){\textbf{(a)}}
  \vskip 0.2in
  \includegraphics[width=0.79\textwidth]{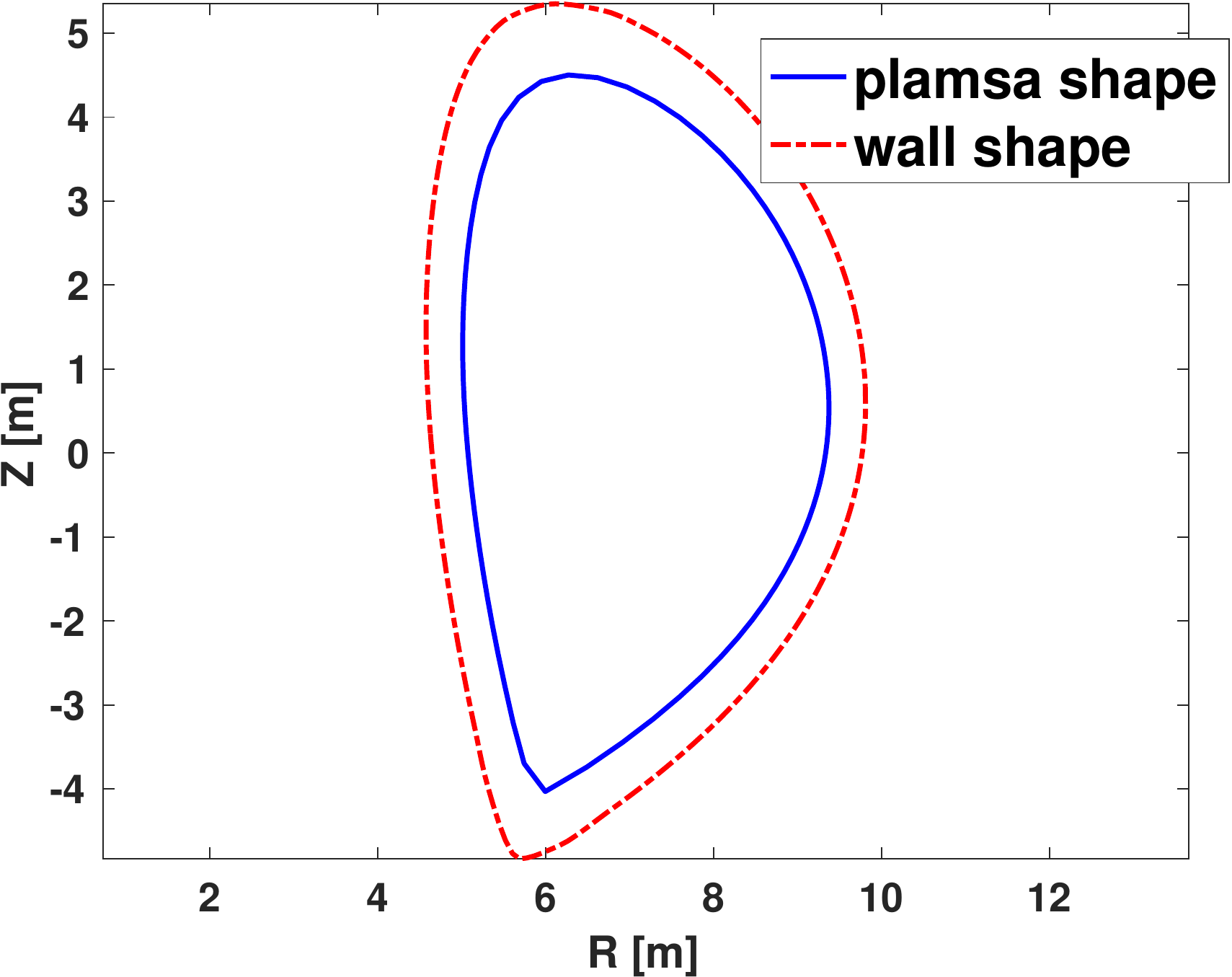}
  \put(-200,160){\textbf{(b)}}
\end{center}
\caption{Two computational models for the designed CFETR wall. (a) Model A includes the plasma boundary (red curve), the blanket (blue curve), and the vacuum vessel (black curve) based on the CFETR engineering design~\cite{libo2019fed}; (b) Model B includes an artificial wall (red dashed curve) conformal to the plasma shape (blue curve).}
\label{rwmfig2}      
\end{figure}

Thus both MARS-F and AEGIS results in Figs.~\ref{rwmfig3} and~\ref{rwmfig4} indicate that the ideal MHD modes in CFETR scenarios 1,3,4 and 5 are unstable in absence of a wall. Next we calculate and compare the instabilities of RWM among these four scenarios, assuming different models of CFETR wall. Based on the CFETR structure design including the first wall, the Tritium breeding module (TBM) blanket, and the vacuum vessels~\cite{libo2019fed}, their sketch and the corresponding plasma shape is plotted in Fig.~\ref{rwmfig2}(a) as one of the wall models considered in this study. The minor radius of vacuum vessel (VV) has been fixed to be double of plasma minor radius away from the plasma boundary, i.e. $d_{VV}=2a$. However, here the TBM resistivity remains unknown, due to the unknown material and structure of the TBM. Thus in another model, an artificial conducting wall with its shape conformal to the plasma shape is assumed, which is shown in Fig.~\ref{rwmfig2}(b).

\subsubsection{Instabilities of RWMs with conformal wall}% for various CFETR scenarios}
In this subsection, we shall assume an artificial conformal wall outside the plasma torus as shown in Fig.~\ref{rwmfig2}(b). The minor radius of wall is denoted as $d_{wall}$. Here, only an ideal plasma is considered without taking into account of plasma flow or resistivity.

\begin{figure}
\begin{center}
  \includegraphics[width=0.49\textwidth]{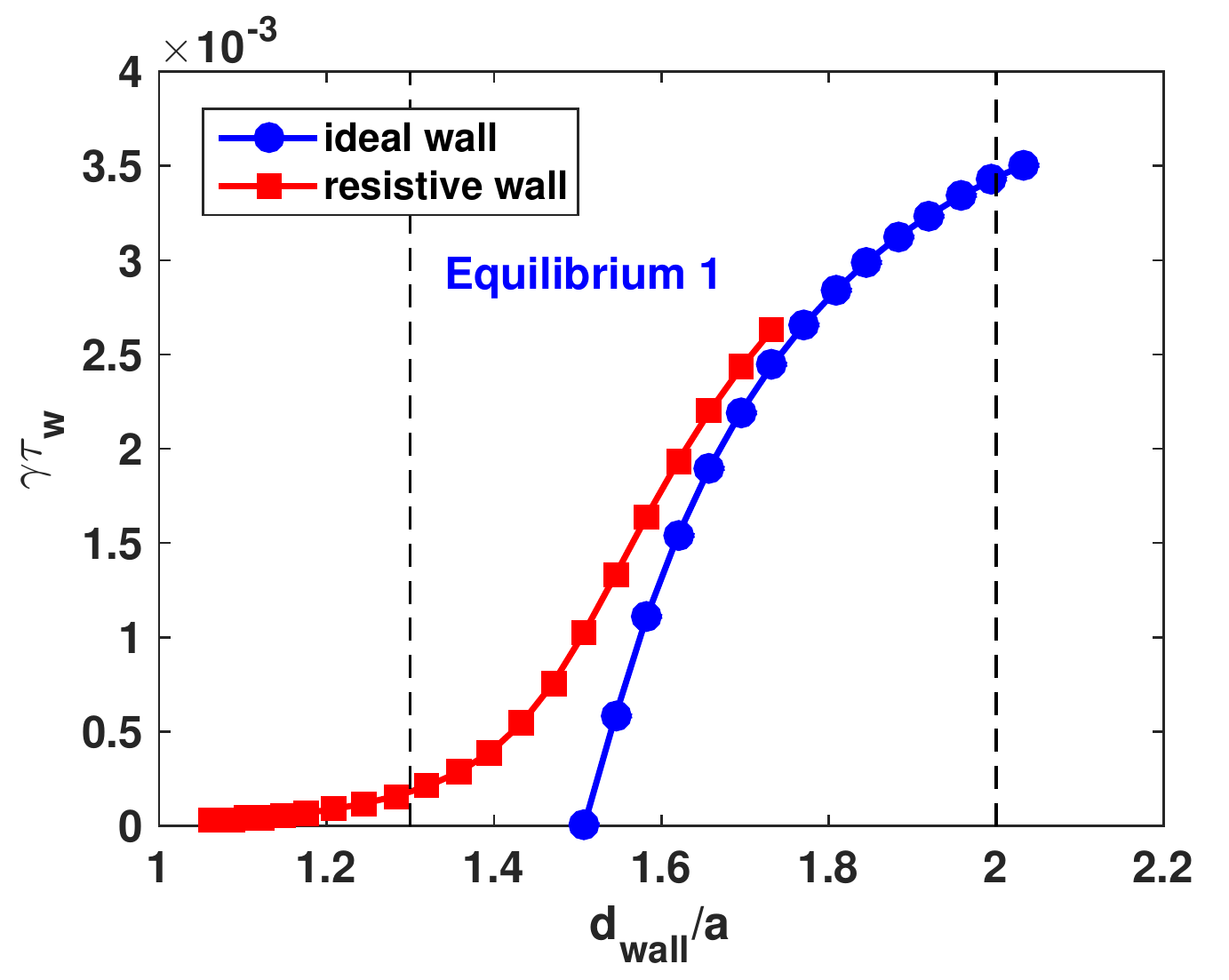}
  \put(-100,60){\textbf{(a)}}
  \includegraphics[width=0.49\textwidth]{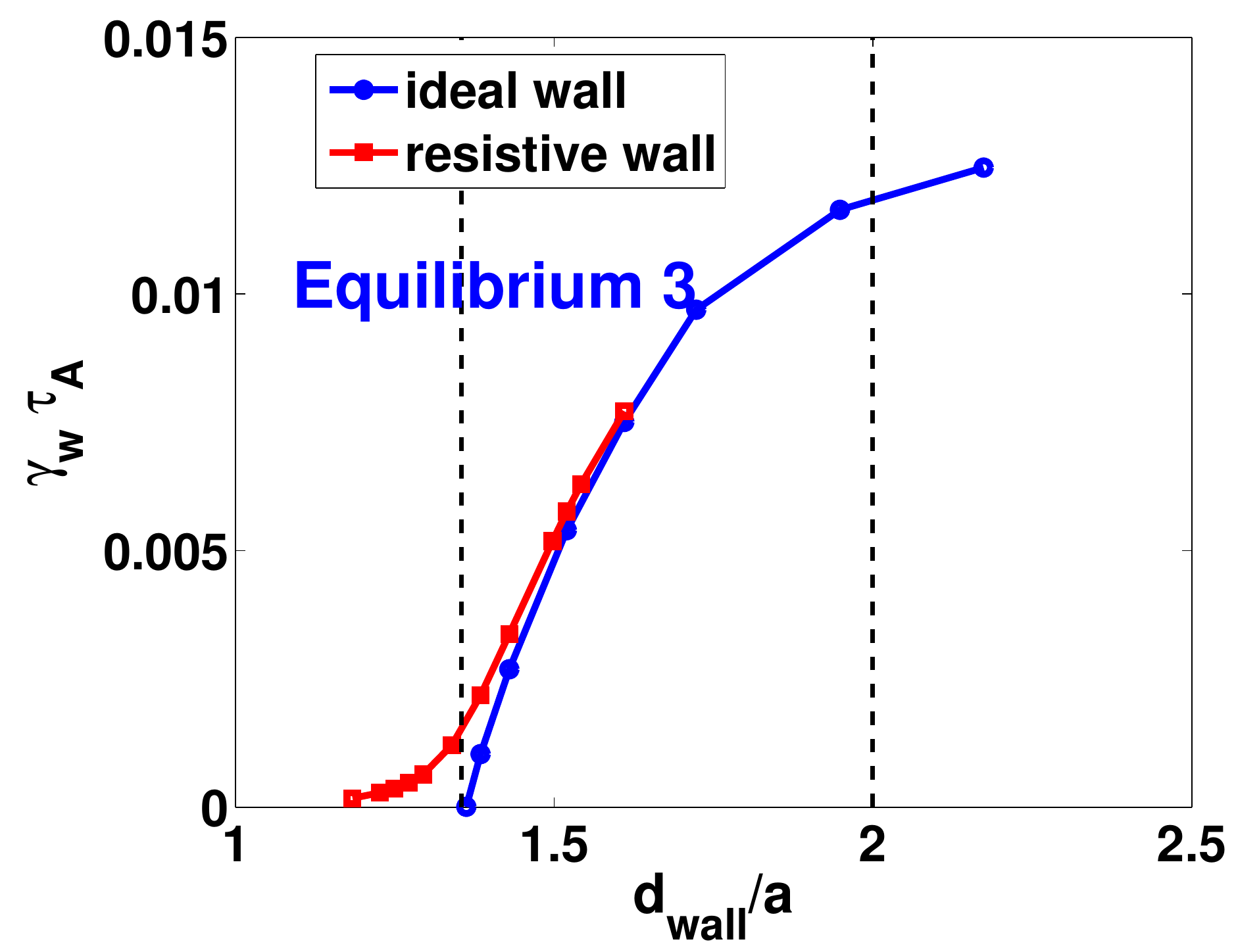}
  \put(-100,60){\textbf{(b)}}
  
  \includegraphics[width=0.49\textwidth]{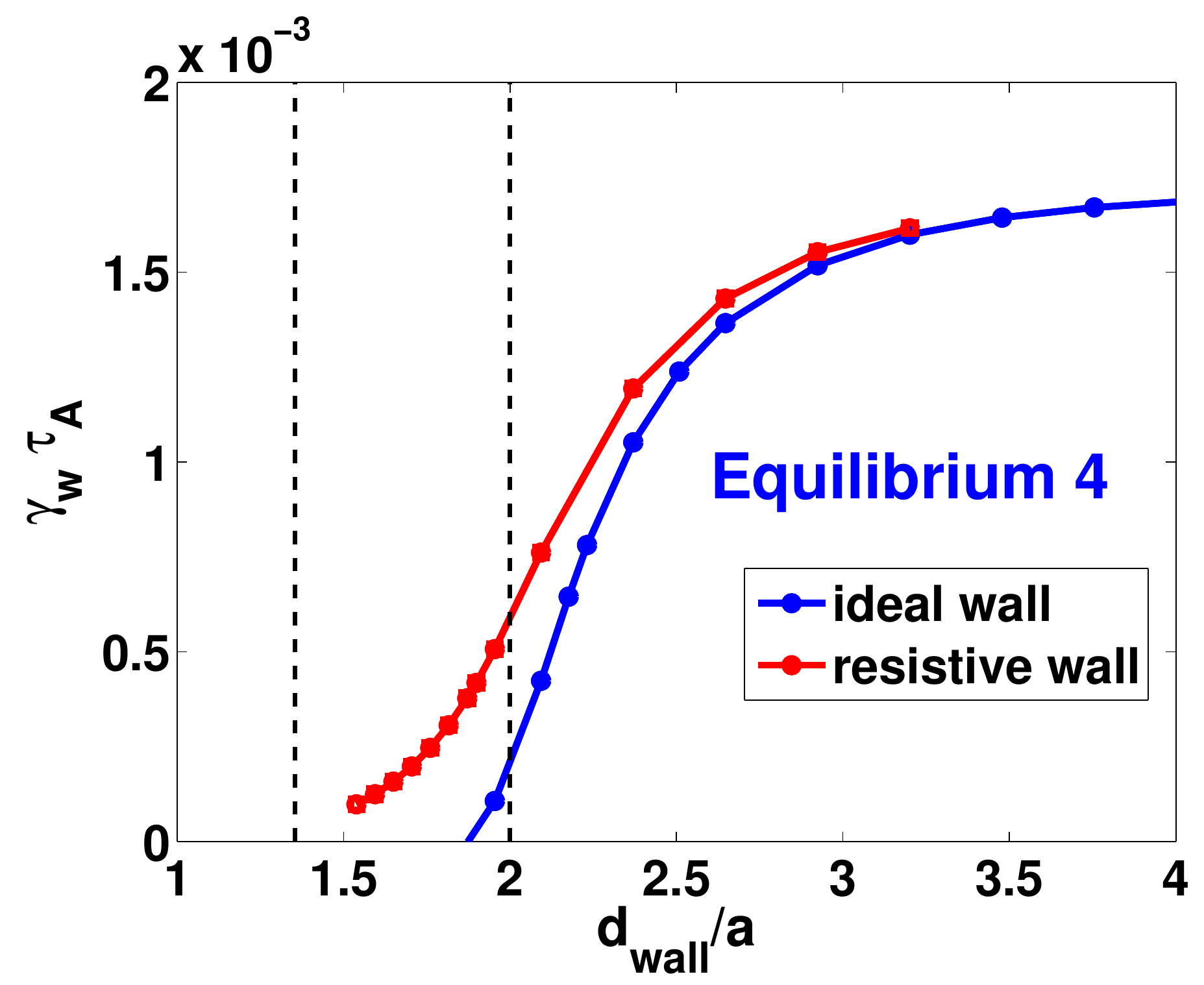}
  \put(-100,60){\textbf{(c)}}
  \includegraphics[width=0.49\textwidth]{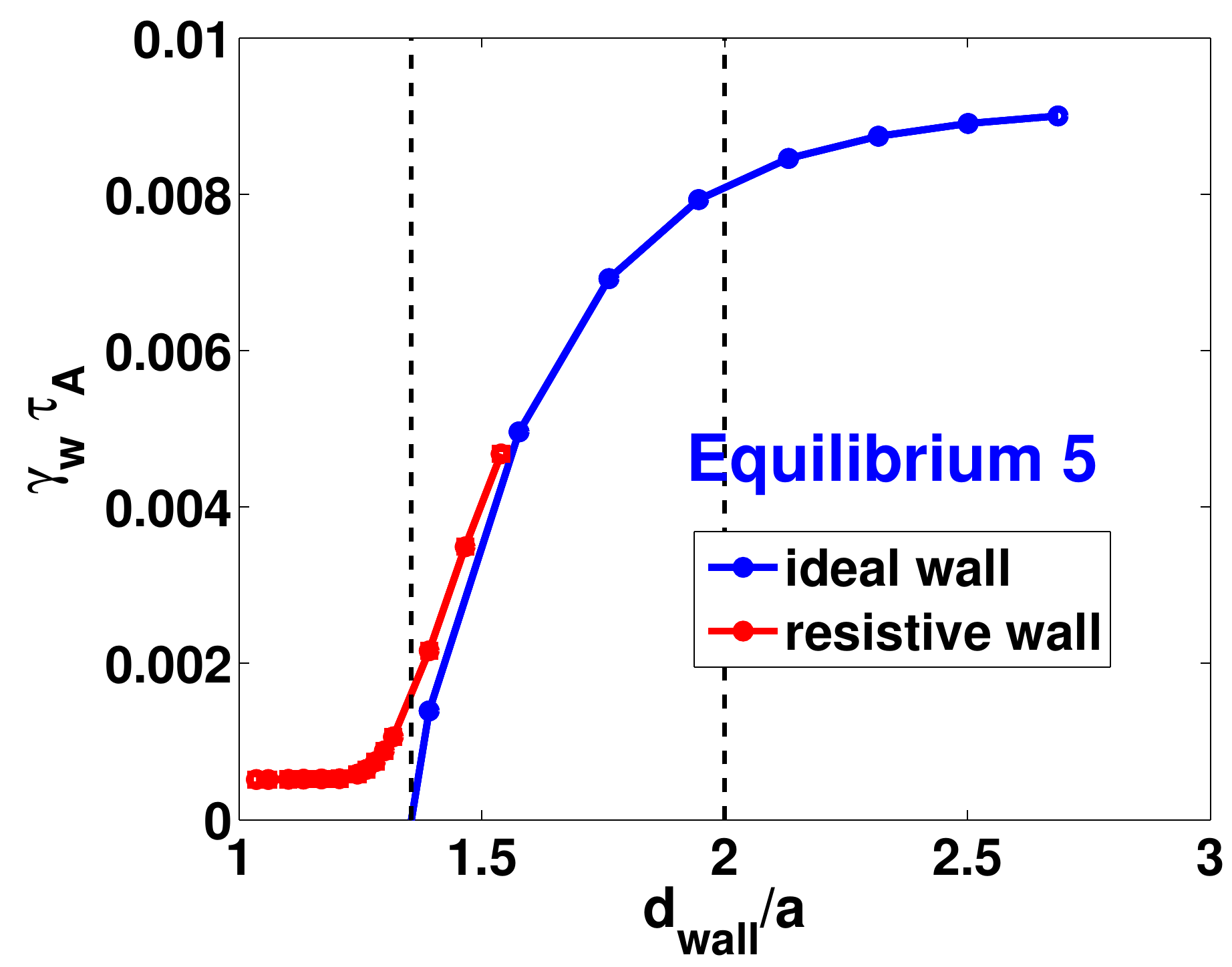}
  \put(-100,60){\textbf{(d)}}
\end{center}
\caption{Growth rates of the $n=1$ ideal MHD modes from MARS-F calculations as functions of the wall minor radius, in presence of an ideal wall (blue curves) or a resistive wall (red curves), where the effective resistive wall time is fixed at $\tau_w=10^4 \tau_A$. Here, the vertical dashed lines denote the minor radius of the TBM and VV designed for CFETR, at $1.3a$ and $2.0a$, respectively.}
\label{rwmfig5}      
\end{figure}

The growth rates of the external kink modes in presence of an ideal wall or resistive wall as functions of the wall minor radius are computed using MARS-F code (Fig.~\ref{rwmfig5}). Here, the effective wall time is estimated to be $\tau_w=10^4 \tau_A$. The vertical dashed lines denote the minor radius of the TBM and VV in CFETR. If the VV is designed to stabilize the ideal external kink modes, it can be found from these results that the VV may be located too far away from the plasma boundary, such that there is nearly no effect of VV on the growth rate of those modes, except for the scenario 4. Therefore, in the following discussion, we shall take this scenario 4 as an example to investigate on the RWM control scheme for CFETR.

%First and general, parts of the perturbed magnetic fields can be penetrated from the conducting wall, due to the presence of the resistivity, compared with using an ideal conducting wall, for all four CFTER scenarios. In other words, with applying these four plasmas as the steady state operation in CFETR, these low-$n$ MHD modes or RWM should be stabilized by other control methods. In addition, Fig.~\ref{rwmfig5} also shows the minor radius of the TBM and VV designed for CFETR. 

\begin{figure}
  \begin{center}
    \includegraphics[width=0.8\textwidth]{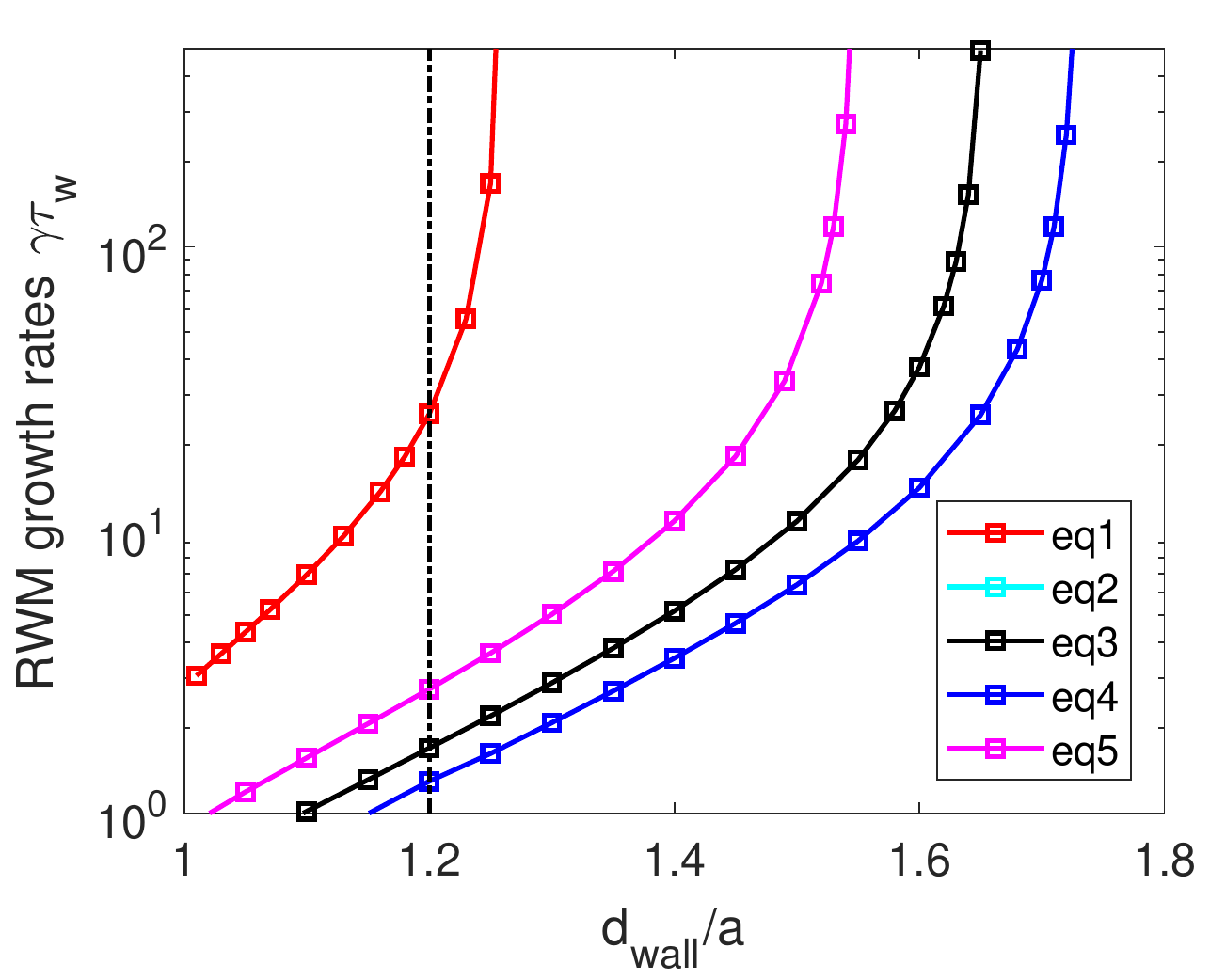}

    \includegraphics[width=0.8\textwidth]{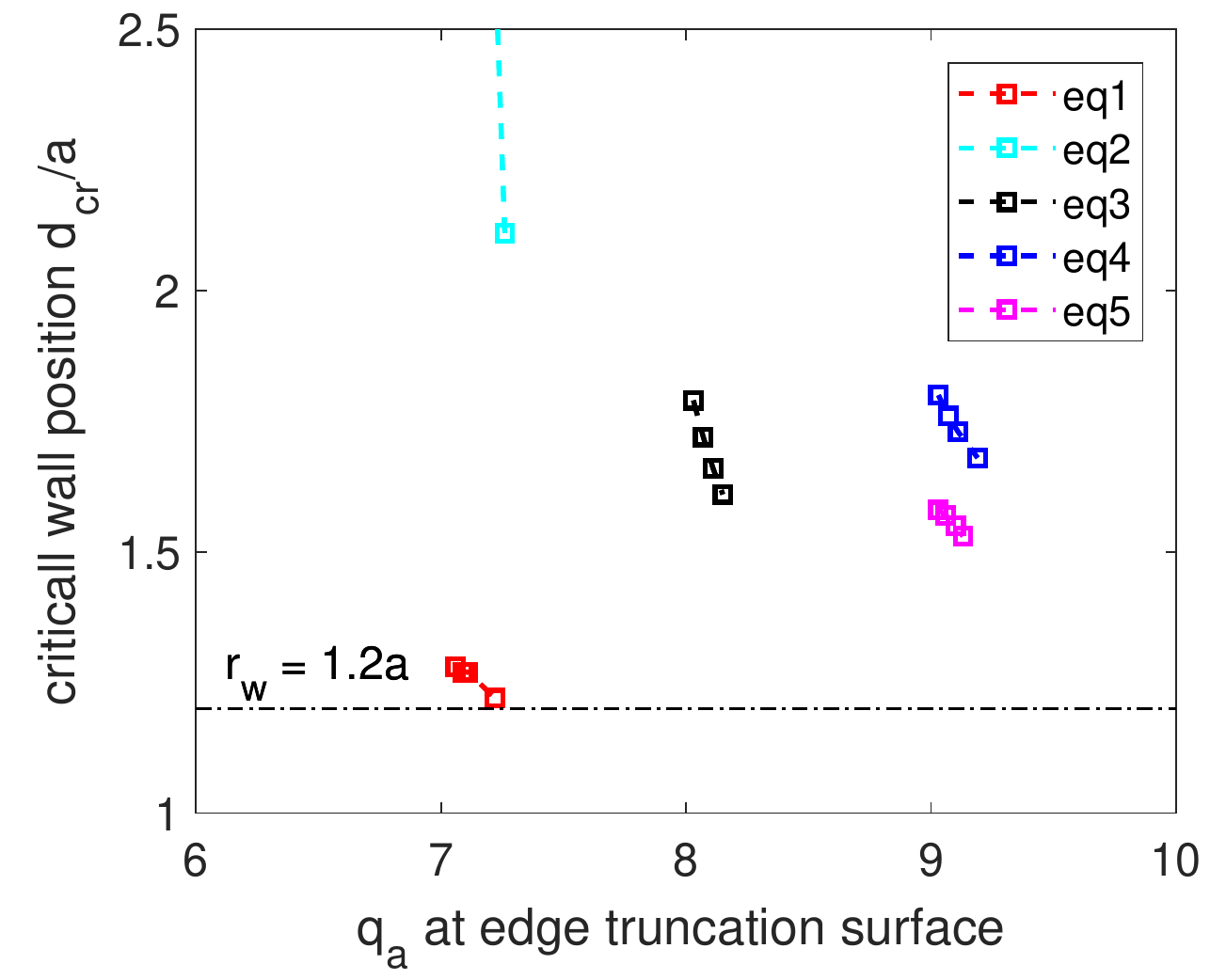}
  \end{center}
  \caption{(a) $n = 1$ RWM growth rates as functions of wall minor radius in presence of a conformal thin resistive wall for various CFETR scenarios (where for equilibrium 2, the $n=1$ ideal MHD mode is stable, thus no growth rate is shown), and (b) The critical wall positions where ideal MHD modes turn unstable as functions of the truncation edge safety factor $q_a$ for various CFETR scenarios from AEGIS calculations.}
\label{rwmfig6}
\end{figure}

\begin{figure}
  \begin{center}
  \includegraphics[width=0.8\textwidth]{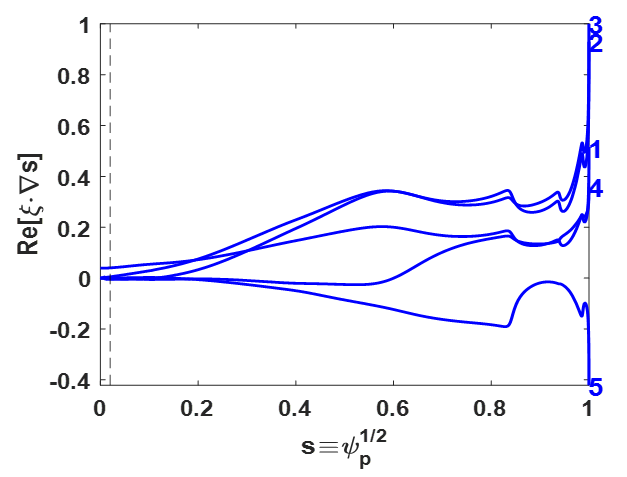}
  \put(-200,180){\textbf{(a)}}
  
  \hspace{0.2in}
  \includegraphics[width=0.85\textwidth]{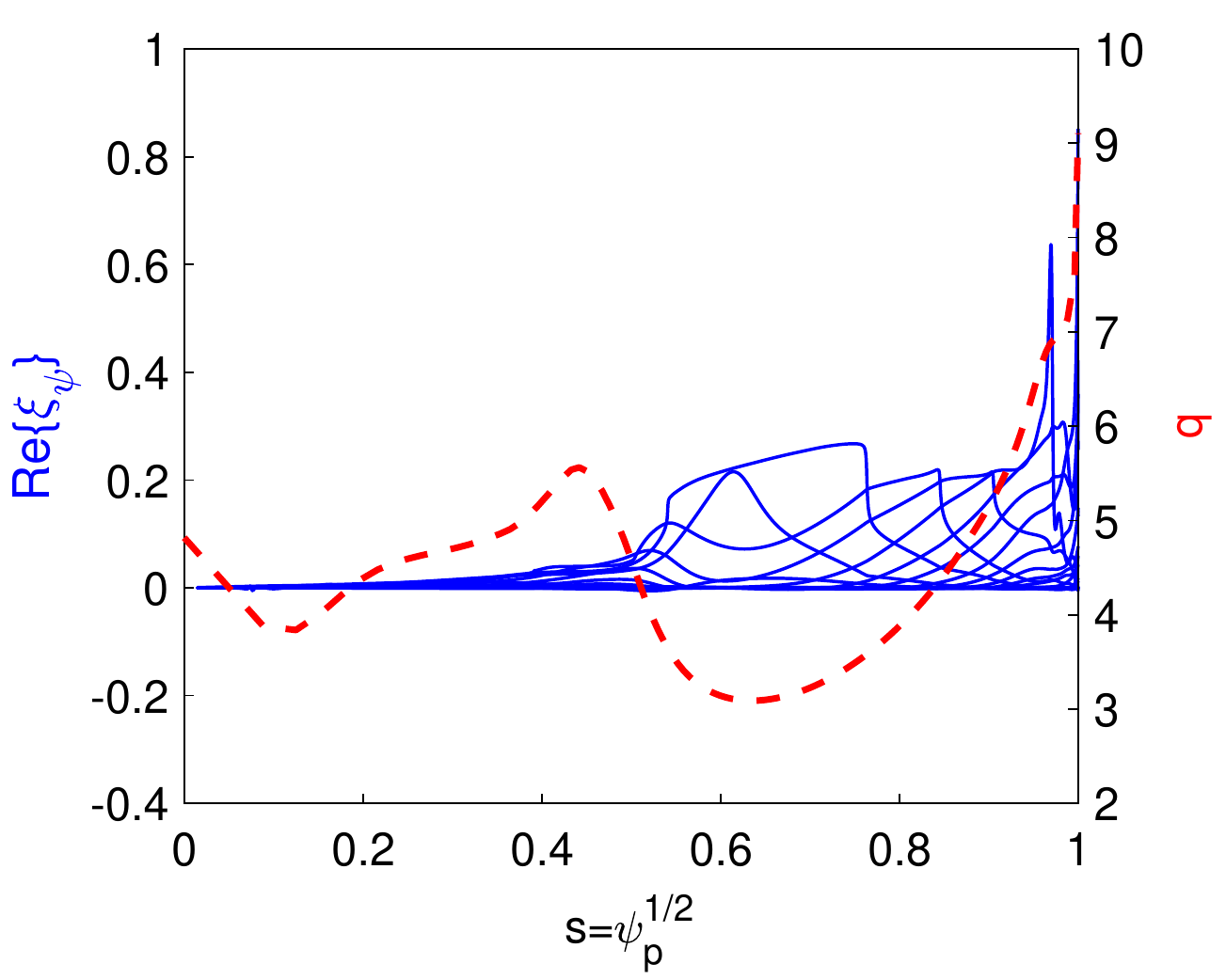}
  \put(-220,190){\textbf{(b)}}
  \end{center}
  \caption{Radial profiles of the real normal component of plasma displacement computed using (a) MARS code (with the geometric configuration shown in Fig.~\ref{rwmfig2}(b)), and (b) AEGIS code for equilibrium 4.}
\label{rwmfig7}      
\end{figure}

%Similar study is also computed by AEGIS code.
The marginal stability boundaries of the five SSO equilibria in presence of the perfectly conducting wall are also obtained from AEGIS calculations, along with the explicit effects of the equilibrium flux surface domain truncation (Fig.~\ref{rwmfig6}). Note that a small fraction of plasma edge region is truncated off in AEGIS computation domain in order to avoid the X-point singularity in the flux coordinate representation of tokamak equilibria, a practice similarly adopted in the MARS-F calculations as well. For a fixed wall location, the ideal MHD mode growth rate in presence of a resistive wall in scenario 4 is the smallest. Furthermore, the radial profiles of the real part of the perturbed normal displacement computed using MARS-F and AEGIS codes are compared for the same target plasma $\beta$ and resistive wall location at $d_{wall}=1.2a$, which show similar global mode structure (Fig.~\ref{rwmfig7}).

\subsubsection{Instabilities of RWMs with designed wall}
In this sub-section, we consider the model for designed wall including both TBM and VV components, and evaluate the contribution of these structures to the growth rate of the ideal MHD modes. Based on the findings from previous subsection, here the analysis is focused on the equilibrium of scenario 4 as an example. In addition, because the VV is so far away from the plasma torus that the TBM has to be also taken into account. However, the exact properties of material of the TBM remains unknown at this stage. A parameter $C$ for the ratio of resistivities between TBM and VV is introduced as $C=\tau_{TBM}/\tau_{VV}$, and the VV wall time is assumed to be $\tau_{VV}=1865.3ms$.

\begin{figure}
  \begin{center}
    \includegraphics[width=0.8\textwidth]{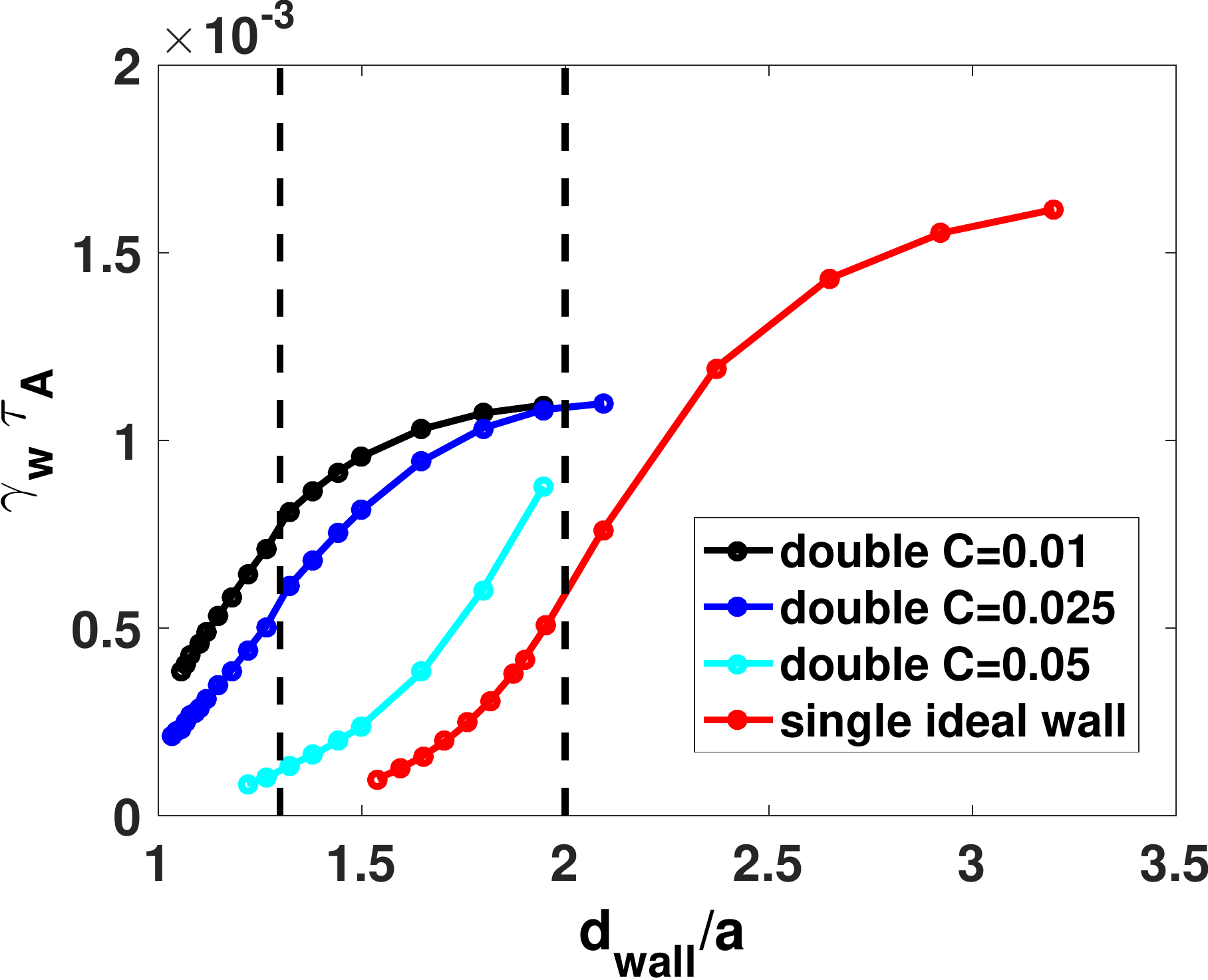}
  \end{center}
  \caption{Growth rates of $n=1$ modes with different ratios of TBM and VV wall time $C=\tau_{TBM}/\tau_{VV}$ as functions of the wall minor radius, where the VV wall time is fixed at $\tau_{VV}=1865.3ms$. Here, the vertical dashed lines denote the minor radii of the TBM and VV locations designed for CFETR ($1.3a$ and $2.0a$ respectively).}
\label{rwmfig8}      
\end{figure}

Fig.~\ref{rwmfig8} compares the RWM growth rates as functions of the TBM location from MARS-F calculations for several different values of the ratio parameter $C$, as well as the case with a single ideal conducting wall, where the dependence of growth rate is on the ideal wall location. The ratio $C$ designed for ITER is $0.05$. As shown in Fig.~\ref{rwmfig8}, the growth rate of ideal MHD modes is higher with lower $C$-value due to higher resistivity of TBM. However, if TBM can achieve to the condition of blanket designed for ITER, this ideal MHD mode can become nearly marginal stable with the synergistic effect of TBM and VV.

\subsubsection{Effect of plasma shaping on instabilities of RWM}
Based on the geometry design for CFETR, we note that the VV boundary is far away from the plasma-vacuum interface, so that the effect of conducting wall on the growth rate of unstable MHD modes is expected weak. In this section, we shall gradually modify the shape of VV in the lower outboard corner. Only CFETR scenario 4 with the designed wall model, is evaluated in this section.

\begin{figure}
  \begin{center}
  \includegraphics[width=0.8\textwidth]{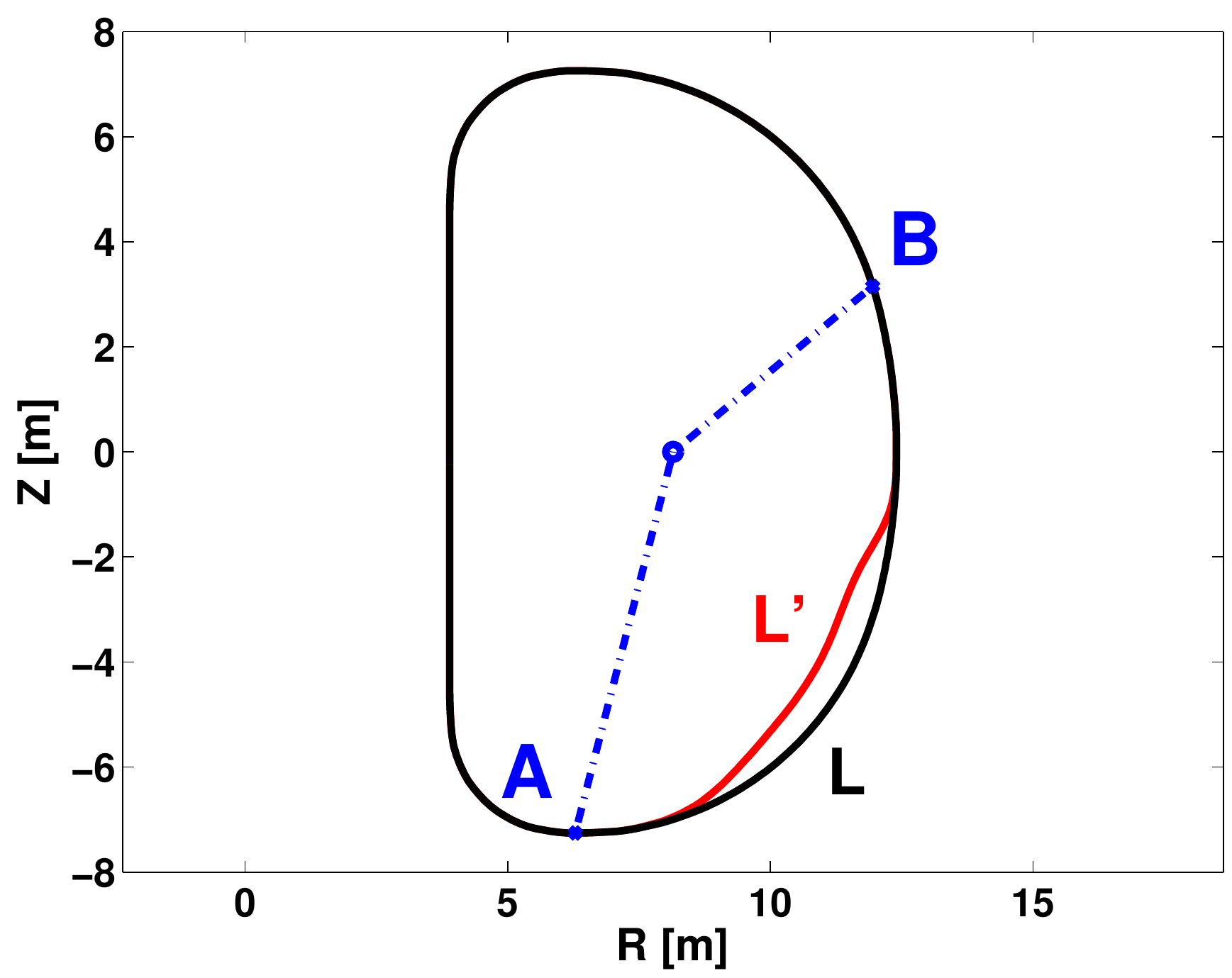}
  \put(-80,70){\textbf{(a)}}
  
  \includegraphics[width=0.8\textwidth]{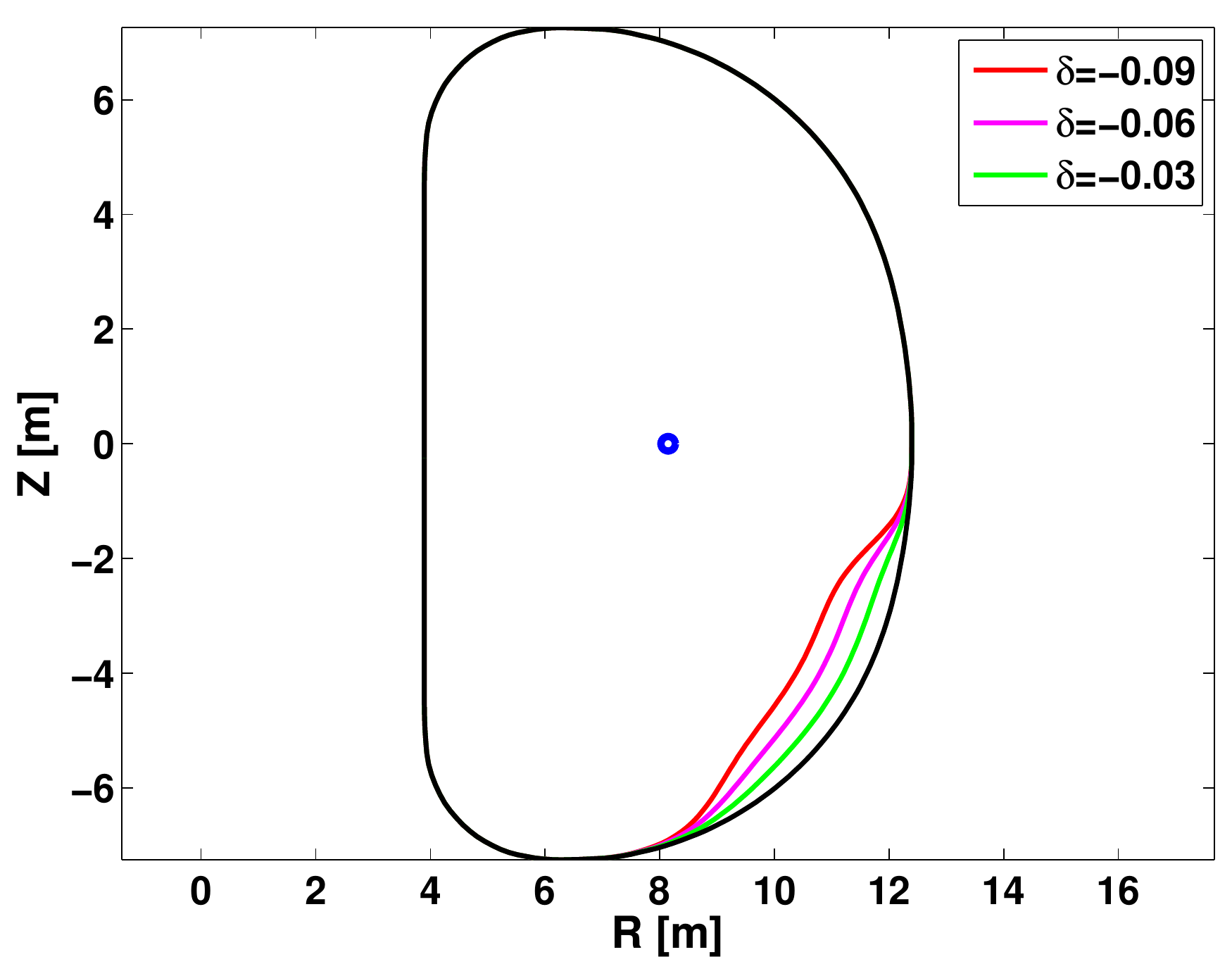}
  \put(-80,70){\textbf{(b)}}

  \includegraphics[width=0.8\textwidth]{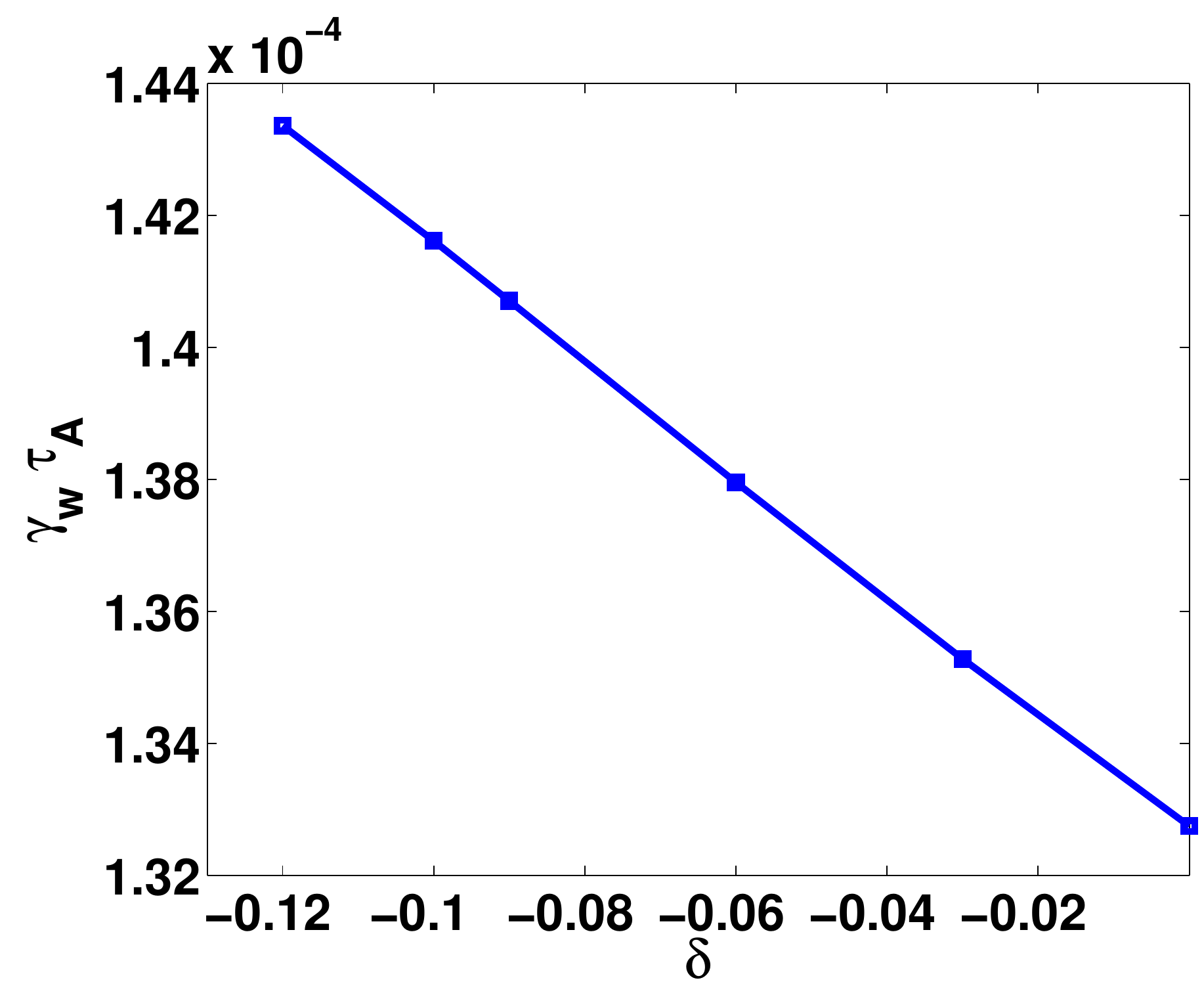}  
  \put(-80,70){\textbf{(c)}}  
  \end{center}
  \caption{(a) A sketch of the modification of the lower low field side quarter of the plasma boundary shape, based on the CFETR design; (b) a family of new VV shape; (c) the RWM growth rate as a function of the shaping parameter $\delta$.}
\label{rwmfig13}      
\end{figure}

Our primary concern is the effect of the lower triangularity of VV shape on the growth rate of the ideal MHD modes. We introduce an analytic model for systematically scanning the lower triangularity, which can be written as
\begin{equation}
\rho^{'}=\rho\left\{1+\delta\exp{\left[-\frac{(\phi-\phi_m)^4}{2\kappa^2}\right]}\right\}, \quad \phi\in[\phi_A,\phi_B]
\end{equation}
where $R^{'}=\rho^{'}cos\phi$ and $Z^{'}=\rho^{'}sin\phi$. A sketch of this model in the $(R,Z)$ plane is shown in Fig.~\ref{rwmfig13}(a). The black curve L is the original VV shape designed for CFETR. The red curve $L^{'}$ is the new shape obtained from varying parameter $\delta$ along the poloidal circumference between points A and B. In this analysis, other parameters are fixed as $\phi_m=0.8$ and $\kappa=0.2$, and only the value of $\delta$ is varied to obtain a family of new VV shapes (Fig.~\ref{rwmfig13}(b)). For the new family of VV shape and TBM with $C=0.05$, the growth rates of the ideal MHD modes are shown to decrease with the magnitude of $\delta$, indicating a stabilizing effect as the VV draws closer to the plasma surface. However, such a stabilizing effect is rather limited (Fig.~\ref{rwmfig13}(c)).

\subsubsection{Stabilization of RWM based on plasma flow}
As mentioned earlier, the designed CFETR scenarios except scenario 2, are unstable to the ideal MHD modes in presence of resistive wall or in absence of wall. In this section, we evaluate the potential stabilization of these unstable RWMs by plasma flow. The conformal resistive wall is assumed to be located at $d_{wall}=1.2a$.

\begin{figure}
  \begin{center}
  \includegraphics[width=0.8\textwidth]{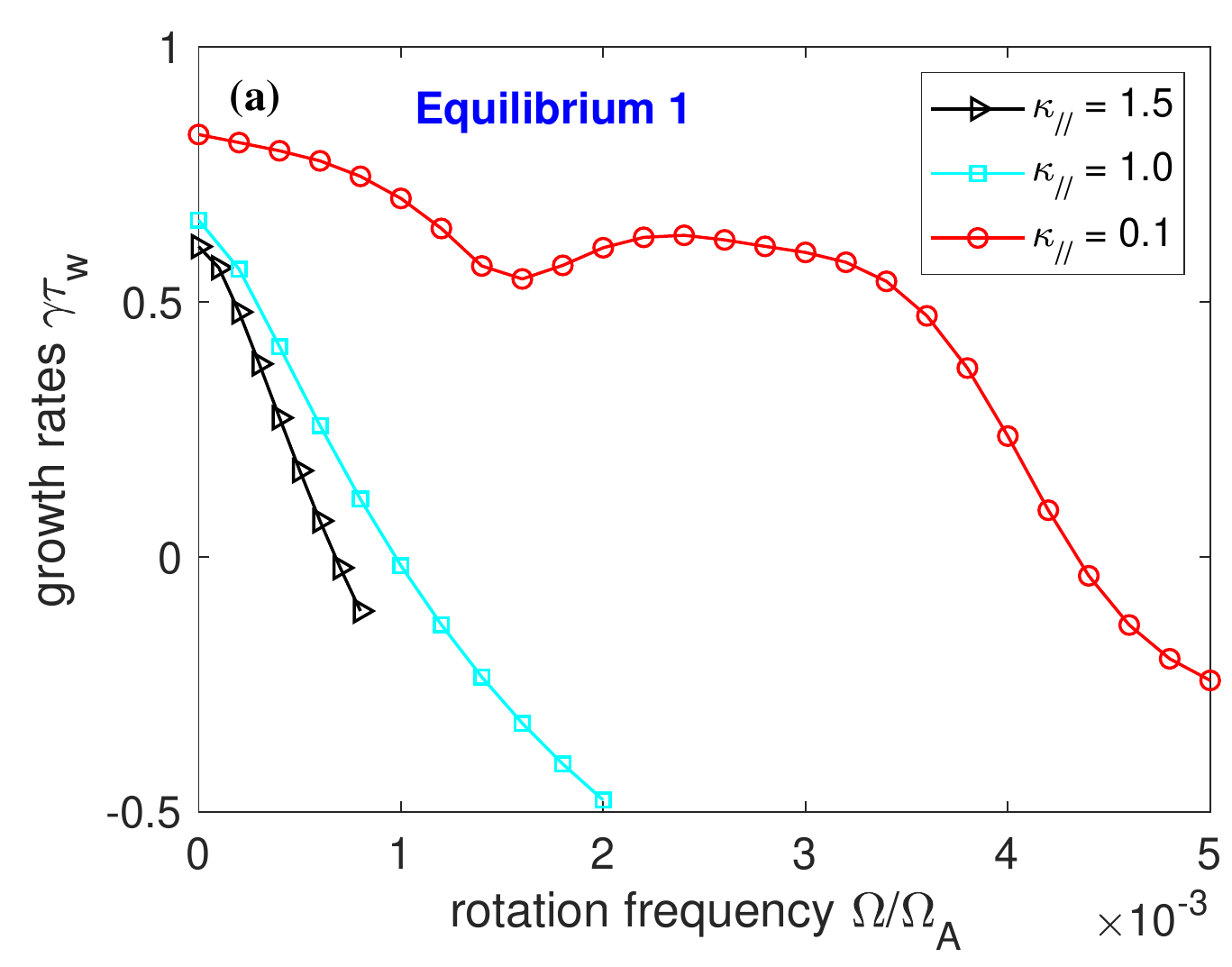}
  %\put(-110,90){\textbf{(a)}}

  \includegraphics[width=0.8\textwidth]{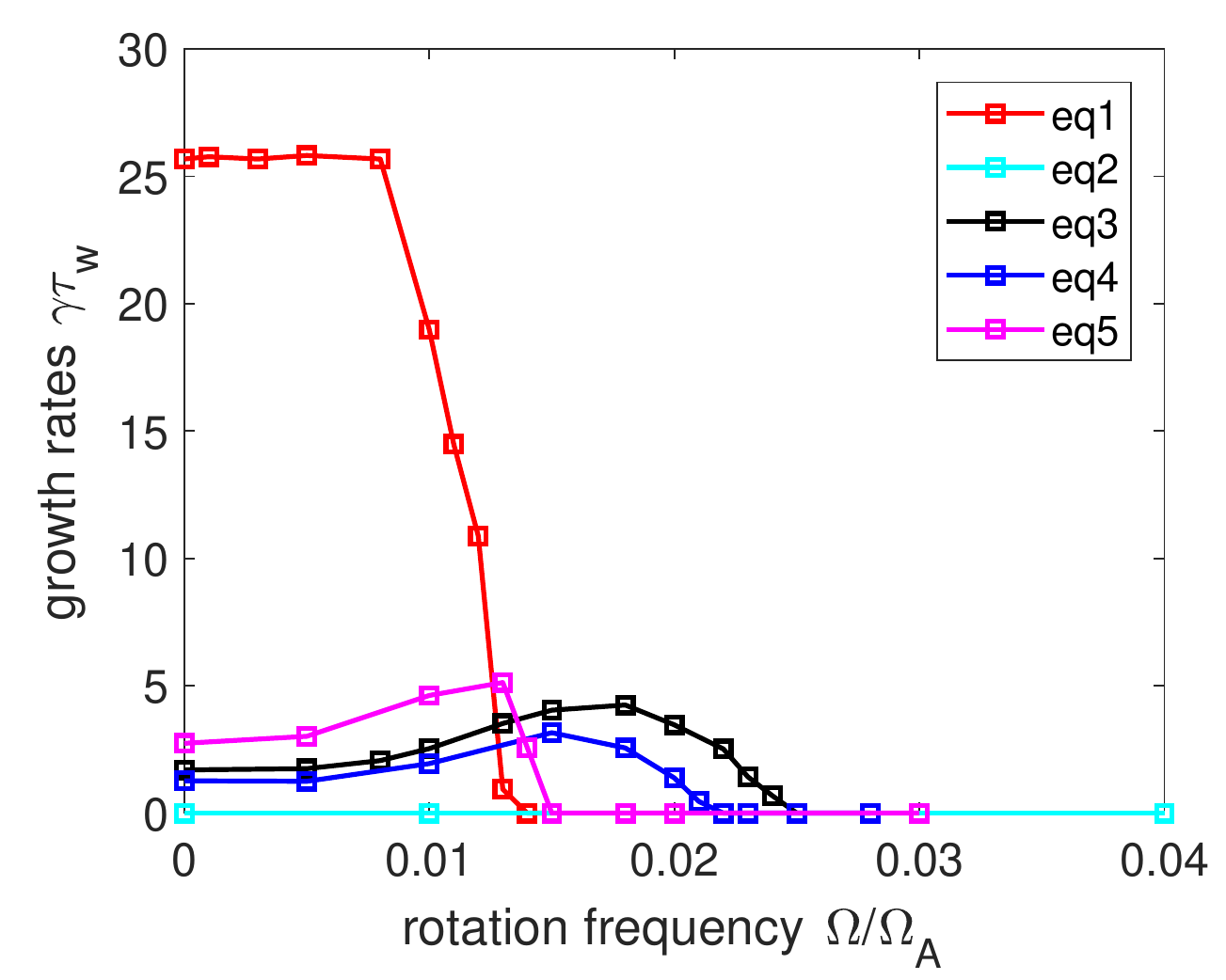}
  \put(-222,192){\textbf{(b)}}
  \end{center}
  \caption{RWM growth rates as functions of the uniform toroidal rotation frequency obtained using (a) MARS-F and (b) AEGIS codes, for various values for the damping coefficient, $\kappa_\parallel$. The rotation frequency is normalized by the Alfv\'en frequency at the magnetic axis.}
\label{rwmfig9}      
\end{figure}

In MARS-F calculations, the equilibrium of scenario 1 along with a uniform rotation profile is considered. The RWM growth rates are shown to decrease with rotation frequency, and such a rotational stabilization is also enhanced with higher ion Landau damping rate $\kappa_{\parallel}$ (Fig.~\ref{rwmfig9}(a)). Similar calculations using AEGIS code for all five CFETR scenarios indicate that the unstable RWMs can all be stabilized by plasma rotation at a few percent (1.5\%-2\%) of Alfv\'en speed (Fig.~\ref{rwmfig9}(b)), where the shear-Alfv\'en continuum resonance in the rotating plasma leads to the stabilization of resistive wall modes. 

%shows the normalized growth rates, $\gamma\tau_w$, of the RWM versus the rotation frequency, $\Omega_0/\Omega_A$. Here $\Omega_A$ (defined as $v_a/R_0$) is the central Alfv\'en frequency. $v_A=B_0/\sqrt{\mu_0\rho_0}$ is the Alfv\'en speed, where $B_0$ is the toroidal filed and $\rho_0$ is the mass density at the center of the plasma. The damping coefficient,  , is scanned.  Whatever, the unstable RWM can be suppressed by the passive control. \\

\subsubsection{Active control of RWM based on feedback}
\begin{figure}
  \begin{center}
    \includegraphics[width=0.8\textwidth]{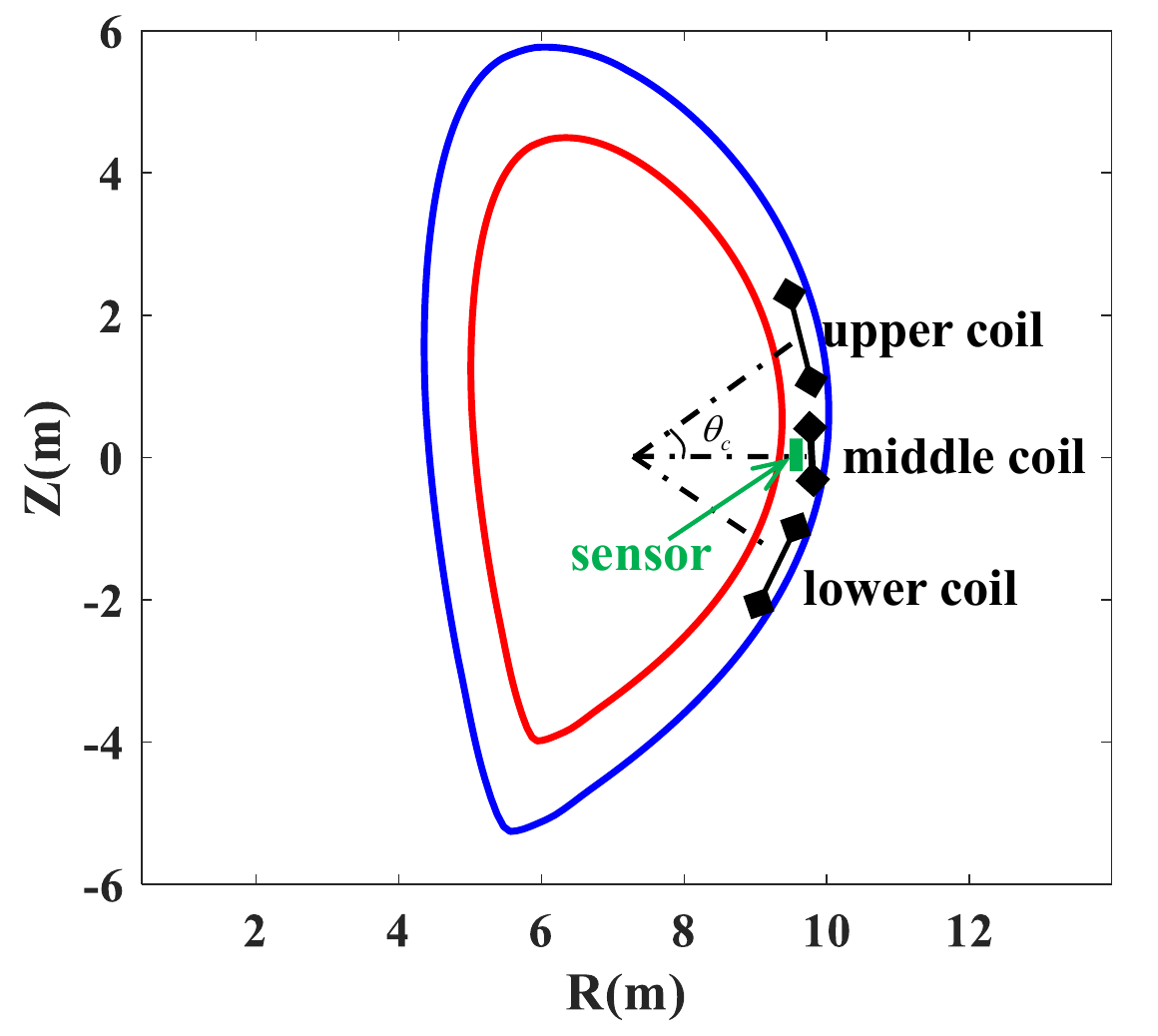}
  \end{center}
  \caption{Geometry of a CFETR configuration including the RWM feedback coils, the plasma boundary, and a single resistive wall ($r_w =1.3$ and $\tau_w=10^4\tau_A$). Two sets of active coils, referred to as the middle and upper-lower symmetric coils, respectively, are located inside the wall. The sensor coil is located inside the wall, measuring the poloidal field perturbation.}
\label{rwmfig10}      
\end{figure}

In addition to the passive stabilization of RWMs by wall design and toroidal rotation, we continue to evaluate active control scheme for the RWM in CFETR based on feedback coils, considering for example the equilibrium of scenario 2 with a conformal conducting wall. Three rows of feedback coils, the upper, middle and lower rows along the toroidal angle, are shown in Fig.~\ref{rwmfig10}, where $\theta_c$, normalized by $\pi$, denotes the poloidal angle of the coil location and the resistive wall is assumed to locate at $r_w=1.3$.

For the feedback relation $M_{sf}I_f=-G\psi_s(t)$ and the transfer function $P(s)=\psi_s/M_{sf}I_f$, the characteristic equation with proportional feedback satisfies $1+GP(s)=0$, where $s$ is the Laplace variable representing the mode eigenvalue. Here $G=|G|e^{i\Phi}$ is feedback gain, and $|G|$ and $\Phi$ represent its amplitude and phase respectively. $\psi_s(t)$ is magnetic signal. And $M_{sf}$ is the free-space mutual inductance between the feedback coil and the sensor loop, used largely to normalize the feedback gain. $I_f$ is the current in active control coil.

\begin{figure}
  \begin{center}
  \includegraphics[width=0.75\textwidth]{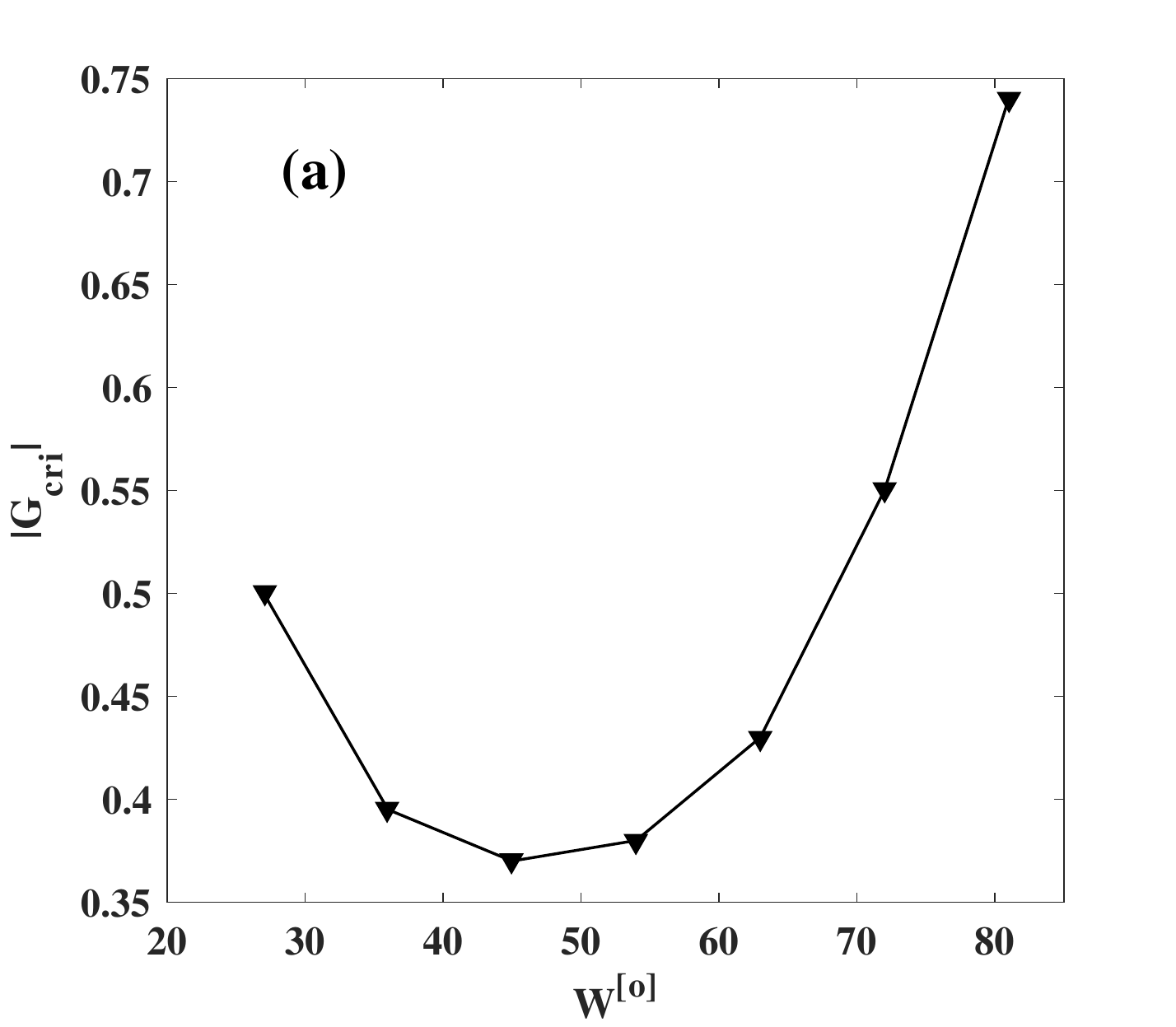}
  \put(-80,70){\textbf{(a)}}
  
  \includegraphics[width=0.75\textwidth]{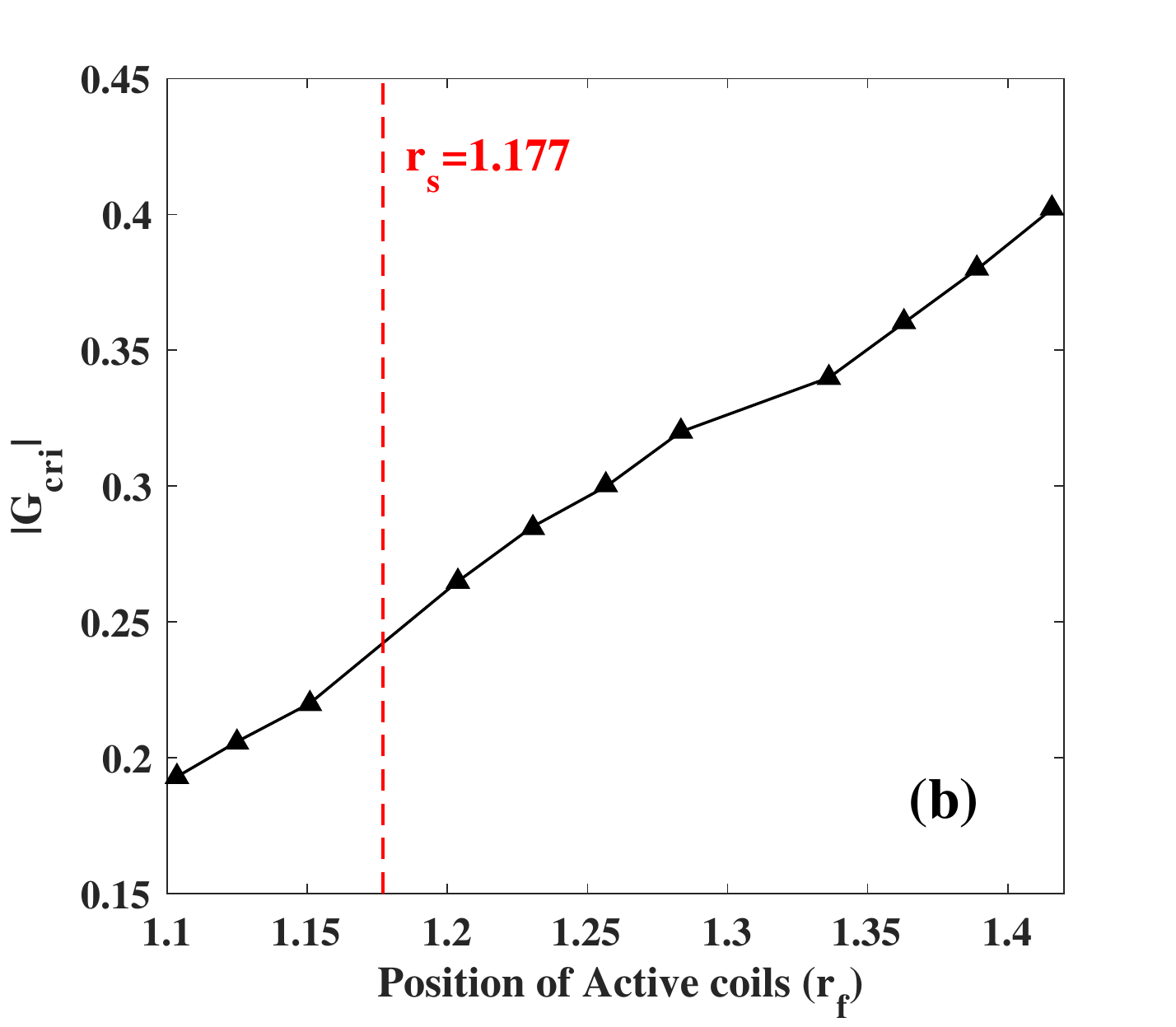}
  \put(-80,70){\textbf{(b)}}
  
  \includegraphics[width=0.75\textwidth]{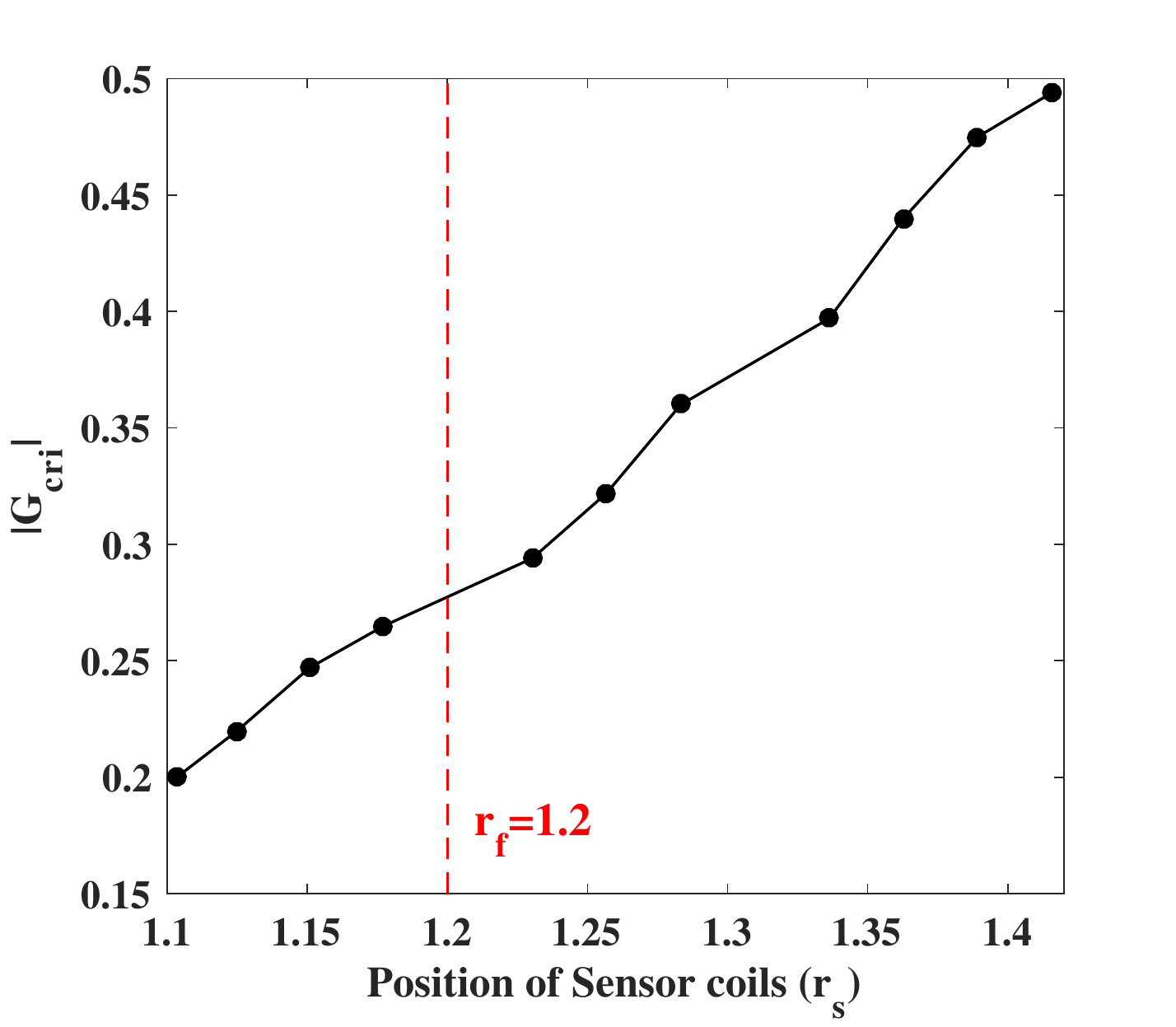}  
  \put(-80,70){\textbf{(c)}}  
  \end{center}
\caption{(a) The critical gain required for stabilization of the RWM as a function of the poloidal covering width of active coil $W$ with fixed $r_f=1.283$ and $r_s=1.257$. (b) and (c) are the critical gains as functions of the positions of active coils ($r_f$) and sensor coils ($r_s$), respectively, with fixed $W=45^\circ$. The feedback system is assumed to be a set of middle control coils with zero polar angle and zero phase angle.}
\label{rwmfig11}      
\end{figure}

\begin{figure}
  \begin{center}
  \includegraphics[width=0.49\textwidth]{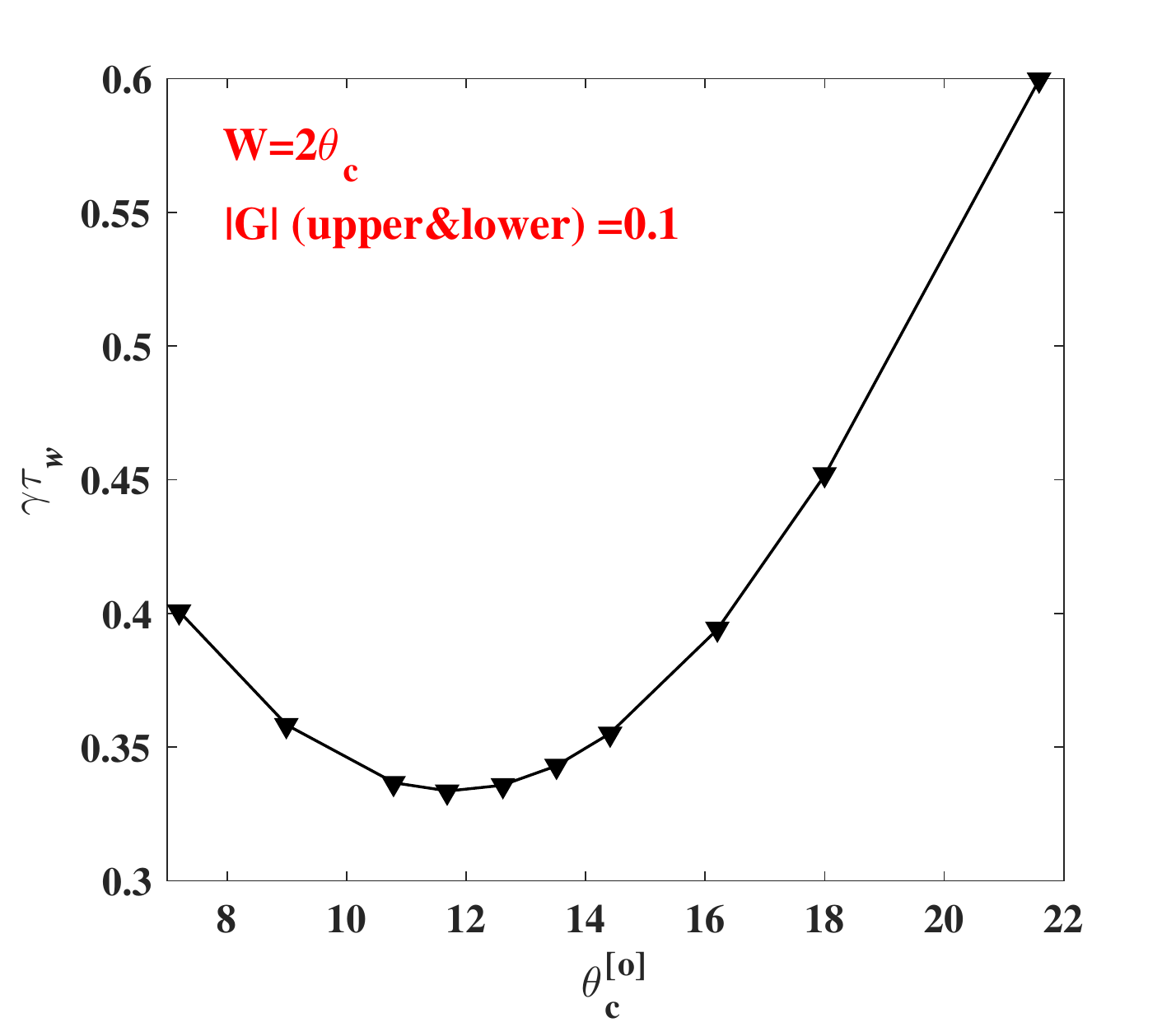}
  \put(-100,60){\textbf{(a)}}
  \includegraphics[width=0.49\textwidth]{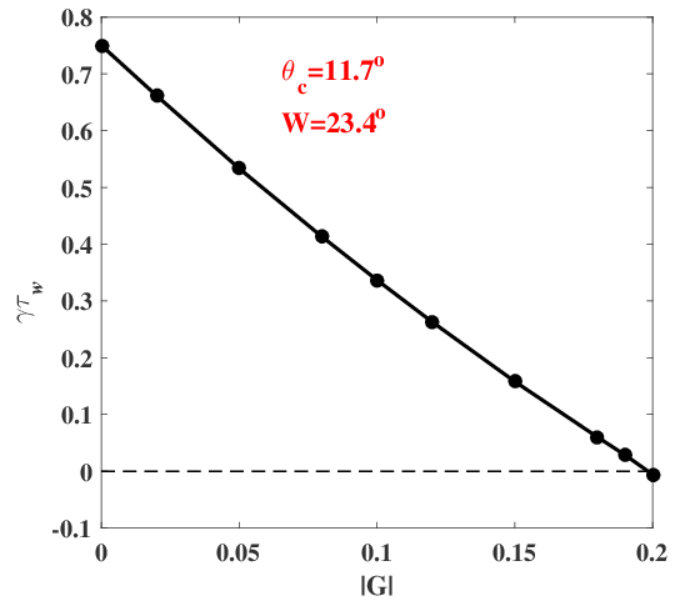}
  \put(-100,60){\textbf{(b)}}
  
  \includegraphics[width=0.49\textwidth]{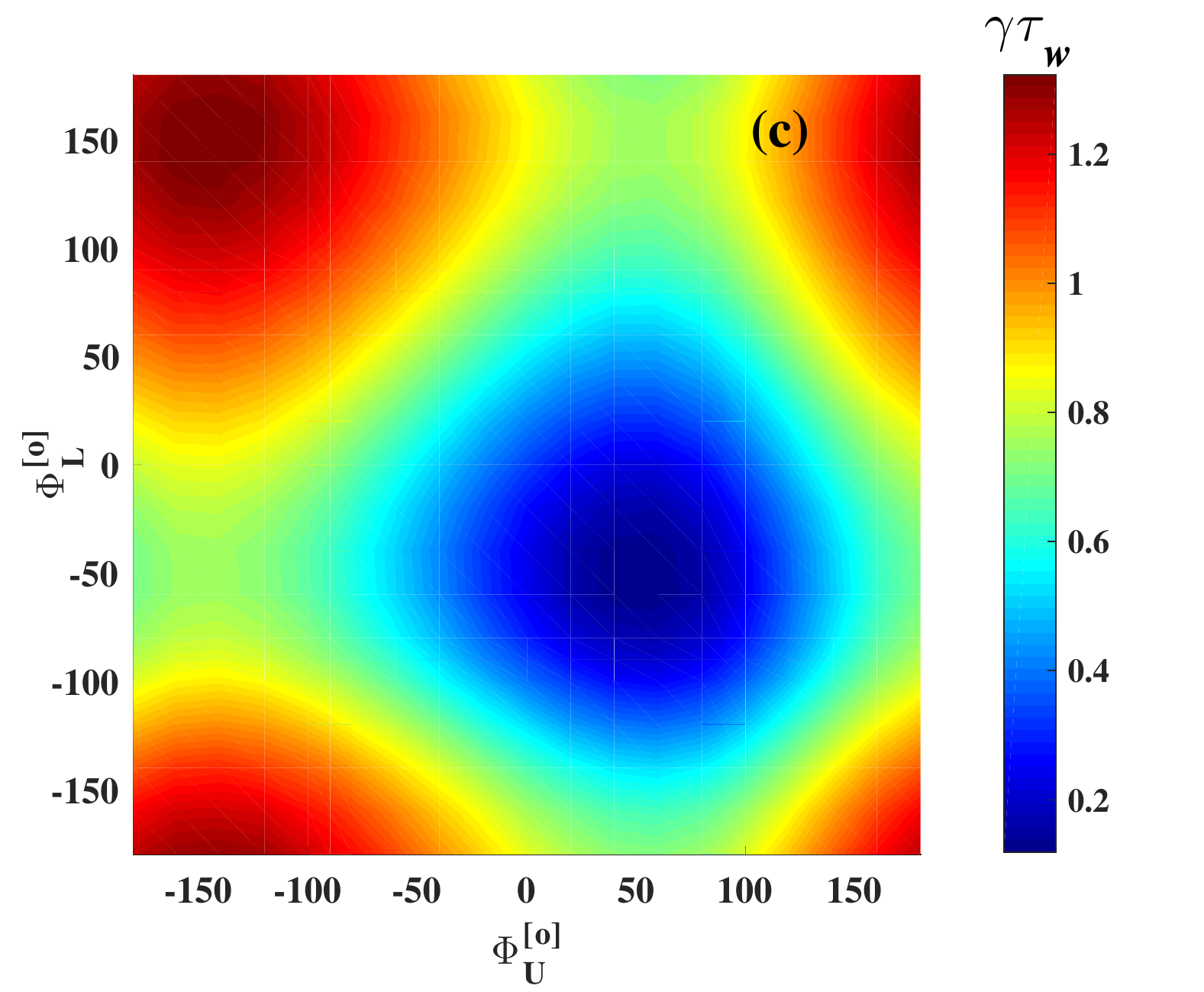}
  %\put(-100,60){\textbf{(c)}}
  \includegraphics[width=0.49\textwidth]{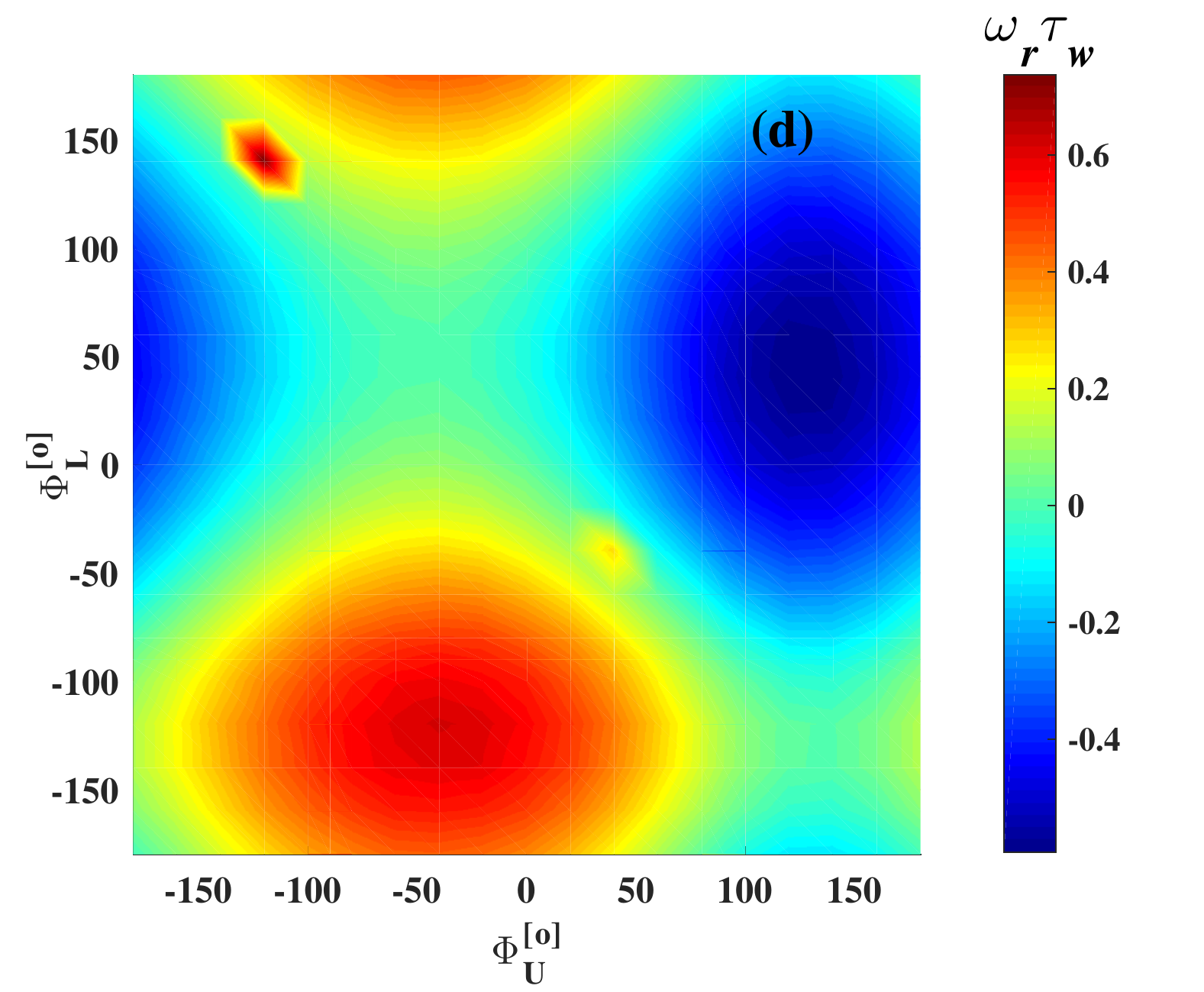}
  %\put(-100,60){\textbf{(d)}}
  \end{center}
\caption{(a) Growth rate of RWM as a function of the poloidal location ($\theta_c$) of the upper and lower symmetric active coils with the fixed gain $|G|=0.1$ and the poloidal covering width $W=2\theta_c$. The optimal poloidal location is $\theta_c=11.7^\circ$. (b) The growth rate of RWM as a function of the gain with fixed poloidal angles of coils. The critical gain is $|G|=0.2$ smaller than the middle active coil. Contours of (c) growth rate and (d) real frequency of RWM in the 2D space of the proportional feedback gain phases $\Psi_U-\Psi_L$, respectively.}
\label{rwmfig12}      
\end{figure}

Analysis indicates that the RWM can be stabilized by such a feedback coil system. Assuming only a set of middle active coils $(\theta_c=0)$ are used, as shown in Fig.~\ref{rwmfig11}, the minimum critical gain is obtained to be $|G|=0.3$, when the poloidal covering width of active coil $W=45^{\circ}$, normalized by $\pi$. And the critical gain decreases as the radial position of active or sensor coils becomes closer to plasma surface. The stabilizing effect of feedback coil is optimal in the absence of phase angle. If only a set of upper and lower symmetric active coils are employed as in Fig.~\ref{rwmfig12}, the optimal poloidal positions for upper and lower active coils are $|\theta_c|=11.7^{\circ}$, in the absence of phase angle, when the gain amplitude is same for both upper and lower active coils $|G|=0.1$ and the poloidal covering width of active coil is $W=2\theta_c$. For the optimal parameter as in Fig~\ref{rwmfig12}(a), the critical gain is $|G|=0.2$. Fig~\ref{rwmfig12}(c-d) show that the RWM stabilization sensitively depends on the feedback gain phase. The best phases for reducing the RWM growth rate are $\Phi_U\sim 50^{\circ}$ and $\Phi_L\sim -50^{\circ}$.

\subsection{CFETR error field tolerance}
\label{errorfiled}
The estimate on error field tolerance for CFETR is based on the design parameters that the major radius is $R=7.62m$, the minor radius $a=2.25m$, and the toroidal field $B_T=6.5T$, and the assumptions that the safety factor at the 95\% of magnetic flux $q{95}\sim 5.5$, the electron density $n_e=8.94\times 10^{19}m^{-3}$, and the rotation frequency $f=2.34\times10^4/(2\pi)Hz$ in the hybrid scenario. For comparison, we assume that in ITER the toroidal field $B_T=5.3T$, the safety factor at the 95\% magnetic flux $q_{95}\sim 3.2$, the rotation frequency $f=3kHz$, and the electron density is same as that in CFETR. We evaluate the error field tolerances for both CFETR and ITER by extrapolation using theoretical scaling laws~\cite{fitzpatrick1998pop,wanghuihui2015pst,cole2007prl,fitzpatrick2012ppcf} and the experimental ones on the basis of the EAST error field penetration threshold data~\cite{wanghuihui2018nf,wanghuihui2020nf,hender1992nf,buttery1999nf,lazzaro2002pop} (Table.~\ref{table_ef}).

According to the theoretical extrapolation~\cite{fitzpatrick1998pop,wanghuihui2015pst,cole2007prl,fitzpatrick2012ppcf}, the error field tolerance ($b_r/B_T$) of CFETR is similar to that of ITER, from both MHD and two-fluid models. Most extrapolation results on the error field tolerance using the experimental scalings~\cite{wanghuihui2018nf,wanghuihui2020nf,hender1992nf,buttery1999nf,lazzaro2002pop} are similar for CFETR and ITER as well. However, when the safety factor at the 95\% magnetic flux $q_{95}$ is taken into account, such as in the case of EAST experimental scaling, the error field tolerance is larger than that in ITER where the $q_{95}\sim 3.2$ in ITER is much lower than the $q_{95}~5.5$ in CFETR. That is to say if $q_{95}$ is higher in CFETR than that in ITER, then the CFETR operation would be less susceptible to error field.
%The correction field coils should be under consideration for CFETR.
At present the correction field coils are designed for ITER. If CFETR operates in the ITER-like operational scenario, then the correction field coils are needed. If CFETR operates in much higher safety factor at the 95\% magnetic flux than that in ITER, then the correction field coils may not be absolutely necessary. 

%\newpage
\subsection{Control of the neoclassical tearing modes by electron cyclotron current drive in CFETR}
\label{NTM-ECCD}
Neoclassical tearing mode (NTM), if uncontrolled, limits the performance of advanced tokamak devices such as CFETR. The NTM induced large magnetic islands can significantly degrade the plasma confinement or even lead to the plasma disruption. Thus, the control of NTMs is necessary for the steady operations of CFETR. To suppress NTMs, extra current drive can be deposited near the magnetic island region to compensate the loss of bootstrap current caused by the flattening of pressure inside the magnetic island. In the tokamak experiment, it has been verified that electron cyclotron current drive (ECCD) is an effective method to for that purpose. Here we numerically investigate the ECCD control schemes for NTMs in the hybrid scenarios of CFETR.

\subsubsection{Benchmark between MD and NIMROD on nonlinear tearing mode in hybrid scenario}
%, which has been further confirmed by the NIMROD and TM8 codes.
%(presented in the following subsections).
%It is shown that the tearing mode grows exponentially from a noise level and saturates in the Rutherford regime. 

For the purpose of investigating tearing instability of CFETR hybrid scenario, we first use polynomials to fit the original equilibrium profiles. Then we use the equilibrium solver NIMEQ to generate a new equilibrium on the NIMROD simulation mesh based on the fitted pressure and safety factor profiles, along with the CFETR boundary shape. Toroidal mode components $n=0-1$ are included in the nonlinear NIMROD simulation here. Plasma parameters are set as following: $S=10^5$, $P_{rm}=0.26$, $\tau_A=5\times10^{-7}s$.

\begin{figure}
\centering
  \includegraphics[width=0.9\textwidth]{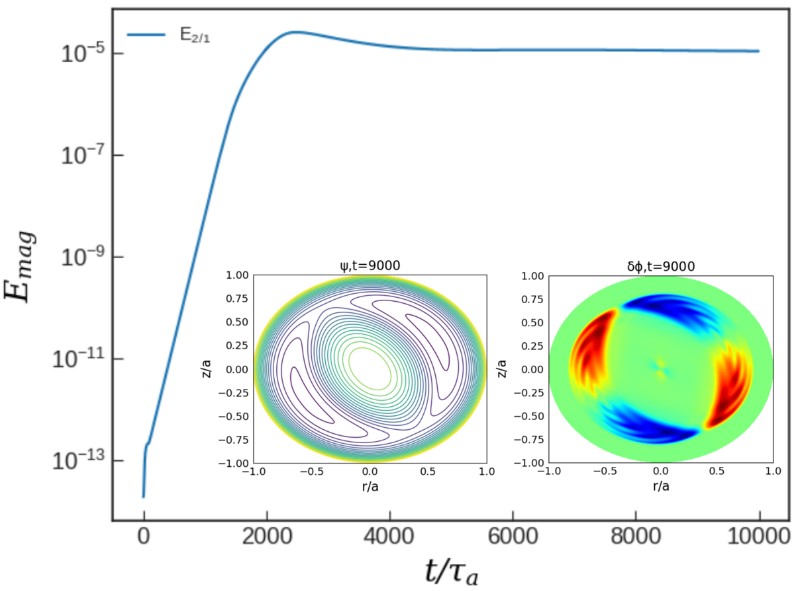}
  \put(-250,200){\textbf{(a)}}
  
  \includegraphics[width=0.95\textwidth]{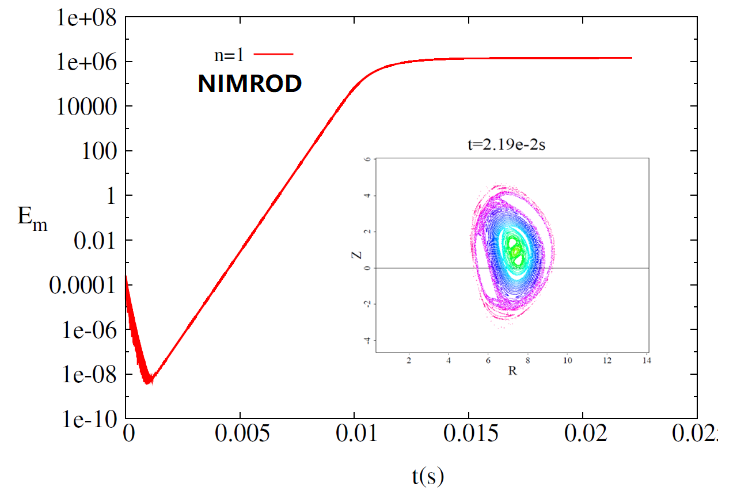}
  \put(-255,185){\textbf{(b)}}
\caption{(a) Nonlinear evolution of the $m/n=2/1$ resistive tearing mode in the hybrid scenario EQ3, together with the corresponding magnetic island and the plasma flow pattern in the saturation phase from the MD simulation. (b) Evolution of $n=1$ component of magnetic energy, together with the Poincare plot in the saturation phase from the NIMROD simulation.}
\label{NTM_bench_nim_md}      
\end{figure}

For the hybrid scenarios EQ3 shown in Fig.~\ref{fig:cfe_hyeq}, the resistive tearing mode is found linearly unstable before nonlinear saturation from the time evolution of the perturbed magnetic energy, as indicated by both MD and NIMROD simulation results in Fig.~\ref{NTM_bench_nim_md}. The corresponding magnetic island and plasma flow patterns in the saturation phase given in the sub-figures of Fig.~\ref{NTM_bench_nim_md}, show that the saturated island is dominated by the m/n=2/1 structure, whose width is near 20\% (33\%) of the minor radius from the MD (NIMROD) simulation result.

%After applying the initial perturbation, $n=1$ component of magnetic energy grows fast and saturates finally. Poincare plot at saturation phase shows a clear $2/1$ magnetic island structure (Fig.~\ref{NTM_bench_nim_md}(b)). The island width is approximately $0.3a$ ($a$ is minor radius), which is similar to MD simulation results. 

\subsubsection{MD simulation results on ECCD control of NTMs}
For the MD simulations, Westerhof-Pratt’s closure relation~\cite{westerhof2014pop} is employed for the ECCD density $j_d$ appearing in the Ohm's law Eq.~(\ref{NTMeq1}), which can be calculated based on the following equations 

\begin{eqnarray}
\frac{\partial j_{d1}}{\partial t}=-j_{sc}-\nu_1j_{d1}+\upsilon_{\parallel,res} \nabla_\parallel j_{d1} \\
\frac{\partial j_{d2}}{\partial t}=-j_{sc}-\nu_2j_{d2}+\upsilon_{\parallel,res} \nabla_\parallel j_{d2} \\
j_d=j_{d1}+j_{d2}
\end{eqnarray}

Here, $\nu_1$ and $\nu_2$ denote the collision rates near the electron cyclotron waves driven 'hole' and 'bulge' at small and high perpendicular velocities respectively. $\upsilon_{\parallel,res}$ is the parallel velocity of the resonant electrons. The source term for ECCD $j_{sc}$ is assumed to be of the Gaussian distribution as
\begin{equation*}
j_{sc}=j_{d0}\exp\{-4[(\frac{r-r_0}{\Delta_{rd}})^2 + (\frac{\chi-\chi_0}{\Delta_\chi})^2] \}
\end{equation*}
where $r_0$ and $\chi_0$ are the center of the Gaussian distribution in the radial and helical angle directions respectively, $j_{d0}$ is the peak value of ECCD source, $\Delta_{rd}$ and $\Delta_\chi$ are the half deposition width of the distribution in the radial and helical angle directions respectively. Unless otherwise stated, the ECCD is aimed at the center of the O-point of the magnetic island. Other ECCD related parameters are set as $\Delta_{rd}=0.05$, $\Delta_\chi=0.2$, $\nu_1=2.5\times10^{-3}$, $\nu_2=0.5\times10^{-3}$ and $\upsilon_{\parallel,res}=2$, all in SI units.

\begin{figure}
  \centering
  \includegraphics[width=0.9\textwidth]{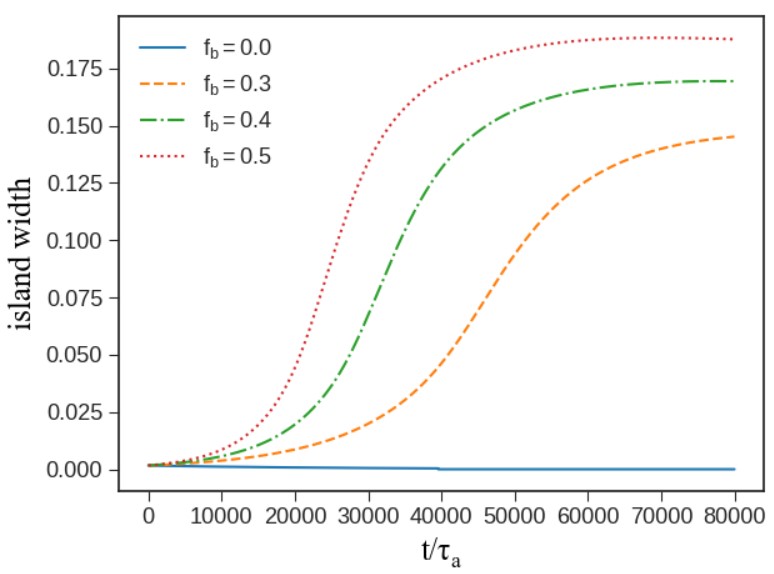}
\caption{Evolution of magnetic island width for various fractions of bootstrap current from the MD simulations.}
\label{eccdfig2}      
\end{figure}

For the hybrid scenario EQ1 shown in Fig.~\ref{fig:cfe_hyeq}, the classical $m/n=2/1$ tearing mode with $f_b=0$ is stable. However, the fraction of bootstrap current for the hybrid scenarios in CFETR is nearly 50\%. Thus, the evolution of magnetic island with various fractions of bootstrap current is calculated and given in Fig.~\ref{eccdfig2}. It is found that the width of the saturated magnetic island is about $0.2a$, i.e. near $40cm$ for the size of CFETR. This big magnetic island is very dangerous and can obviously lead to the major disruption of discharge. Therefore, ECCD must be employed to control the growth of the NTMs for CFETR.

\begin{figure}
  \centering
  \includegraphics[width=0.9\textwidth]{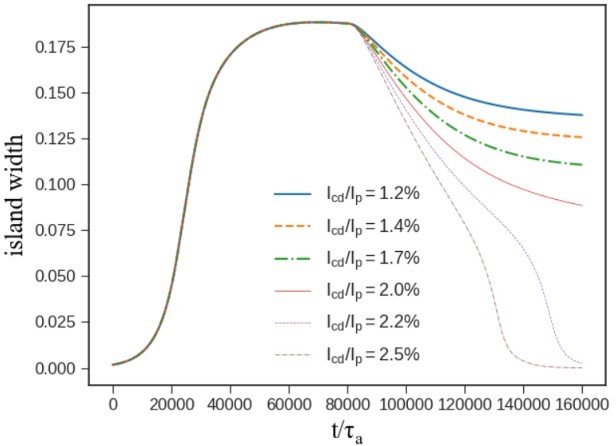}
  \put(-150,180){\textbf{(a)}}
  
  \includegraphics[width=0.9\textwidth]{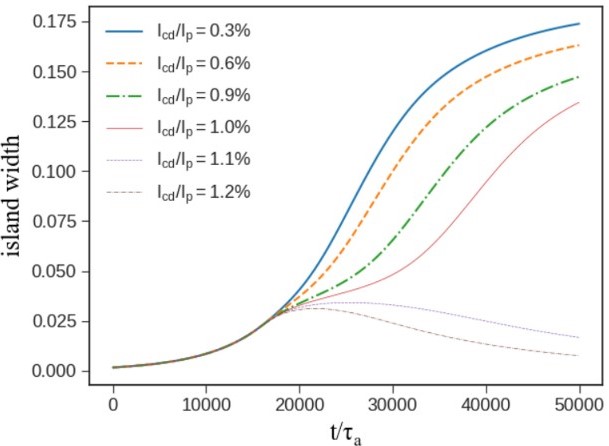}
  \put(-150,180){\textbf{(b)}}
  
  \includegraphics[width=0.9\textwidth]{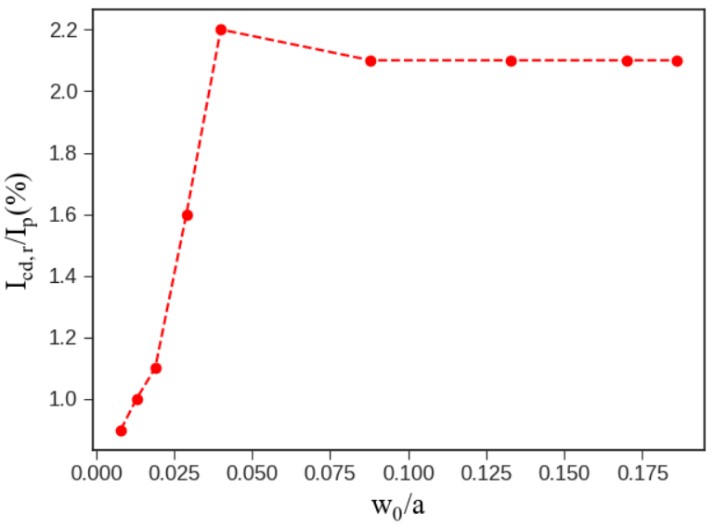}
  \put(-150,180){\textbf{(c)}}
  \caption{Evolution of magnetic island width from the MD simulations for different ECCD amplitudes, with ECCD being turned on at (a) $t/\tau_a=80000$ and (b) $t/\tau_a=15000$, respectively. (c) Dependence of the driven current required for the suppression of NTM on the magnetic island width at the time when ECCD is turned on.}
\label{eccdfig3}      
\end{figure}

If the ECCD is turned on after the magnetic island is saturated, as shown in Fig.~\ref{eccdfig3}(a), it is found that the magnetic island can be suppressed completely when the driven current $I_{cd}$ is larger than 2\% of the total plasma current ($I_p=13 MA$). However, if the ECCD is turned on before the magnetic island is saturated, the strength of the driven current $I_{cd}$ required for the suppression of NTM can be reduced. For instance, the required $I_{cd}$ is near 1\% of $I_p$, when the ECCD is turned on at $t/\tau_a=15000$, as shown in Fig.~\ref{eccdfig3}(b). In fact, the required $I_{cd}$ can be less than 1\% of $I_p$, if the ECCD can be turned on before $t/\tau_a= 15000$ when the magnetic island width is smaller than $0.025a$ or $5cm$. Dependence of the driven current for the suppression of NTM on the magnetic island width is given in Fig.~\ref{eccdfig3}(c). It can be clearly observed that the required $I_{cd}$ is proportional to the width of magnetic island at the time when the ECCD is turned on. However, the required $I_{cd}$ for NTM suppression levels out when the island size increases above a certain threshold. %The numerical result is quantitively similar to the MRE results obtained in Fig XX (Wang X. J.).   \\

To summarize, MD simulations find that the saturated island width in the hybrid scenarios of CFETR is about $0.2a$. The required ECCD for the suppression of the saturated NTM is just above 2\% of the total plasma current $I_p=13 MA$. However, if the ECCD is turned on when the magnetic island width is less than a critical value, the required ECCD for the suppression of NTM can be reduced. Both the numerical and theoretical results indicate that the required ECCD is proportional to the size of magnetic island at the time when ECCD is turned on in the small magnetic regime, and becomes independent of island size in the large magnetic island regime.

\subsubsection{TM8 simulation results on ECCD control of NTMs}
The evolution of neoclassical tearing mode and its stabilization by electron cyclotron current drive for the hybrid scenario of CFETR are also numerically performed using the reduced MHD code TM8. The parameters of our numerical calculations are set as the following if not mentioned elsewhere: a localized distribution of current density from ECCD is applied at the resonant surface with $w_{cd}/a=0.04$, $\chi_{\perp}=12.5a^2/\tau_R$ and $\chi_{\parallel}/\chi_{\perp}=10^8$, where $a$ is the minor radius, $w_{cd}$ the half width of driven current, $\tau_R=a^2\mu_0/\eta$ the resistive time, $\chi_{\parallel}$ and $\chi_{\perp}$ are the parallel and perpendicular heat transport coefficients, respectively. The Lundquist number $S=\tau_R/\tau_A$ is taken to be $5.56\times 10^6$, where $\tau_A$ is the Alfv\'en time. And $r_s=0.51a$ is the minor radius of the $q=2$ surface. The local bootstrap current density fraction at the resonant surface is set to be $j_b/j_p=0.3$ initially. The inverse aspect ratio $\varepsilon=a/R\simeq 0.31$ as designed.

\begin{figure}
  \centering
  \includegraphics[width=0.9\textwidth]{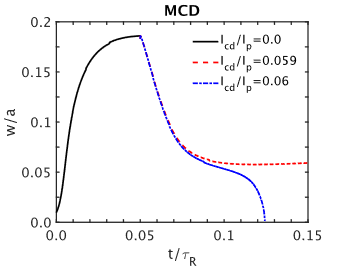}
  \put(-150,180){\textbf{(a)}}
  
  \includegraphics[width=0.9\textwidth]{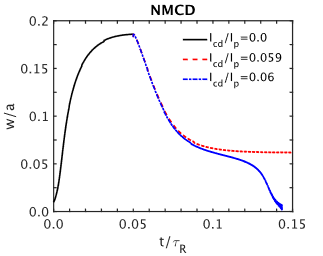}
  \put(-150,180){\textbf{(b)}}
  \caption{Time evolution of the normalized island width $w/a$ from the TM8 simulations in absence (solid curve) or presence (dashed and dot-dashed curves) of ECCD for (a) MCD and (b) NMCD with $I_{cd}/I_p=0.059$ (dashed curve) and $0.06$ (dot-dashed curves).}
  \label{tmfig1}      
\end{figure}

%stabilization of the $m/n=2/1$ neoclassical tearing mode by ECCD is studied. 
For the hybrid scenario, the time evolution of the normalized 2/1 magnetic island width, $w/a$, is shown in Fig.~\ref{tmfig1}. The solid curve is for the case without applying ECCD. The modulated current drive (MCD), which is in phase with the island's O-point, is applied at $t/\tau_R=0.05$ with $I_{cd}/I_p=0.059$ (dashed curve) and $I_{cd}/I_p=0.06$ (dot-dashed curve), where $I_{cd}$ is the current driven by ECW, and $I_p$ is the plasma current. Similar to Fig.~\ref{tmfig1}a, the case for non-modulated current drive (NMCD) is also shown in Fig.~\ref{tmfig1}b. It can be seen that there is a threshold in the driven current for mode stabilization.

\begin{figure}
  \centering
  \includegraphics[width=0.9\textwidth]{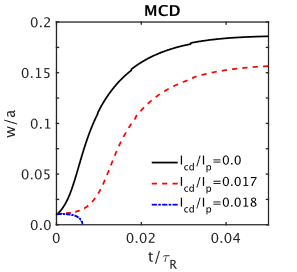}
  \put(-100,180){\textbf{(a)}}
  
  \includegraphics[width=1.1\textwidth]{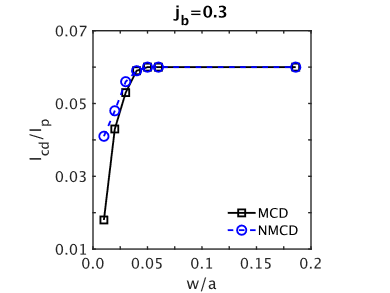}
  \put(-150,180){\textbf{(b)}}
  \caption{(a) Time evolution of the island width from the TM8 simulations in absence (solid curve) or presence (dashed curves) of ECCD. MCD is turned on when the island width $w=0.01a$ is reached, with $I_{cd}/I_p=0.017$ (dashed curve) and $0.018$ (dot-dashed curve). (b) The required $I_{cd}/I_p$ for mode stabilization as a function of the normalized island width. The solid (dashed) curve is for MCD (NMCD).}
  \label{tmfig3}    
\end{figure}

When ECCD is applied during the island growth before nonlinear saturation, less driven current is expected to be required for mode stabilization. The time evolution of the island width is shown in Fig.~\ref{tmfig3}a with MCD applied when the island width $w=0.01a$ is reached. The solid curve is for the case without ECCD. The dashed and dot-dashed curves are for $I_{cd}/I_p=0.017$ and $0.018$, respectively, which indicate that less driven current is required for the stabilization of a smaller magnetic island. The required $I_{cd}/I_p$ for fully stabilizing the 2/1 mode is shown as a function of the island width in Fig.~\ref{tmfig3}b, in which ECCD is applied when the island width $w$ grows to the value shown by the horizontal axis. The solid (dashed) curve is for MCD (NMCD). For both MCD and NMCD, when applied before the nonlinear mode saturation, the required driven current for mode stabilization increases with $w$, suggesting the advantage of applying ECCD earlier in time. For a smaller island width, the MCD scheme is much more effective than NMCD for stabilizing the NTM.

\subsection{Disruption mitigation simulation with massive neon injection on CFETR}
\begin{figure}
  \centering
  \includegraphics[width=0.9\textwidth]{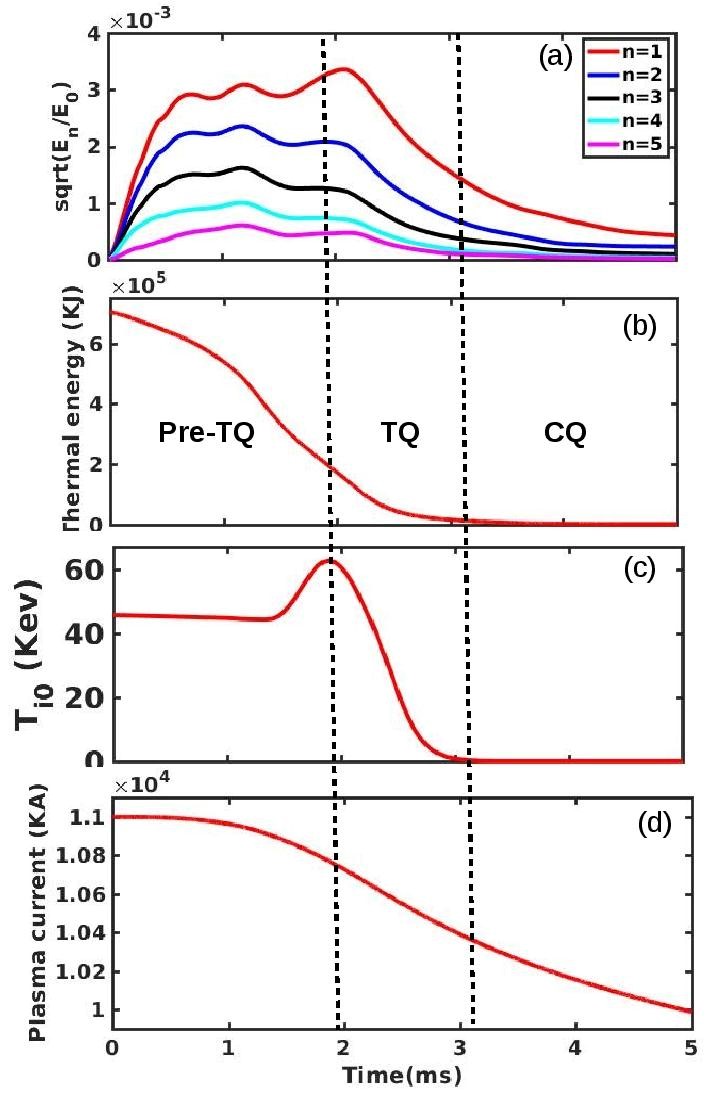}
  \caption{Time evolution of (a) normalized magnetic energy of each primary toroidal component (in unit $(W_{mag,n}/W_{mag,0})^{1/2}$), (b) plasma thermal energy (kJ), (c) core temperature (keV), and (d) plasma current (kA) during an MGI process from a NIMROD simulation.}
  \label{mgifig1}      
\end{figure}

\begin{figure}
  \centering
  \includegraphics[width=0.9\textwidth]{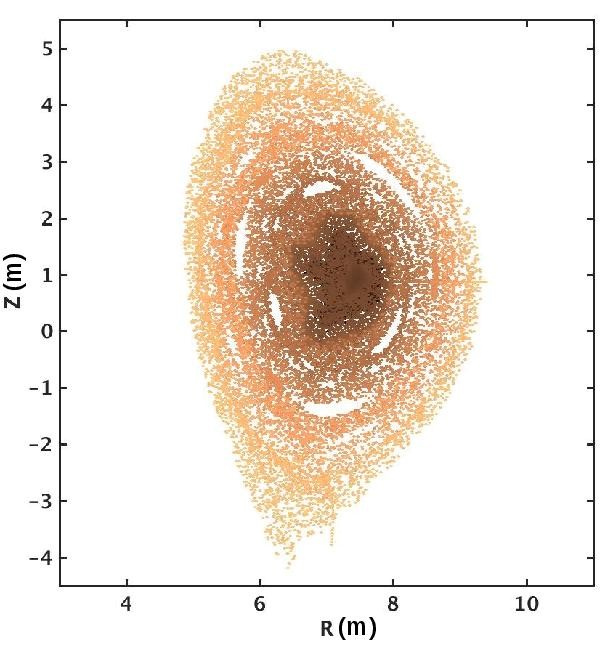}
\caption{Poincare plot at the end of the TQ during an MGI process from a NIMROD simulation.}
`\label{mgifig2}      
\end{figure}

Simulation evaluation of the disruption mitigation scheme with massive neon injection on CFETR using the 3D nonlinear MHD code NIMROD, which incorporates a radiation and atomic physics model taken from the KPRAD code, has been performed and the main findings are reported here. The time evolution of the several key discharge parameters are shown in Fig.~\ref{mgifig1}. During the pre-thermal quench (pre-TQ) phase before $t=2ms$, the thermal energy is dissipated gradually due to radiation cooling by the injected impurity (Fig.~\ref{mgifig1}(b)), however, the core electron temperature remains relatively unchanged and even increases to a peak value by the end of the pre-TQ phase (Fig.~\ref{mgifig1}(c). Meanwhile, MHD activity grows to nonlinear saturation and the $n = 1$ mode dominates from beginning (Fig.~\ref{mgifig1}(a)). The thermal quench (TQ) phase starts when the core electron temperature starts to collapse at $t = 2 ms$. By the end of TQ phase at $t = 3.1ms$, the thermal energy is almost totally dissipated and the core electron temperature drops to zero. Soon after the start of TQ phase, the $n = 1$ mode reaches its peak magnitude and the magnetic field becomes completely stochastic (Fig.~\ref{mgifig2}), leading to loss of plasma confinement entirely. The current quench (CQ) sets on after $t=3.1ms$, i.e the end of TQ phase.

\begin{figure}
  \centering
  \includegraphics[width=\textwidth]{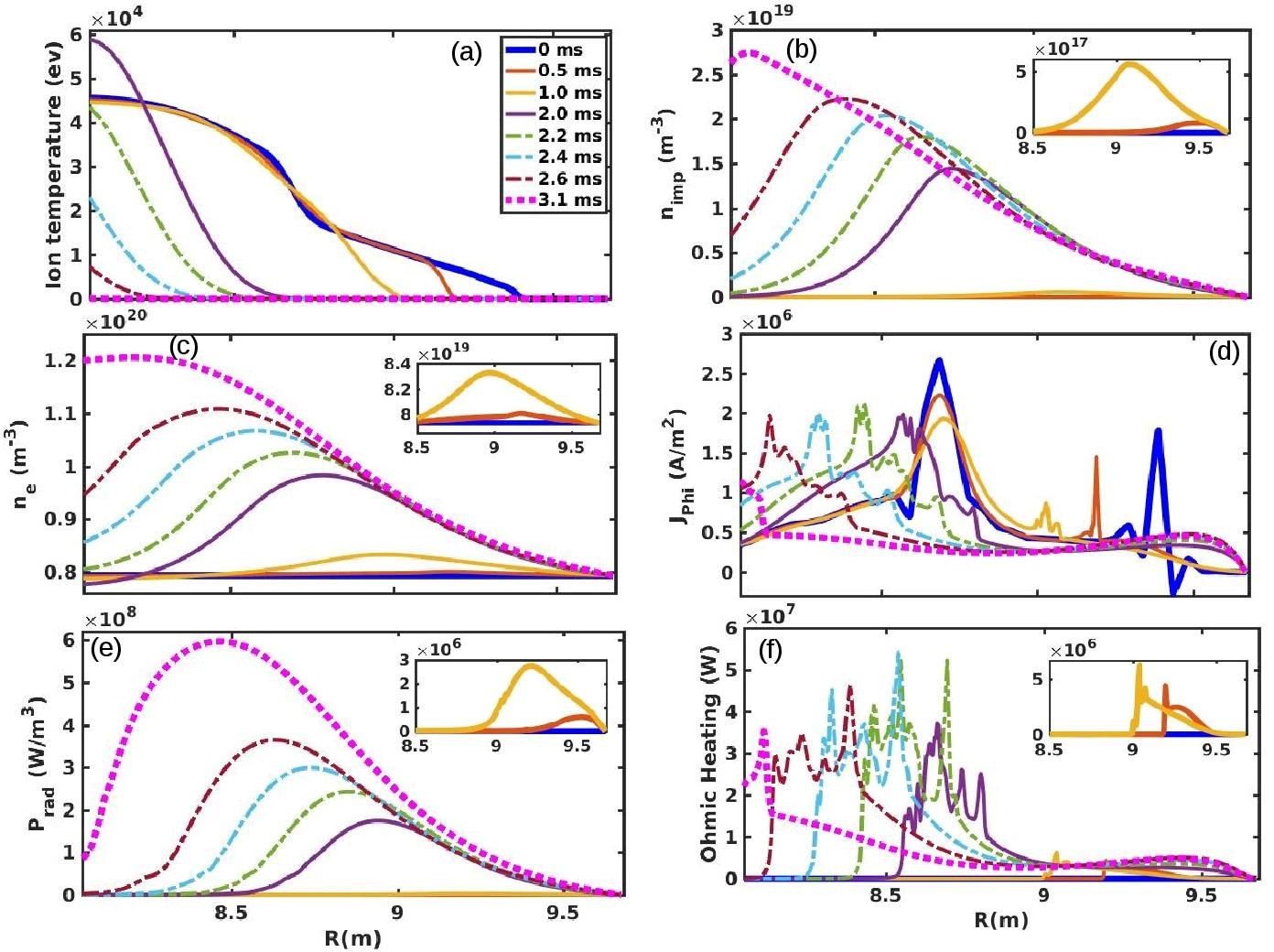}
\caption{Profile evolution of (a) ion temperature, (b) impurity number density, (c) electron density, (d) toroidal current density, (e) radiated power density, and (f) Ohmic heating power during an MGI process from a NIMROD simulation.}
\label{mgifig3}      
\end{figure}

Profile evolution of ion temperature, impurity number density, electron number density, toroidal current density, radiation power and Ohmic heating during the neon injection process are obtained respectively from the simulation (Fig.~\ref{mgifig3}(a)-(f)). The impurity density gradually penetrates into the core region from boundary as shown in Fig.~\ref{mgifig3}(b), and the corresponding total electron density increases accordingly as a result of impurity ionization (Fig.~\ref{mgifig3}(c)). The radiation power profile rises and shifts towards the core region along with the impurity penetration over time (Fig.~\ref{mgifig3}(e)), leading to the drop of the core ion temperature profile as shown in Fig.~\ref{mgifig3}(a). At the same time, the plasma resistivity increases and the current density profile contracts as a consequence (Fig.~\ref{mgifig3}(d)), which contributes to a strong localized deposition of Ohmic heating power at the cold region (Fig.~\ref{mgifig3}(f)).

\section{Analyses of MHD stability of HFRC}
\label{HFRCmhd}
For FRC plasma, the tilt mode and the rotational mode are the most dominant MHD instabilities. Thus it is necessary to evaluate these instabilities during the magnetic compression for the HFRC design. The linear calculation results from the NIMROD code~\cite{sovinec2004jcp} based on a single-fluid MHD model show that for the designed parameters of the FRC neutron source, the growth rates of the dominant MHD modes are in the characteristic Alfv\'{e}n time scale of about 0.3$\mu s$, which is much shorter than the achievable time scale of the magnetic compression current ramp, and thus may seriously hinder the progress of adiabatic compression. However, due to the reversal of the magnetic field in the FRC plasma, there is a wide range of weak and even null magnetic fields. Therefore the finite Larmor radius (FLR) effects of thermal and energetic ions and the related two-fluid and kinetic effects can hardly be ignored.

\begin{figure}[htp]
\centering
\includegraphics[width=\columnwidth]{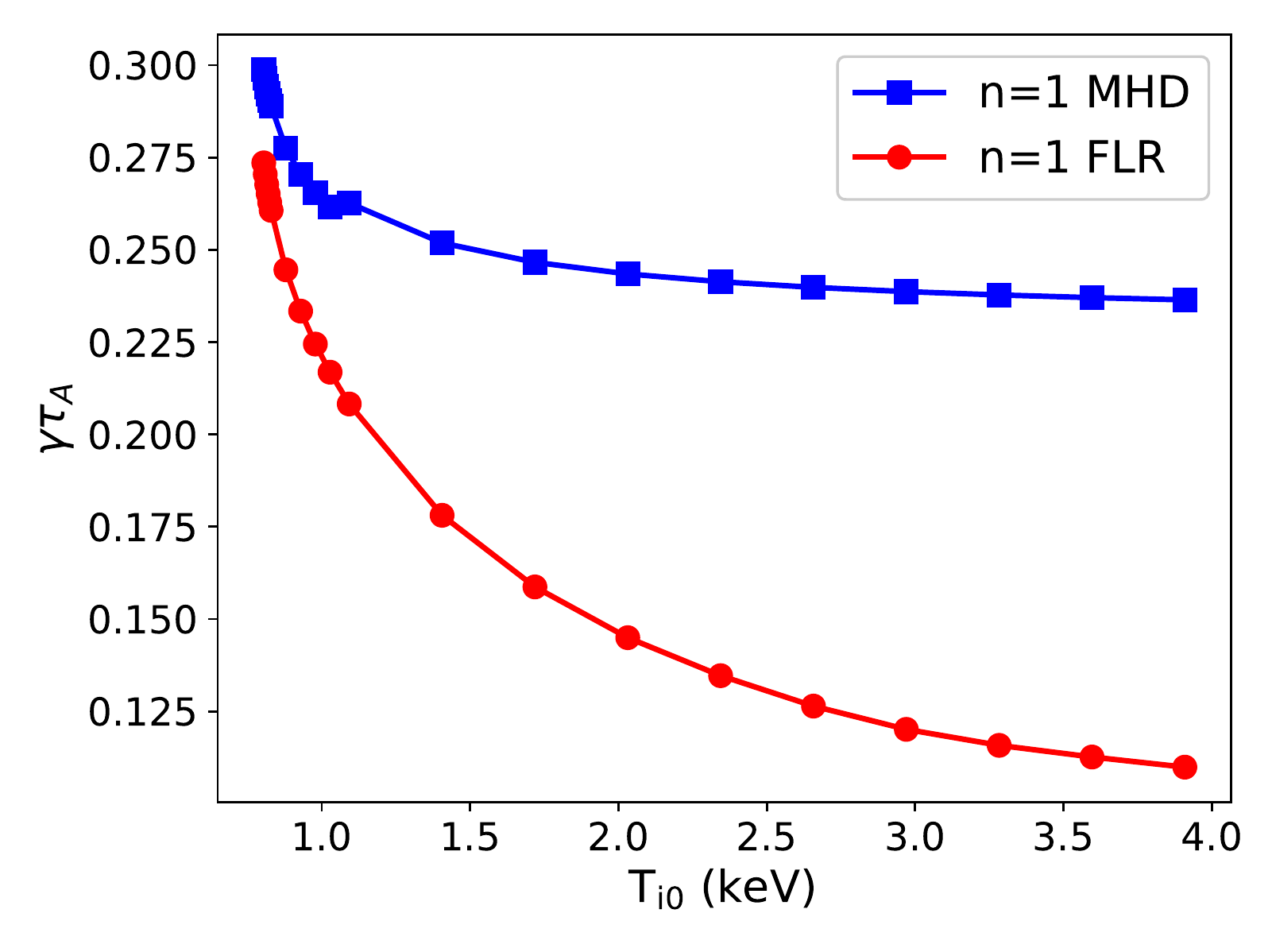}
\put(-50,190){\textbf{(a)}}

\includegraphics[width=\columnwidth]{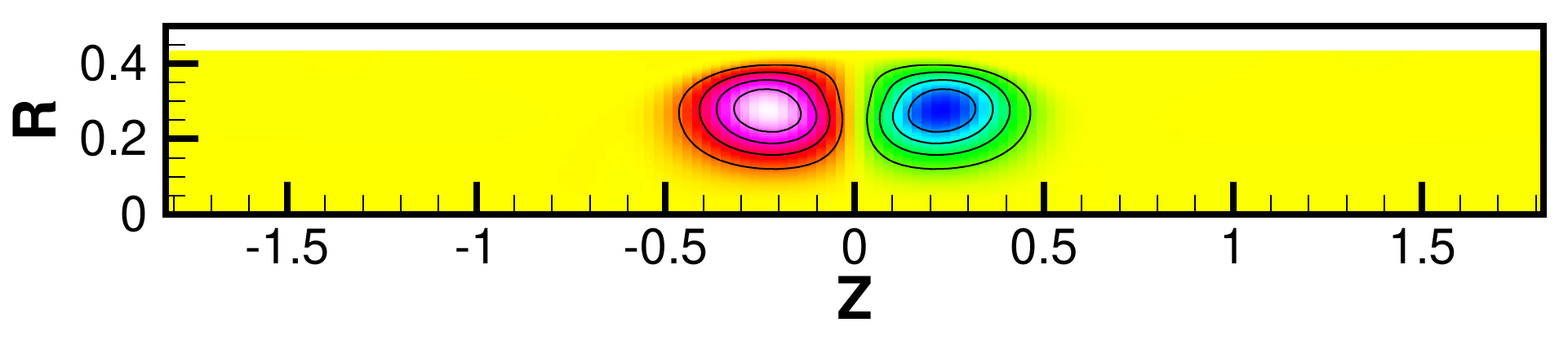}
\put(-50,50){\textbf{(b)}}

\includegraphics[width=\columnwidth]{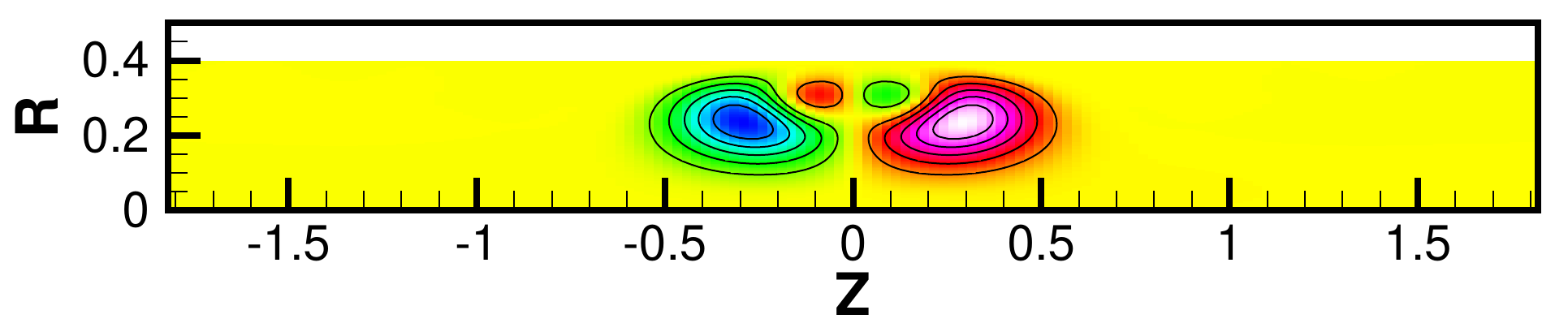}
\put(-50,50){\textbf{(c)}}
\caption{$\mathrm{n}=1$ mode: (a) growth rates as functions of ion temperature $\left(\tau_{A}=1 \times 10^{-7} s\right)$, and the distribution of perturbed pressure in the $\mathrm{R}-\mathrm{Z}$ plane from (b) single-fluid and (c) two-fluid model NIMROD calculations.}
\label{fig:n1mode}
\end{figure}

\begin{figure}[htp]
\centering
\includegraphics[width=\columnwidth]{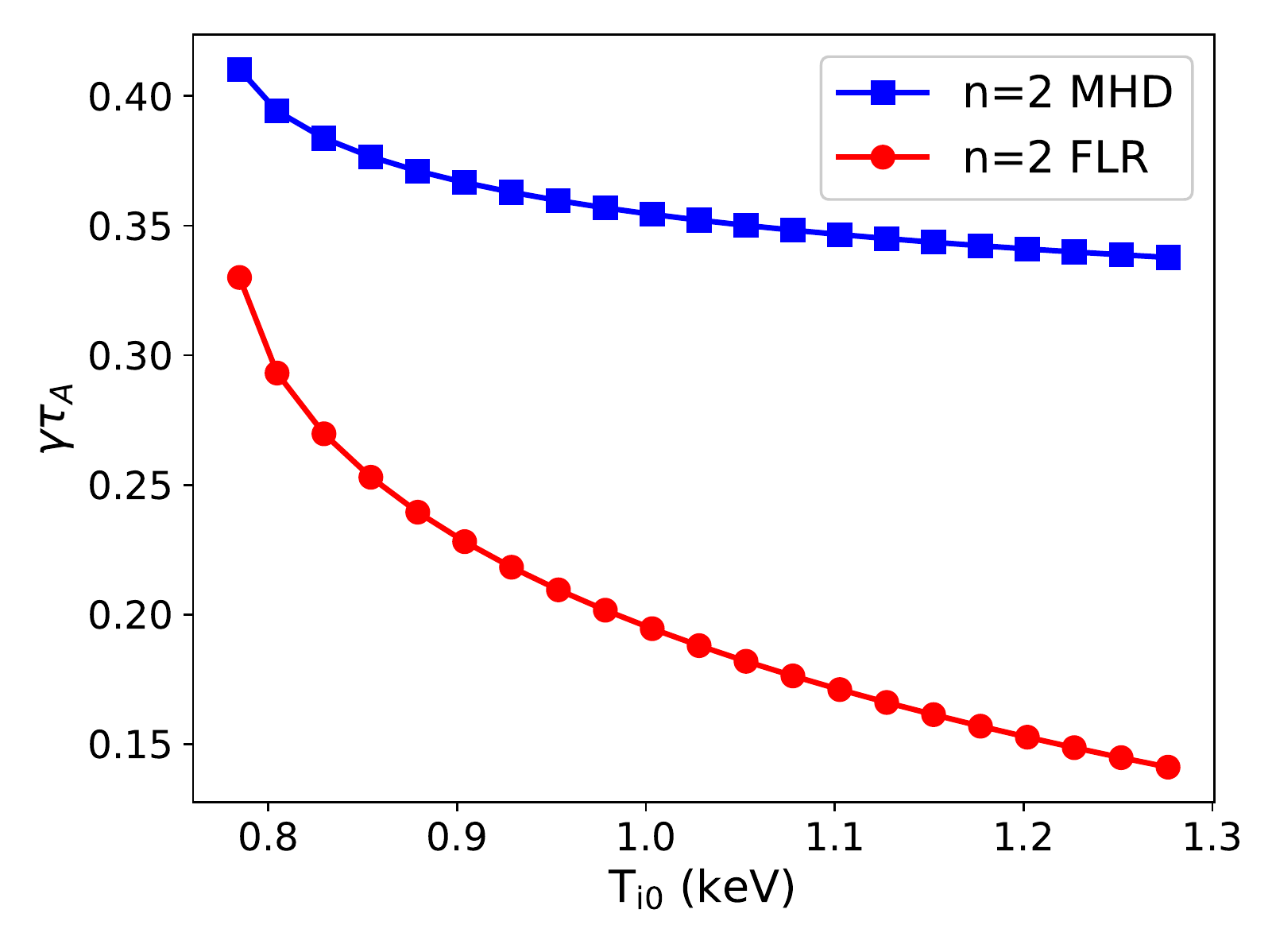}
\put(-50,190){\textbf{(a)}}

\includegraphics[width=\columnwidth]{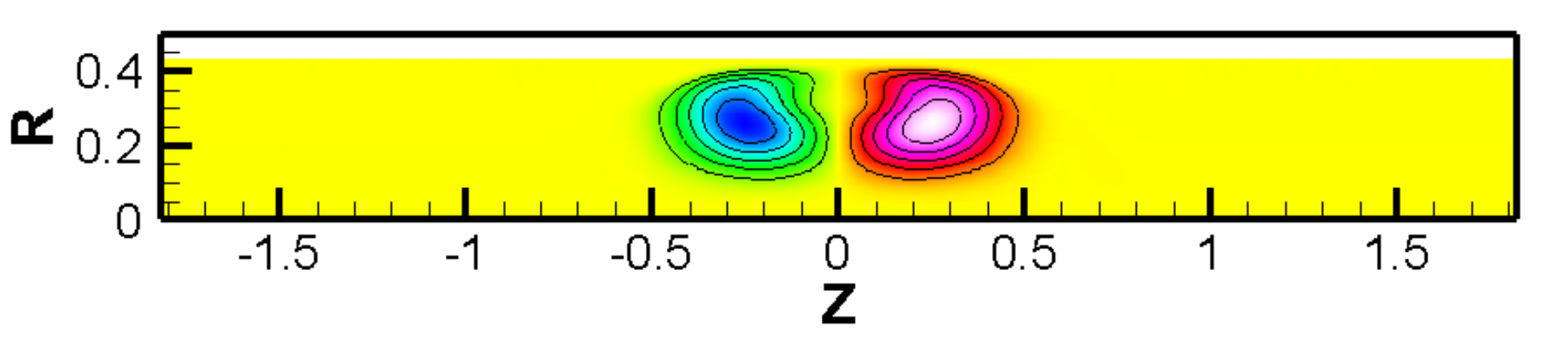}
\put(-50,50){\textbf{(b)}}

\includegraphics[width=\columnwidth]{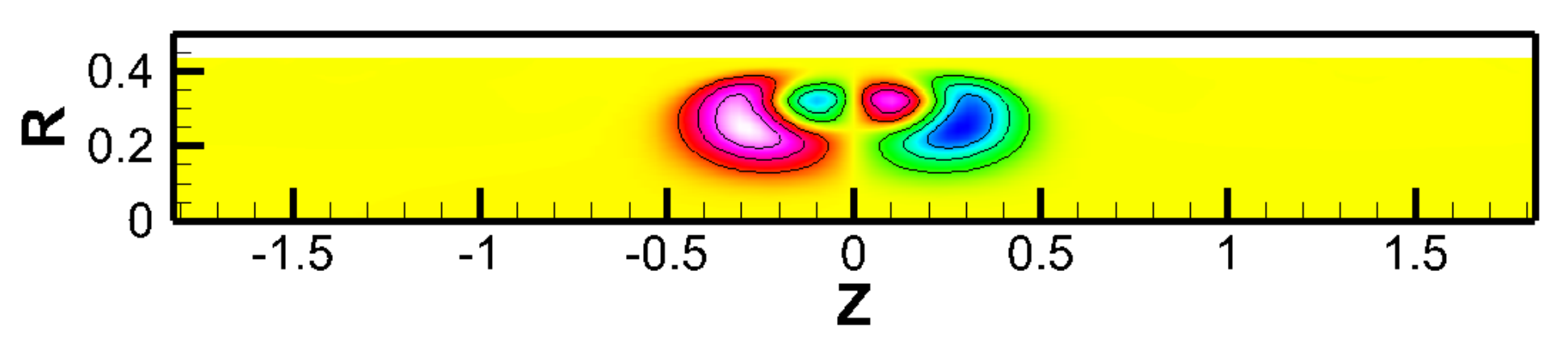}
\put(-50,50){\textbf{(c)}}
\caption{$\mathrm{n}=2$ mode: (a) growth rates as functions of ion temperature $\left(\tau_{A}=1 \times 10^{-7} \mathrm{~s}\right)$, and the distribution of perturbed pressure in the $\mathrm{R}-\mathrm{Z}$ plane from (b) single-fluid and (c) two-fluid model NIMROD calculations.}
\label{fig:n2mode}
\end{figure}

Calculations using the linear two-fluid MHD model in the NIMROD code show that as the plasma temperature of an FRC increases, the stabilizing effects brought in by the FLR of ions can become more dominant. The growth rates of main MHD modes can be substantially reduced, and the model structure is also significantly different from the single-fluid model (Figs.~\ref{fig:n1mode}-\ref{fig:n2mode}). If the kinetic effects of thermal ions and externally injected fast particles are more fully considered in the hybrid kinetic MHD model, the FLR effects of thermal ions and fast particles are expected to completely suppress the primary MHD instabilities in the FRC. The team of C-2 series of FRC device has reported the experimental evidence of suppressing instability using neutral beam injection, which has been confirmed in NIMROD simulations~\cite{guo_achieving_2015}.

In addition, based on previous experimental studies and after considering the kinetic effect correction, the empirical criterion for the stability of FRC operation is typically obtained as: $S/\kappa<3.5$, where $\kappa$ is the ratio of the axial dimension to the radial dimension of the FRC separatrix surface, i.e. the elongation of the separatrix surface $\kappa=l_s/R_s$, and $S=R_s/\delta_i$ is the ratio of the radius of the separatrix surface to the ion skin depth $\delta_i$~\cite{Barnes2003a}. If the magnetic compression process is slow enough, that is, the compression characteristic time is much longer than the equilibrium relaxation time of the FRC plasma, and the process can be considered approximately quasi-static. In the meantime, the compression process is fast enough in comparison to the transport process and thus can be considered adiabatic. Under these approximations, the empirical stability criterion for the FRC plasma magnetic compression can be obtained as:
\begin{equation}
%\begin{aligned}
\frac{S}{\kappa}=\frac{2 \sigma^{2} R_{s}^{2}}{cn_{i}\sigma^{\frac{2(3-\epsilon-\gamma)}{\gamma}}\left(1-\frac{\sigma^{2}}{2}\right)^{-\frac{1+\epsilon-\gamma \epsilon}{\gamma}}} \times
\sqrt{\frac{e^{2}l_{s} \sigma^{-\frac{2(3-\varepsilon)}{\gamma}}\left(1-\frac{\sigma^{2}}{2}\right)^{-\frac{(1+\varepsilon)(\gamma-1)}{\gamma}}}{\epsilon_{0} m_{i}}}<3.5
%\end{aligned}
\end{equation}
 
\begin{figure}[htb]
\centering
\includegraphics[width=\columnwidth]{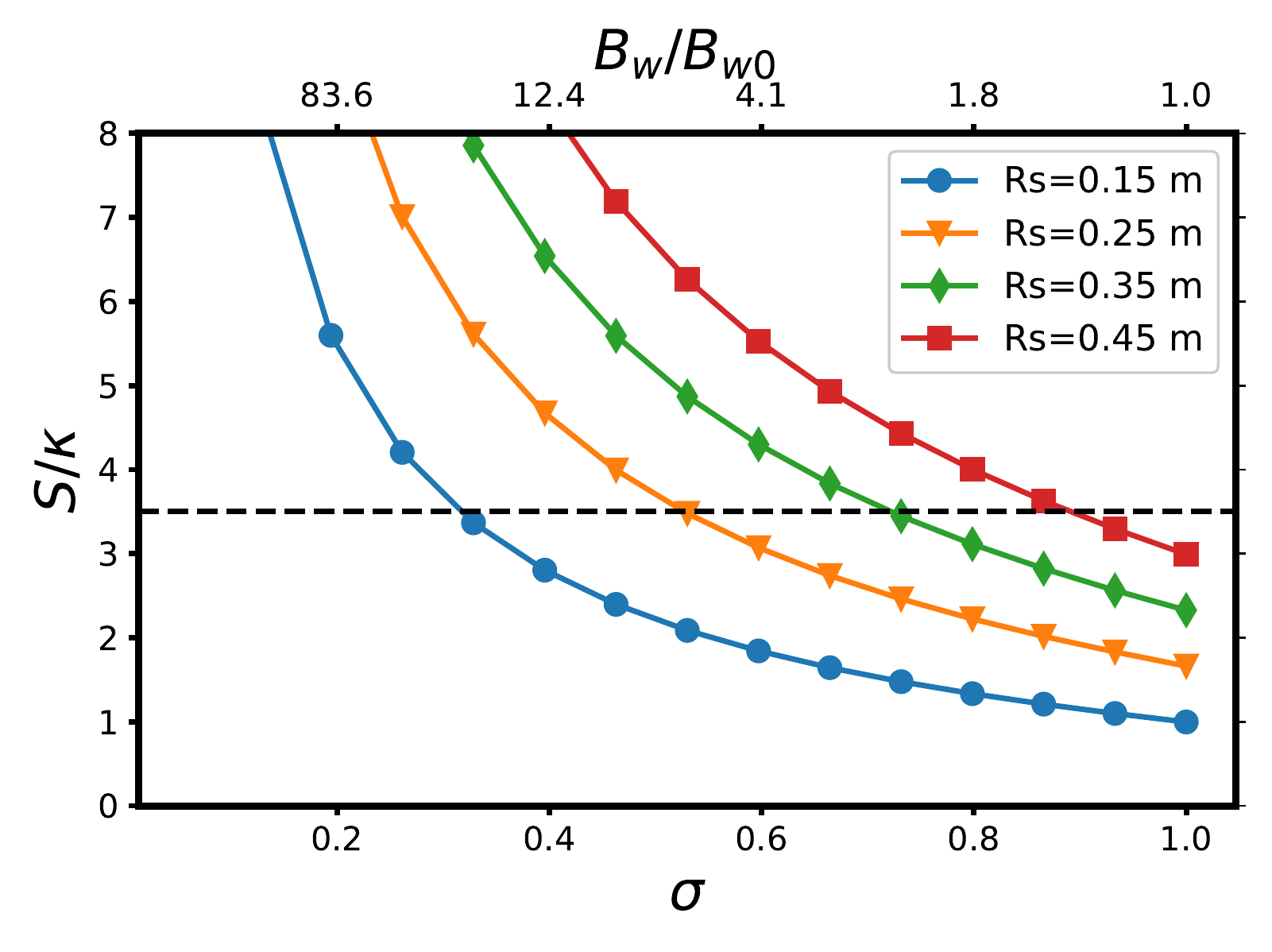}%
\caption{For different initial plasma radius $R_{s}$ and axial length $l_s$ with the fixed elongation ratio $\kappa=2.5/0.45$, $S/\kappa$ as functions of the compression ratios for the plasma radius (i.e. $\sigma$) and the magnetic field strength at wall (i.e. $B_w/B_{w0}$). The black dashed line indicates the critical value of empirical stability condition $S/\kappa=3.5$, below which the FRC system is stable.}
\label{fig:sk}
\end{figure}

Consider the design parameters for the HFRC plasma at the beginning of magnetic compression: $R_s=0.45m$, $n_i=3.0\times {10}^{19}m^{-3}$, $l_s=2.5m$, $\gamma=\frac{5}{3}$, $\epsilon=-0.25$, it can be seen that $S/\kappa$ gradually increases during the compression process (Fig.~\ref{fig:sk}). When $S/\kappa$ is greater than 3.5, FRC's stability condition can be violated, and unstable MHD modes may grow. When the initial equilibrium plasma radius $R_s$ decreases, the stability window of the compression process becomes broader. For $R_s=0.25m$, the radial compression ratio can drop to around $0.53$, and equivalently the magnetic field compression ratio at wall can almost reach $5.7$. According to the above analysis, under the quasi-static and adiabatic approximations, the magnetic compression process of the HFRC plasma should have a reasonable range of MHD stable operation regime. 

\section{Summary and discussion}
\label{summary}
In summary, CFETR and HFRC physical and engineering designs have provided unprecedented opportunities to the advancement in MHD theory and simulations. Comprehensive efforts on the assessment of MHD stability of CFETR and HFRC baseline scenarios have led to following preliminary progresses that may further benefit engineering designs.

For CFETR, the ECCD power and current required for the full stabilization on NTM have been predicted in this work, as well as the corresponding modulated magnetic island width. A thorough investigation on RWM stability for CFETR is performed. For 80\% of the SSO scenarios, active control methods may be required for RWM stabilization. The process of disruption mitigation with massive neon injection on CFETR is simulated. The time scale of and consequences of plasma disruption on CFETR are estimated, which are found equivalent to ITER. Major MHD instabilities such as NTM and RWM remain challenge to steady state tokamak operation. On this basis, next steps on CFETR MHD study are planned. Further analysis on NTM control with ECCD system will be processed with TM8 code and NIMROD code, along with TORAY code, in order to provide more detailed and quantitative information on the required ECCD current amplitude and distribution and optimized injection angle for NTM stabilization. More careful prediction on RWM stability boundaries with kinetic effects included will be performed using MARS-K and AEGIS-K codes. On the other hand, the design and feasibility analyses on the RWM active and feedback control systems are also necessary for the unstable RWM scenarios. A new task of simulation on the disruption mitigation based on the shattered pellet injection (SPI) of impurity gas is expected to start, so that we can evaluate on the corresponding time scale, gas injection depth, MHD modes, and current distribution in vacuum chamber.

For HFRC, the MHD stability of the plasma heating scheme through adiabatic compression has been evaluated under quasi-static approximation. The dominant $n=1$ and $n=2$ instabilities can grow on the ideal MHD time scale in single fluid MHD model as shown from NIMROD calculations. Two-fluid MHD calculations using NIMROD find FLR stabilizing effects on both the dominant $n=1$ and $n=2$ modes. Energetic-particle stabilization of tilt mode was previously demonstrated in C-2 experiments and NIMROD simulations. With the strong stabilization on major MHD instabilities from two-fluid and energetic particle effects, FRC may promise to be an alternative route to compact magnetic fusion ignition. To explore such a potential, we plan on further perform analyses of the MHD instabilities in HFRC during the dynamic magnetic compression process.

\section{Acknowledgments}
%This work was supported by the National Natural Science Foundation of China (NSFC) (Grant Nos. 11905251, 11475225, 11805054, 11875098, 11905067, 11847219, 11775221 and 51821005), the National Key Research and Development Program of China (under contract Nos. 2017YFE0300500, 2017YFE0300501, 2017YFE0301100, 2017YFE0301104 and 2017YFE0301805), the Fundamental Research Funds for the Central Universities at Huazhong University of Science and Technology Grant No. 2019kfyXJJS193 and at Donghua University Grant No. 2233019G-10, the U.S. DOE Grant Nos. DE-FG02-86ER53218 and DE-SC0018001, and the China Postdoctoral Science Foundation under Grant No.2019M652931. \\
This work was supported by the National Key Research and Development Program of China (under contract Nos. 2017YFE0301805, 2017YFE0300500, 2017YFE0300501, 2017YFE0301100, 2017YFE0301104, and 2019YFE03050004), the National Natural Science Foundation of China (NSFC) (Grant Nos. 11905251, 11475225, 11805054, 11875098, 11905067, 11847219, 11775221 and 51821005), the Fundamental Research Funds for the Central Universities at Huazhong University of Science and Technology Grant No. 2019kfyXJJS193 and Donghua University Grant No.~2233019G-10, the U.S. DOE Grant Nos. DE-FG02-86ER53218 and DE-SC0018001, and the China Postdoctoral Science Foundation under Grant No.2019M652931. This research used the computing resources from the Supercomputing Center of University of Science and Technology of China, and the ShenMa High Performance Computing Cluster at the Institute of Plasma Physics, Chinese Academy of Sciences. The authors are very grateful for the supports from J-TEXT team, the NIMROD team, and the developers of AEGIS, MARS, and TM8 codes.

\begin{appendices}
\renewcommand\thesection{Appendix \arabic{section}}

\section{MARS code}
\label{sec:mars}
MARS-F code~\cite{liuyueqiang2000pop} is based on the single fluid, linearized resistive MHD model,
\begin{equation}
(\gamma+in\Omega)\boldsymbol{\xi}=\boldsymbol{v}+(\boldsymbol{\xi}\cdot\boldsymbol{\nabla})R^2\nabla\phi \\
\end{equation}
\begin{equation}
\begin{split}
\rho(\gamma+in\Omega)\boldsymbol{v}= & \quad -\boldsymbol{\nabla}p+\boldsymbol{j}\times\boldsymbol{B}+\boldsymbol{J}\times\boldsymbol{b} \\
& \quad -\rho[2\Omega\hat{Z}\times\boldsymbol{v}+(\boldsymbol{v}\cdot\boldsymbol{\nabla}\Omega)R^2\boldsymbol{\nabla}\phi]-\boldsymbol{\nabla}\cdot\boldsymbol{\Pi} \\
\end{split}
\end{equation}
\begin{eqnarray}
(\gamma+in\Omega)\boldsymbol{b}=\boldsymbol{\nabla}\times(\boldsymbol{v}\times\boldsymbol{B}-\eta\boldsymbol{j})+(\boldsymbol{j}\cdot\boldsymbol{\nabla}\Omega)R^2\boldsymbol{\nabla\phi} \\
(\gamma+in\Omega)\boldsymbol{p}=-\boldsymbol{v}\cdot\boldsymbol{\nabla}P-\Gamma{P}\boldsymbol{\nabla}\cdot\boldsymbol{v} \\
\boldsymbol{j}=\boldsymbol{\nabla}\times\boldsymbol{b} 
\end{eqnarray}
where $R$ and $\phi$ are the plasma major radius and geometric toroidal angle, and $\hat{Z}$ is unit vectors along the vertical direction in the poloidal plane, respectively. The variables $(\boldsymbol{\xi},\boldsymbol{v},\boldsymbol{j},\boldsymbol{b},\rho,p)$ represent the plasma perturbed displacement, velocity, current, magnetic field, density and pressure, respectively. The corresponding equilibrium quantities are denoted by $(\boldsymbol{J},\boldsymbol{B},P)$. $\Omega$ is the angular frequency of the plasma flow along the toroidal angle, and $n$ is the toroidal harmonic number. $\boldsymbol{\Pi}$ is a viscous stress tensor, which is associated with the viscous force damping, such as the parallel sound wave damping.

Although the growth rate of the RWM is very slow, it eventually sets the upper limit on plasma pressure for the long pulse or steady-state advanced tokamak operations. MARS code is designed to compute the growth rate of the RWM and how to control it by the passive (the plasma rotation and drift kinetic resonances) and active (the feedback control system) methods.

MARS code has been benchmarked and extensively applied to model RWM and compare with the experimental observation. For examples, a kinetic version of MARS found low-rotation threshold when applied to model a DIII-D discharge with balanced beam injection, agreeing with experimental observations~\cite{reimerdes2007prl}. MARS-F and its coupling to CARIDDI (CarMa) found quantitative agreement between the computed RWM growth rate and the experiments in RFX~\cite{villone2008prl}. MARS-F has also modeled the resonant field amplification for a series of JET plasmas which agrees with experimental measurements~\cite{liuyueqiang2009ppcf}. \cite{okabayashi2011pop} has compared the unstable RWM regime obtained using MARS-K with that in DIII-D experiments, revealing the impact of energetic particle losses and toroidal rotation drop in destabilizing the mode. Finally, the MARS-K modeling of stable RWM induced resonant field amplification quantitatively agrees with DIII-D experiments~\cite{wang2015prl}.

\section{AEGIS code}
\label{sec:aegis}
The Adaptive Eigenfunction Independent Solution shooting (AEGIS) code employs the adaptive shooting method in the radial direction and Fourier decomposition in the poloidal direction~\cite{zheng2006jcp}. Therefore, the AEGIS code has high resolution near the singular surfaces for the study of MHD instabilities. The AEGIS code has been used to study the linear behaviors of RWMs in ITER and the earlier smaller-sized design of CFETR scenarios~\cite{zheng2005prl,hanrui2020ppcf}. 

The following perpendicular MHD equation was solved in AEGIS,
\begin{equation}
-\rho_m(\omega+n\Omega+i\gamma_p)^2 \boldsymbol{\xi_{\perp}} = \delta\boldsymbol{J}\times\boldsymbol{B}+\delta\boldsymbol{B}\times\boldsymbol{J}-\boldsymbol{\nabla}\delta P\nonumber
\end{equation}
where $\rho_m$ is the total apparent mass density, $\omega$ the mode frequency, $n$ the toroidal mode number, $\Omega$ the toroidal rotation frequency, $\gamma_p$ is a small parameter used to heal the numerical singularity while calculating the Alfv\'en damping, $\xi$ is the fluid displacement, with subscript $\perp$ denoting the perpendicular component to the magnetic field, and $\boldsymbol{J}$, $\boldsymbol{B}$, and $P$ are the equilibrium current density, magnetic field, and plasma pressure, respectively.

\section{MD code}
The reduced MHD model implemented in the MD code is given as follows~\cite{weilai2016nf}
\begin{eqnarray}
\frac{\partial\psi}{\partial t} = [\psi,\phi]-\partial_z\phi-S^{-1}_A (j-j_b-j_d)+E_{z0} \label{NTMeq1}\\
\frac{\partial u}{\partial t} = [u,\phi]+[j,\psi]+\partial_z j+R^{-1}\nabla^2_\perp u \\
\frac{\partial p}{\partial t}=[p,\phi]+\chi_\parallel\nabla^2_\parallel p + \chi_\perp\nabla^2_\perp p +S_0 \label{NTMeq3} 
\end{eqnarray}
where $\psi$ and $\phi$ are the magnetic flux and electrostatic potential, $j=-\nabla^2_\perp\psi$ and $u=\nabla^2_\perp\phi$ are the current density and vorticity in the axial direction, respectively. The bootstrap current density is proportional to the pressure gradient as in $j_b=-f\frac{\sqrt{\varepsilon}}{B_\theta}\frac{\partial p}{\partial r}$, with $f$ measuring the strength of bootstrap current fraction, which is defined as $f_b=\int_0^a j_brdr/\int_0^a j_zrdr$. $S_A=\tau_\eta/\tau_A$ and $R=\tau_\nu/\tau_A$ are the magnetic Reynolds number and kinematic Reynolds number, respectively, where $\tau_\eta=a^2\mu_0/\eta$, $\tau_\nu=a^2/\nu$ and $\tau_A=\sqrt{\mu_0\rho}a/B_0$ are the resistive diffusion time, the viscous diffusion time, and the Alfv\'{e}n time, respectively. $\chi_\parallel$ and $\chi_\perp$ are the parallel and perpendicular transport coefficients. The source terms $E_{z0}=S_A^{-1}(j_0-j_{b0})$ and $S_0=-\chi_\perp\nabla^2_\perp p_0$ in equations (\ref{NTMeq1}) and (\ref{NTMeq3}) are chosen to balance the diffusion of equilibrium Ohm current and pressure, respectively. The length, time and velocity are normalized by the plasma minor radius $a$, Alfv\'en time $\tau_A$  and Alfv\'en velocity $V_A=B_0/\sqrt{\mu_0\rho}$ respectively. The Poisson bracket is defined as $[f,g]=\hat{z}\cdot\nabla f\times\nabla g$.

\section{TM8 code}
The TM8 code has been used to model the physics of the ECCD stabilization of NTM~\cite{yuqingquan2004pop2}, the drift-tearing modes~\cite{yuqingquan2010nf}, the double tearing modes~\cite{yuqingquan1996pop}, the mode coupling~\cite{yuqingquan2000nf}, the stochastic field~\cite{yuqingquan2006pop}, the resonant magnetic perturbation~\cite{yuqingquan2000prl}, and the error field~\cite{yuqingquan2008nf}. The corresponding simulation results compare well with experiments.

The reduced MHD model implemented in the TM8 code includes the Ohm’s law, the plasma vorticity equation, and the plasma pressure evolution equation~\cite{yuqingquan2000pop}
\begin{eqnarray}
\frac{\partial\psi}{\partial{t}}+\boldsymbol{v}\cdot\boldsymbol{\nabla}\psi=E-\eta(j_p-j_b-j_d)  \\
\rho(\frac{\partial}{\partial t}+\boldsymbol{v}\cdot\boldsymbol{\nabla})\nabla^2\Theta=\boldsymbol{e}_t\cdot(\boldsymbol{\nabla}\psi\times\boldsymbol{\nabla}j_p)+\rho\mu\nabla^4\Theta \\
\frac{3}{2}(\frac{\partial}{\partial t}+\boldsymbol{v}\cdot\boldsymbol{\nabla})p=\boldsymbol{\nabla}\cdot(\chi_{\parallel}\nabla_{\parallel}p)+\boldsymbol{\nabla}\cdot(\chi_{\perp}\nabla{\perp}{p})+Q
\end{eqnarray}
where $\boldsymbol{v}=\boldsymbol{\nabla}\Theta\times\boldsymbol{e}_t$, $\Theta$ is the stream function, $\boldsymbol{e}_t$ the unit vector in the toroidal direction, and $j_p=-\nabla^2\psi-2nB_{0t}/(mR)$, $j_b=-c_b\frac{\sqrt{\varepsilon}}{B_{\theta}}\frac{\partial p}{\partial r}$, and $j_d$ are the plasma current density, the bootstrap current density, and the current density driven by electron cyclotron wave (ECW) in the $\boldsymbol{e}_t$ direction, respectively. \\

\section{NIMROD/KPRAD code}
%NIMROD code
In the NIMROD code~\cite{sovinec2004jcp}, the $3D$ extended MHD model is coupled with an atomic and radiation physics model from the KPRAD code~\cite{Izzo2008,Izzo2013,KPRAD}, and the implemented equations for the coupled impurity-MHD model are as follows:
% MHD model equations
\begin{eqnarray}
\rho \frac{d\vec{V}}{dt} = - \nabla p + \vec{J} \times \vec{B} + \nabla \cdot (\rho \nu \nabla \vec{V})
\label{eq:momentum}
\\
\frac{d n_e}{dt} + n_e \nabla \cdot \vec{V} = \nabla \cdot (D \nabla n_e) + S_{ion/rec}
\label{eq:contiune1}
\\
\frac{d n_i}{dt} + n_i \nabla \cdot \vec{V} = \nabla \cdot (D \nabla n_i) + S_{ion/3-body}
\label{eq:contiune2}
\\
\frac{d n_Z}{dt} + n_Z \nabla \cdot \vec{V} = \nabla \cdot (D \nabla n_Z) + S_{ion/rec}
\label{eq:contiune3}
\\
n_e \frac{d T_e}{dt} = (\gamma - 1)[n_e T_e \nabla \cdot \vec{V} + \nabla \cdot \vec{q_e} - Q_{loss}]
\label{eq:temperature}
\\
\vec{q}_e = -n_e[\kappa_{\parallel} \hat{b} \hat{b} + \kappa_{\perp} (\mathcal{I} - \hat{b} \hat{b})] \cdot \nabla T_e
\label{eq:heat_flux}
\\
\vec{E} + \vec{V} \times \vec{B} = \eta \vec{j}
\label{eq:ohm}
\end{eqnarray}
% explanation equation variables
Here, $n_i$, $n_e$, and $n_Z$ are the main ion, electron, and impurity ion number density respectively, $\rho$, $\vec{V}$, $\vec{J}$, and $p$ the plasma mass density, velocity, current density, and pressure respectively, $T_e$ and $\vec{q}_e$ the electron temperature and heat flux respectively, $D$, $\nu$, $\eta$, and $\kappa_{\parallel} (\kappa_{\perp})$ the plasma diffusivity, kinematic viscosity, resistivity, and parallel (perpendicular) thermal conductivity respectively, $\gamma$ the adiabatic index, $S_{ion/rec}$ the density source from ionization and recombination, $S_{ion/3-body}$ also includes contribution from 3-body recombination, $Q_{loss}$ the energy loss, $\vec{E} (\vec{B})$ the electric (magnetic) field, $\hat{b}=\vec{B}/B$, and $\mathcal{I}$ the unit dyadic tensor.

%movement equation
All particle species share a single temperature $T=T_e$ and fluid velocity $\vec{V}$, which assumes instant thermal equilibration among the main ions, the impurity ions, and the electrons. Pressure $p$ and mass density $\rho$ in momentum equation (\ref{eq:momentum}) include contributions from the impurity species.
%continuity equation
Each charge state of impurity ion density is tracked in the KPRAD
module and used to update the source/sink terms in the continuity
equations due to ionization and recombination. Both convection and
diffusion terms are included in each continuity equations where all
the diffusivities are assumed same.  Quasi-neutrality is maintained
through the condition $n_e=n_i+\sum Z n_{z}$, where $Z$ is the charge
of impurity ion.
% KPRAD
The energy loss term $Q_{loss}$ in equation (\ref{eq:temperature}) is calculated from KPRAD module based on a coronal model, which includes contributions from bremsstrahlung, line radiation, ionization, recombination, Ohmic heating, and intrinsic impurity radiation.
Anisotropic thermal conductivities are temperature dependent, i.e. $\kappa_{\parallel} \propto T^{5/2}$ and $\kappa_{\perp} \propto T^{-1/2}$.
Similarly, the temperature dependence of resistivity $\eta$ is included through the Spitzer model.

\end{appendices}

%\section{References}
\bibliographystyle{spphys}       % APS-like style for physics
\bibliography{cfec_mhd}   % name your BibTeX data base

\begin{thebibliography}{10}
\providecommand{\url}[1]{{#1}}
\providecommand{\urlprefix}{URL }
\expandafter\ifx\csname urlstyle\endcsname\relax
  \providecommand{\doi}[1]{DOI \discretionary{}{}{}#1}\else
  \providecommand{\doi}{DOI \discretionary{}{}{}\begingroup
  \urlstyle{rm}\Url}\fi

\bibitem{aymar2002ppcf}
R.~Aymar, P.~Barabaschi, Y.~Shimomura, Plasma Physics and Controlled Fusion
  \textbf{44}(5), 519 (2002).
\newblock \doi{10.1088/0741-3335/44/5/304}

\bibitem{wan2014}
B.~{Wan}, S.~{Ding}, J.~{Qian}, G.~{Li}, B.~{Xiao}, G.~{Xu}, IEEE Transactions
  on Plasma Science \textbf{42}(3), 495 (2014).
\newblock \doi{10.1109/TPS.2013.2296939}

\bibitem{chan2015}
V.~Chan, A.~Costley, B.~Wan, A.~Garofalo, J.~Leuer, Nuclear Fusion
  \textbf{55}(2), 023017 (2015).
\newblock \doi{10.1088/0029-5515/55/2/023017}

\bibitem{shi2016}
N.~Shi, V.~Chan, Y.~Wan, J.~Li, X.~Gao, M.~Ye, Fusion Engineering and Design
  \textbf{112}, 47  (2016).
\newblock \doi{https://doi.org/10.1016/j.fusengdes.2016.07.017}

\bibitem{wanyuanxi2017nf}
Y.~Wan, J.~Li, Y.~Liu, X.~Wang, V.~Chan, C.~Chen, X.~Duan, P.~Fu, X.~Gao,
  K.~Feng, S.~Liu, Y.~Song, P.~Weng, B.~Wan, F.~Wan, H.~Wang, S.~Wu, M.~Ye,
  Q.~Yang, G.~Zheng, G.~Zhuang, Q.L. and, Nuclear Fusion \textbf{57}(10),
  102009 (2017).
\newblock \doi{10.1088/1741-4326/aa686a}

\bibitem{song2014}
Y.T. {Song}, S.T. {Wu}, J.G. {Li}, B.N. {Wan}, Y.X. {Wan}, P.~{Fu}, M.Y. {Ye},
  J.X. {Zheng}, K.~{Lu}, X.~{Gao}, S.M. {Liu}, X.F. {Liu}, M.Z. {Lei}, X.B.
  {Peng}, Y.~{Chen}, IEEE Transactions on Plasma Science \textbf{42}(3), 503
  (2014).
\newblock \doi{10.1109/TPS.2014.2299277}

\bibitem{lijiangang2019jfe}
J.~Li, Y.~Wan, Journal of Fusion Energy \textbf{38}(1), 113 (2019).
\newblock \doi{10.1007/s10894-018-0165-2}

\bibitem{zhuang2019}
G.~Zhuang, G.~Li, J.~Li, Y.~Wan, Y.~Liu, X.~Wang, Y.~Song, V.~Chan, Q.~Yang,
  B.~Wan, X.~Duan, P.~Fu, B.X. and, Nuclear Fusion \textbf{59}(11), 112010
  (2019).
\newblock \doi{10.1088/1741-4326/ab0e27}

\bibitem{nehl2019jfe}
C.L. Nehl, R.J. Umstattd, W.R. Regan, S.C. Hsu, P.B. McGrath, Journal of Fusion
  Energy \textbf{38}(5), 506 (2019)

\bibitem{BETHE}
Bethe website.
\newblock https://arpa-e.energy.gov/?q=arpa-e-programs/bethe

\bibitem{INFUSE}
Infuse website.
\newblock https://infuse.ornl.gov

\bibitem{steinhauer2011pop}
L.C. Steinhauer, Physics of Plasmas \textbf{18}(7), 070501 (2011).
\newblock \doi{10.1063/1.3613680}

\bibitem{guohouyang2015nc}
H.Y. Guo, M.W. Binderbauer, T.~Tajima, R.D. Milroy, L.C. Steinhauer, X.~Yang,
  E.G. Garate, H.~Gota, S.~Korepanov, A.~Necas, T.~Roche, A.~Smirnov, E.~Trask,
  Nature Communications \textbf{6}(1) (2015).
\newblock \doi{10.1038/ncomms7897}

\bibitem{tuszewski1988nf}
M.~Tuszewski, Nuclear Fusion \textbf{28}(11), 2033 (1988).
\newblock \doi{10.1088/0029-5515/28/11/008}

\bibitem{Jian2017}
X.~Jian, J.~Chen, V.S. Chan, G.~Zhuang, G.~Li, Z.~Deng, N.~Shi, G.~Xu, G.M.
  Staebler, W.~Guo, Nuclear Fusion \textbf{57}(4), 046012 (2017).
\newblock \doi{10.1088/1741-4326/aa5bd5}

\bibitem{Chen2017}
J.~Chen, X.~Jian, V.S. Chan, Z.~Li, Z.~Deng, G.~Li, W.~Guo, N.~Shi, X.~Chen,
  C.P. Team, Plasma Physics and Controlled Fusion \textbf{59}(7), 075005
  (2017).
\newblock \doi{10.1088/1361-6587/aa6d20}

\bibitem{howell14a}
E.~Howell, C.~Sovinec, Computer Physics Communications \textbf{185}(5), 1415
  (2014).
\newblock \doi{https://doi.org/10.1016/j.cpc.2014.02.008}

\bibitem{lihl20a}
H.~Li, P.~Zhu, Computer Physics Communications \textbf{260}, 107264 (2021).
\newblock \doi{https://doi.org/10.1016/j.cpc.2020.107264}

\bibitem{troyon1984ppcf}
F.~Troyon, R.~Gruber, H.~Saurenmann, S.~Semenzato, S.~Succi, Plasma Physics and
  Controlled Fusion \textbf{26}(1A), 209 (1984).
\newblock \doi{10.1088/0741-3335/26/1a/319}

\bibitem{liuyueqiang2000pop}
Y.Q. Liu, A.~Bondeson, C.M. Fransson, B.~Lennartson, C.~Breitholtz, Physics of
  Plasmas \textbf{7}(9), 3681 (2000).
\newblock \doi{10.1063/1.1287744}

\bibitem{zheng2006jcp}
L.J. Zheng, M.~Kotschenreuther, Journal of Computational Physics
  \textbf{211}(2), 748  (2006).
\newblock \doi{https://doi.org/10.1016/j.jcp.2005.06.009}

\bibitem{lutjens96a}
H.~Lütjens, A.~Bondeson, O.~Sauter, Computer Physics Communications
  \textbf{97}(3), 219  (1996).
\newblock \doi{https://doi.org/10.1016/0010-4655(96)00046-X}

\bibitem{libo2019fed}
B.~Li, L.~Liu, Y.~Guo, L.~Xue, S.~Chen, Y.~Wang, Y.~Liu, X.~Song, S.~Wang,
  X.~Song, D.~Chen, Y.~Huang, J.~Zhang, Fusion Engineering and Design
  \textbf{148}, 111295 (2019).
\newblock \doi{https://doi.org/10.1016/j.fusengdes.2019.111295}

\bibitem{fitzpatrick1998pop}
R.~Fitzpatrick, Physics of Plasmas \textbf{5}(9), 3325 (1998)

\bibitem{wanghuihui2015pst}
H.~Wang, Z.X. Wang, Y.H. Ding, B.~Rao, Plasma Science and Technology
  \textbf{17}(7), 539 (2015)

\bibitem{cole2007prl}
A.~Cole, C.~Hegna, J.~Callen, Physical Review Letters \textbf{99}(6), 065001
  (2007)

\bibitem{fitzpatrick2012ppcf}
R.~Fitzpatrick, Plasma Physics and Controlled Fusion \textbf{54}(9), 094002
  (2012)

\bibitem{wanghuihui2018nf}
H.H. Wang, Y.W. Sun, T.H. Shi, Q.~Zang, Y.Q. Liu, X.~Yang, S.~Gu, K.Y. He,
  X.~Gu, J.P. Qian, et~al., Nuclear Fusion \textbf{58}(5), 056024 (2018)

\bibitem{wanghuihui2020nf}
H.~Wang, Y.~Sun, et~al, submitted to Nuclear Fusion  (2020)

\bibitem{hender1992nf}
T.~Hender, R.~Fitzpatrick, A.~Morris, P.~Carolan, R.~Durst, T.~Edlington,
  J.~Ferreira, S.~Fielding, P.~Haynes, J.~Hugill, et~al., Nuclear Fusion
  \textbf{32}(12), 2091 (1992)

\bibitem{buttery1999nf}
R.~Buttery, M.~De'Benedetti, D.A. Gates, Y.~Gribov, T.~Hender, R.~La~Haye,
  P.~Leahy, J.~Leuer, A.~Morris, A.~Santagiustina, et~al., Nuclear Fusion
  \textbf{39}(11Y), 1827 (1999)

\bibitem{lazzaro2002pop}
E.~Lazzaro, R.~Buttery, T.~Hender, P.~Zanca, R.~Fitzpatrick, M.~Bigi,
  T.~Bolzonella, R.~Coelho, M.~DeBenedetti, S.~Nowak, et~al., Physics of
  Plasmas \textbf{9}(9), 3906 (2002)

\bibitem{westerhof2014pop}
E.~Westerhof, J.~Pratt, Physics of Plasmas \textbf{21}(10), 2349 (2014)

\bibitem{sovinec2004jcp}
C.~Sovinec, A.~Glasser, T.~Gianakon, D.~Barnes, R.~Nebel, S.~Kruger,
  D.~Schnack, S.~Plimpton, A.~Tarditi, M.~Chu, Journal of Computational Physics
  \textbf{195}(1), 355 (2004).
\newblock \doi{10.1016/j.jcp.2003.10.004}.
\newblock 00000

\bibitem{guo_achieving_2015}
H.Y. Guo, M.W. Binderbauer, T.~Tajima, R.D. Milroy, L.C. Steinhauer, X.~Yang,
  E.G. Garate, H.~Gota, S.~Korepanov, A.~Necas, T.~Roche, A.~Smirnov, E.~Trask,
  Nature Communications \textbf{6}, 6897 (2015).
\newblock \doi{10.1038/ncomms7897}

\bibitem{Barnes2003a}
D.C. Barnes, E.V. Belova, R.C. Davidson,  (Lyon France, 2003).
\newblock Number: IAEA-CSP--19/CD

\bibitem{reimerdes2007prl}
H.~Reimerdes, A.M. Garofalo, G.L. Jackson, M.~Okabayashi, E.J. Strait, M.S.
  Chu, Y.~In, R.J. La~Haye, M.J. Lanctot, Y.Q. Liu, G.A. Navratil, W.M.
  Solomon, H.~Takahashi, R.J. Groebner, Physical Review Letters \textbf{98},
  055001 (2007).
\newblock \doi{10.1103/PhysRevLett.98.055001}

\bibitem{villone2008prl}
F.~Villone, Y.~Liu, R.~Paccagnella, T.~Bolzonella, G.~Rubinacci, Physical
  Review Letters \textbf{100}(25), 255005 (2008)

\bibitem{liuyueqiang2009ppcf}
Y.~Liu, I.T. Chapman, S.~Saarelma, M.P. Gryaznevich, T.C. Hender, D.F.H. and,
  Plasma Physics and Controlled Fusion \textbf{51}(11), 115005 (2009).
\newblock \doi{10.1088/0741-3335/51/11/115005}

\bibitem{okabayashi2011pop}
M.~Okabayashi, G.~Matsunaga, J.S. deGrassie, W.W. Heidbrink, Y.~In, Y.Q. Liu,
  H.~Reimerdes, W.M. Solomon, E.J. Strait, M.~Takechi, N.~Asakura, R.V. Budny,
  G.L. Jackson, J.M. Hanson, R.J. La~Haye, M.J. Lanctot, J.~Manickam,
  K.~Shinohara, Y.B. Zhu, Physics of Plasmas \textbf{18}(5), 056112 (2011).
\newblock \doi{10.1063/1.3575159}

\bibitem{wang2015prl}
Z.R. Wang, M.J. Lanctot, Y.~Liu, J.K. Park, J.E. Menard, Physical Review
  Letters \textbf{114}(14), 145005 (2015)

\bibitem{zheng2005prl}
L.J. Zheng, M.~Kotschenreuther, M.S. Chu, Physical Review Letters \textbf{95},
  255003 (2005).
\newblock \doi{10.1103/PhysRevLett.95.255003}

\bibitem{hanrui2020ppcf}
R.~Han, P.~Zhu, D.~Banerjee, S.~Cheng, X.~Yan, L.~Zheng, Plasma Physics and
  Controlled Fusion  (2020)

\bibitem{weilai2016nf}
L.~Wei, Z.X. Wang, J.~Wang, X.~Yang, Nuclear Fusion \textbf{56}(10), 106015.1
  (2016)

\bibitem{yuqingquan2004pop2}
Q.~Yu, X.~Zhang, S.~G{\"u}nter, Physics of Plasmas \textbf{11}(5), 1960 (2004)

\bibitem{yuqingquan2010nf}
Q.~Yu, Nuclear Fusion \textbf{50}(2), 025014 (2010)

\bibitem{yuqingquan1996pop}
Q.~Yu, Physics of Plasmas \textbf{3}(8), 2898 (1996)

\bibitem{yuqingquan2000nf}
Q.~Yu, S.~G{\"u}nter, K.~Lackner, A.~Gude, M.~Maraschek, Nuclear Fusion
  \textbf{40}(12), 2031 (2000)

\bibitem{yuqingquan2006pop}
Q.~Yu, Physics of Plasmas \textbf{13}(6), 062310 (2006)

\bibitem{yuqingquan2000prl}
Q.~Yu, S.~G{\"u}nter, K.~Lackner, Physical Review Letters \textbf{85}(14), 2949
  (2000)

\bibitem{yuqingquan2008nf}
Q.~Yu, S.~G{\"u}nter, Nuclear Fusion \textbf{48}(6), 065004 (2008)

\bibitem{yuqingquan2000pop}
Q.~Yu, S.~G{\"u}nter, G.~Giruzzi, K.~Lackner, M.~Zabiego, Physics of Plasmas
  \textbf{7}(1), 312 (2000)

\bibitem{Izzo2008}
V.A. Izzo, D.G. Whyte, R.S. Granetz, P.B. Parks, E.M. Hollmann, L.L. Lao, J.C.
  Wesley, Physics of Plasmas \textbf{15}(5) (2008).
\newblock \doi{10.1063/1.2841526}

\bibitem{Izzo2013}
V.A. Izzo, Physics of Plasmas \textbf{20}(5) (2013).
\newblock \doi{10.1063/1.4803896}

\bibitem{KPRAD}
D.G. Whyte, T.E. Evans, A.G. Kellman, D.A. Humphreys, A.W. Hyatt, T.C.
  Jernigan, R.L. Lee, S.L. Luckhardt, P.B. Parks, M.J. Schaffer, P.L. Taylor,
  Proceedings of the 24th European Conference on Controlled Fusion and Plasma
  Physics, 9–14 June 1996, Berchtesgaden, Germany(European Physical Society,
  Geneva, 1997) \textbf{21A}, p. 1137 (1997)

\end{thebibliography}

%% Non-BibTeX users please use
%\begin{thebibliography}{}
%%
%% and use \bibitem to create references. Consult the Instructions
%% for authors for reference list style.
%\end{thebibliography}

%Fig 1
\newpage
\clearpage

\newpage
%---------
%table 1
%---------
%\newpage
%\begin{center}
\begin{table}
\begin{tabular}{lllllllll}
\hline\noalign{\smallskip}
equilibrium & $R_0$ & $B_0$ & $a$ & $q_{min}$ & $q_0$ & $q_{95}$ & $q_a$ & $\beta_N$  \\
\noalign{\smallskip}\hline\noalign{\smallskip}
1 & 7.2 & 6.53 & 2.2358 & 1.2443 & 2.3663 & 6.0235 & 7.45 & 2.3799 \\
2 & 7.2 & 6.53 & 2.1990 & 1.2049 & 1.9571 & 5.4053 & 7.45 & 2.3277 \\
3 & 7.2 & 6.53 & 2.2117 & 3.0493 & 4.6612 & 6.8730 & 9.05 & 3.0751 \\
4 & 7.2 & 6.53 & 2.1833 & 3.2566 & 6.2540 & 5.7156 & 7.2912 & 2.3781 \\
5 & 7.2 & 6.53 & 2.2273 & 2.4089 & 5.4913 & 7.3602 & 9.0954 & 2.8720 \\
\noalign{\smallskip}\hline
\end{tabular}
\caption{Summary of key design parameters for the five CFETR SSO scenarios considered in this study.}
%\label{rwmeqtable}       % Give a unique label
\label{tableRWMeq}
\end{table}
%\end{center}
\clearpage

%Fig. 2
\newpage
\clearpage

%HFRC-figures
%-----------------------------------------------------------------------------------
%Fig 3
%-----------------------------------------------------------------------------------
\newpage

%Fig 4
\newpage
\clearpage

%-----------------------------------------------------------------------------------
%Fig 5 
%-----------------------------------------------------------------------------------
\newpage
\clearpage

%Fig 6
\newpage
\clearpage

%Fig 7
\newpage

%Fig 8
\newpage
\clearpage

%Fig 9
\newpage
\clearpage

%Fig 10
\newpage
\clearpage

%Fig 11
\newpage
\clearpage

%Fig 12
\newpage
\clearpage

%Fig 13
\newpage
\clearpage

%Fig 14
\newpage
\clearpage

%Fig 15
\newpage
\clearpage

%\begin{figure}[htbp]
%%\centering
%\includegraphics[width=0.7\textwidth]{ntm_nimfig1.png}
%\caption{Evolution of $n=1$ component of magnetic energy in NIMROD simulation.}
%\label{ntm_nimfig1}
%\end{figure}
%
%\begin{figure}[htbp]
%%\centering
%\includegraphics[width=0.7\textwidth]{ntm_nimfig2.png}
%\caption{Poincare plot at saturation phase in NIMROD simulation.}
%\label{ntm_nimfig2}
%\end{figure}

%\newpage
%% For two-column wide figures use
%\begin{figure}
%% Use the relevant command to insert your figure file.
%% For example, with the graphicx package use
%  \includegraphics[width=0.9\textwidth]{errorfield.jpg}
%\caption{Extrapolation of error field tolerance towards ITER and CFETR using theoretical and experimental scalings. SOC indicates the assumption using saturated ohmic confinement (energy confinement time independent with electron density, we assumed viscosity diffusion time approaches to energy confinement time here), whereas LOC indicates the assumption using linear ohmic confinement (energy confinement time linear dependent with electron density).}
%\label{fig_ef}       % Give a unique label
%\end{figure}
%\clearpage

\newpage
\begin{table}[]
\centering
\begin{tabular}{|c|c|l|l|c|c|}
\hline
\multicolumn{2}{|c|}{\textbf{Extrapolation}}                      & \textbf{Physical   Regime/Device} & \multicolumn{1}{c|}{\textbf{scaling(br/BT)}}                                                           & \textbf{ITER(1e-4/BT)} & \textbf{CFETR(1e-4/BT)} \\ \hline
\multirow{12}{*}{\textbf{Theory}} & \multirow{6}{*}{\textbf{SOC}} & \textbf{visco-resistive}          & ne\textasciicircum{}(7/12)BT\textasciicircum{}(-7/6)R0\textasciicircum{}(1/2)f0                        & \textbf{1.20}          & \textbf{1.30}           \\ \cline{3-6} 
                                  &                               & \textbf{Rutherford}               & ne\textasciicircum{}(3/5)BT\textasciicircum{}(-6/5)R0\textasciicircum{}(11/15)f0                       & \textbf{1.61}          & \textbf{1.82}           \\ \cline{3-6} 
                                  &                               & \textbf{transition}                & ne\textasciicircum{}(1/2)BT\textasciicircum{}(-1)R0\textasciicircum{}(1/3)f0\textasciicircum{}(4/5)    & \textbf{0.91}          & \textbf{0.94}           \\ \cline{3-6} 
                                  &                               & \textbf{Waelbroeck}               & ne\textasciicircum{}(7/16)BT\textasciicircum{}(-7/8)R0\textasciicircum{}(1/3)f0\textasciicircum{}(5/8) & \textbf{0.85}          & \textbf{0.87}           \\ \cline{3-6} 
                                  &                               & \textbf{polarization}             & neBT\textasciicircum{}(-9/5)R0\textasciicircum{}(-1/4)                                                 & \textbf{0.37}          & \textbf{0.24}           \\ \cline{3-6} 
                                  &                               & \textbf{Sci+NTV}                  & neBT\textasciicircum{}(-13/10)R0                                                                       & \textbf{3.12}          & \textbf{2.94}           \\ \cline{2-6} 
                                  & \multirow{6}{*}{\textbf{LOC}} & \textbf{visco-resistive}          & BT\textasciicircum{}(-7/6)R0\textasciicircum{}(1/2)f0                                                  & \textbf{0.35}          & \textbf{0.38}           \\ \cline{3-6} 
                                  &                               & \textbf{Rutherford}               & ne\textasciicircum{}(-1/60)BT\textasciicircum{}(-6/5)R0\textasciicircum{}(11/15)f0                     & \textbf{0.44}          & \textbf{0.50}           \\ \cline{3-6} 
                                  &                               & \textbf{transition}                & BT\textasciicircum{}(-1)R0\textasciicircum{}(1/3)f0\textasciicircum{}(4/5)                             & \textbf{0.32}          & \textbf{0.33}           \\ \cline{3-6} 
                                  &                               & \textbf{Waelbroeck}               & BT\textasciicircum{}(-7/8)R0\textasciicircum{}(1/3)f0\textasciicircum{}(5/8)                           & \textbf{0.34}          & \textbf{0.35}           \\ \cline{3-6} 
                                  &                               & \textbf{polarization}             & neBT\textasciicircum{}(-9/5)R0\textasciicircum{}(-1/4)                                                 & \textbf{0.37}          & \textbf{0.24}           \\ \cline{3-6} 
                                  &                               & \textbf{Sci+NTV}                  & ne\textasciicircum{}(1/2)BT\textasciicircum{}(-13/10)R0                                                & \textbf{1.09}          & \textbf{1.03}           \\ \hline
\multicolumn{2}{|c|}{\multirow{6}{*}{\textbf{Experiment}}}        & \textbf{EAST}                     & ne\textasciicircum{}(0.55)BT\textasciicircum{}(-1.0)q95\textasciicircum{}(1.66)f0                      & \textbf{0.62}          & \textbf{1.54}           \\ \cline{3-6} 
\multicolumn{2}{|c|}{}                                            & \textbf{JET-2000}                 & ne\textasciicircum{}(0.58)BT\textasciicircum{}(-1.274)f0\textasciicircum{}(0.5)                        & \textbf{1.29}          & \textbf{1.10}           \\ \cline{3-6} 
\multicolumn{2}{|c|}{}                                            & \textbf{JET-98}                   & ne\textasciicircum{}(0.97)BT\textasciicircum{}(-1.2)f0                                                 & \textbf{2.84}          & \textbf{2.76}           \\ \cline{3-6} 
\multicolumn{2}{|c|}{}                                            & \textbf{COMPASS-C}                & ne\textasciicircum{}(0.55)BT\textasciicircum{}(-2.2)f0                                                 & \textbf{0.19}          & \textbf{0.12}           \\ \cline{3-6} 
\multicolumn{2}{|c|}{}                                            & \textbf{COMPASS-D}                & ne\textasciicircum{}(1.0)BT\textasciicircum{}(-2.9)f0                                                  & \textbf{0.16}          & \textbf{0.09}           \\ \cline{3-6} 
\multicolumn{2}{|c|}{}                                            & \textbf{DIII-D}                   & ne\textasciicircum{}(0.99)BT\textasciicircum{}(-0.96)f0                                                & \textbf{2.02}          & \textbf{1.68}           \\ \hline
\end{tabular}
\caption{Extrapolation of error field tolerance towards ITER and CFETR using theoretical and experimental scalings. SOC indicates the assumption using saturated Ohmic confinement (energy confinement time independent with electron density, we assumes that the viscous diffusion time approaches to the energy confinement time here), whereas LOC indicates the assumption using linear Ohmic confinement (energy confinement time linear dependent with electron density).}
\label{table_ef}
\end{table}
\clearpage

%Fig 16
\newpage
\clearpage

%\newpage
%\begin{figure}
%  \includegraphics[width=0.7\textwidth]{eccdfig1_eqhy1.jpg}
%\caption{Safety factor profile and pressure profile for the hybrid scenarios EQ1.}
%\label{eccdfig1}      
%\end{figure}

%Fig 17
\newpage
\clearpage

%Fig 18
\newpage
\clearpage

%\newpage
%\begin{figure}
%  \includegraphics[width=0.7\textwidth]{eccdfig4_NTMsuppression.jpg}
%\caption{Dependence of the driven current for the suppression of NTM on the magnetic island width.}
%\label{eccdfig4}      
%\end{figure}
%\clearpage

%\newpage
%\begin{figure}
%  \includegraphics[width=0.7\textwidth]{eccdfig5_eqhy3.jpg}
%\caption{Safety factor profile and pressure profile for the hybrid scenarios EQ3.}
%\label{eccdfig5}      
%\end{figure}

%Fig 19
\newpage

%\newpage
%\begin{figure}
%  \includegraphics[width=0.7\textwidth]{TM8fig2.png}
%  \caption{Similar to Fig.~\ref{tmfig1}, time evolution of $w/a$ for NMCD.}
%  \label{tmfig2}      
%\end{figure}
%\clearpage

%Fig 20
\newpage

%\newpage
%\begin{figure}
%  \includegraphics[width=0.7\textwidth]{TM8fig4.png}
%  \caption{Required $I_{cd}/I_p$ for mode stabilization versus the normalized island width. The solid (dashed) curve is for MCD (NMCD).}
%  \label{tmfig4}      
%\end{figure}
%\clearpage

%Fig 21
\newpage
\clearpage

%Fig 22
\newpage
\clearpage

%Fig 23
\newpage
\clearpage

%Fig 24
\newpage
\clearpage

%Fig 25
\newpage
\clearpage

%Fig 26
\newpage
\clearpage

%\begin{acknowledgements}
%\end{acknowledgements}


\end{document}